\documentclass[longauth]{aa}

\usepackage{changepage}
\usepackage{graphicx}
\usepackage{float}
\usepackage{amsmath}
\usepackage[varg]{txfonts}
\usepackage{booktabs,threeparttable,siunitx}
\usepackage{natbib}
\usepackage{scalerel}

\usepackage[table]{xcolor}
\usepackage{tikz}
\usetikzlibrary{positioning}
\usepackage{tabularx}

\usepackage[pdfencoding=auto,psdextra]{hyperref}
\newcolumntype{L}{>{\raggedright\arraybackslash}X}

\allowdisplaybreaks[2]
\newcommand{\Mvir}{M_{\rm vir}}
\newcommand{\rvir}{r_{\rm vir}}

\newcommand{\rh}{r_{\rm h}}
\newcommand{\fb}{f_{\rm b}}

\newcommand{\fcga}{f_{\rm cga}}
\newcommand{\fsga}{f_{\rm sga}}

\usepackage[nameinlink,noabbrev]{cleveref}

\Crefname{section}{Section}{Sections}
\crefname{section}{Sect.}{Sects.}
\Crefname{figure}{Figure}{Figures}
\crefname{figure}{Fig.}{Figs.}
\Crefname{table}{Table}{Tables}
\crefname{table}{Table}{Tables}
\Crefname{equation}{Equation}{Equations}
\crefname{equation}{Eq.}{Eqs.}
\newcommand{\crefnp}[1]{%
  \crefformat{equation}{Eq.~##2##1##3}%
  \crefmultiformat{equation}{Eqs.~##2##1##3}{, and~##2##1##3}{, ##2##1##3}{, and~##2##1##3}%
  \cref{#1}%
  \crefformat{equation}{Eq.~(##2##1##3)}%
  \crefmultiformat{equation}{Eqs.~(##2##1##3)}{, and~(##2##1##3)}{, (##2##1##3)}{, and~(##2##1##3)}%
}

\defcitealias{2016A&A...595A.111C}{CB16}

\usepackage{placeins}

\usepackage{dblfloatfix}
\usepackage{caption}

\tikzset{
  inputnode/.style={thick, draw=black, fill=blue!20, circle, minimum size=22, inner sep=1pt},
  hiddennode/.style={thick, draw=black, fill=green!20, circle, minimum size=22, inner sep=1pt},
  outputnode/.style={thick, draw=black, fill=pink!20, circle, minimum size=22, inner sep=1pt}
}

\hypersetup{
    colorlinks=true,
    linkcolor=blue,
    filecolor=magenta,
    urlcolor=blue,
    citecolor=blue
}
\makeatletter
\renewcommand*\aa@pageof{, page \thepage{} of \pageref*{LastPage}}
\makeatother

\usepackage[utf8]{inputenc}

\usepackage{euclid}

\begin{document}

\title{\Euclid preparation}
\subtitle{CV. CosmoPostProcess: A simulation calibrated framework for weak lensing selection bias in richness-selected galaxy clusters}

\newcommand{\orcid}[1]{} 
\author{Euclid Collaboration: R.~Ingrao\thanks{\email{ROBERTO.INGRAO@phd.units.it}}\inst{\ref{aff1},\ref{aff2},\ref{aff3},\ref{aff4}}
\and M.~Costanzi\orcid{0000-0001-8158-1449}\inst{\ref{aff1},\ref{aff3},\ref{aff2}}
\and T.~Castro\orcid{0000-0002-6292-3228}\inst{\ref{aff3},\ref{aff4},\ref{aff2},\ref{aff5},\ref{aff6}}
\and A.~Saro\orcid{0000-0002-9288-862X}\inst{\ref{aff1},\ref{aff2},\ref{aff3},\ref{aff4},\ref{aff6}}
\and S.~Borgani\orcid{0000-0001-6151-6439}\inst{\ref{aff1},\ref{aff2},\ref{aff3},\ref{aff4},\ref{aff6}}
\and L.~Baumont\orcid{0000-0002-1518-0150}\inst{\ref{aff1},\ref{aff3},\ref{aff2}}
\and M.~Aguena\orcid{0000-0001-5679-6747}\inst{\ref{aff3}}
\and C.~Murray\orcid{0000-0002-4668-1273}\inst{\ref{aff7}}
\and S.~Bhargava\orcid{0000-0003-3851-7219}\inst{\ref{aff8}}
\and S.~Grandis\orcid{0000-0002-4577-8217}\inst{\ref{aff9}}
\and E.~Munari\orcid{0000-0002-1751-5946}\inst{\ref{aff3},\ref{aff2}}
\and B.~Altieri\orcid{0000-0003-3936-0284}\inst{\ref{aff10}}
\and S.~Andreon\orcid{0000-0002-2041-8784}\inst{\ref{aff11}}
\and N.~Auricchio\orcid{0000-0003-4444-8651}\inst{\ref{aff12}}
\and C.~Baccigalupi\orcid{0000-0002-8211-1630}\inst{\ref{aff2},\ref{aff3},\ref{aff4},\ref{aff13}}
\and M.~Baldi\orcid{0000-0003-4145-1943}\inst{\ref{aff14},\ref{aff12},\ref{aff15}}
\and S.~Bardelli\orcid{0000-0002-8900-0298}\inst{\ref{aff12}}
\and P.~Battaglia\orcid{0000-0002-7337-5909}\inst{\ref{aff12}}
\and A.~Biviano\orcid{0000-0002-0857-0732}\inst{\ref{aff3},\ref{aff2}}
\and E.~Branchini\orcid{0000-0002-0808-6908}\inst{\ref{aff16},\ref{aff17},\ref{aff11}}
\and M.~Brescia\orcid{0000-0001-9506-5680}\inst{\ref{aff18},\ref{aff19}}
\and S.~Camera\orcid{0000-0003-3399-3574}\inst{\ref{aff20},\ref{aff21},\ref{aff22}}
\and V.~Capobianco\orcid{0000-0002-3309-7692}\inst{\ref{aff22}}
\and C.~Carbone\orcid{0000-0003-0125-3563}\inst{\ref{aff23}}
\and J.~Carretero\orcid{0000-0002-3130-0204}\inst{\ref{aff24},\ref{aff25}}
\and M.~Castellano\orcid{0000-0001-9875-8263}\inst{\ref{aff26}}
\and G.~Castignani\orcid{0000-0001-6831-0687}\inst{\ref{aff12}}
\and S.~Cavuoti\orcid{0000-0002-3787-4196}\inst{\ref{aff19},\ref{aff27}}
\and A.~Cimatti\inst{\ref{aff28}}
\and C.~Colodro-Conde\inst{\ref{aff29}}
\and G.~Congedo\orcid{0000-0003-2508-0046}\inst{\ref{aff30}}
\and L.~Conversi\orcid{0000-0002-6710-8476}\inst{\ref{aff31},\ref{aff10}}
\and Y.~Copin\orcid{0000-0002-5317-7518}\inst{\ref{aff32}}
\and F.~Courbin\orcid{0000-0003-0758-6510}\inst{\ref{aff33},\ref{aff34},\ref{aff35}}
\and H.~M.~Courtois\orcid{0000-0003-0509-1776}\inst{\ref{aff36}}
\and H.~Degaudenzi\orcid{0000-0002-5887-6799}\inst{\ref{aff37}}
\and G.~De~Lucia\orcid{0000-0002-6220-9104}\inst{\ref{aff3}}
\and F.~Dubath\orcid{0000-0002-6533-2810}\inst{\ref{aff37}}
\and X.~Dupac\inst{\ref{aff10}}
\and S.~Escoffier\orcid{0000-0002-2847-7498}\inst{\ref{aff38}}
\and M.~Farina\orcid{0000-0002-3089-7846}\inst{\ref{aff39}}
\and R.~Farinelli\inst{\ref{aff12}}
\and S.~Farrens\orcid{0000-0002-9594-9387}\inst{\ref{aff40}}
\and S.~Ferriol\inst{\ref{aff32}}
\and F.~Finelli\orcid{0000-0002-6694-3269}\inst{\ref{aff12},\ref{aff41}}
\and P.~Fosalba\orcid{0000-0002-1510-5214}\inst{\ref{aff42},\ref{aff43}}
\and M.~Frailis\orcid{0000-0002-7400-2135}\inst{\ref{aff3}}
\and E.~Franceschi\orcid{0000-0002-0585-6591}\inst{\ref{aff12}}
\and M.~Fumana\orcid{0000-0001-6787-5950}\inst{\ref{aff23}}
\and K.~George\orcid{0000-0002-1734-8455}\inst{\ref{aff44}}
\and B.~Gillis\orcid{0000-0002-4478-1270}\inst{\ref{aff30}}
\and C.~Giocoli\orcid{0000-0002-9590-7961}\inst{\ref{aff12},\ref{aff15}}
\and J.~Gracia-Carpio\orcid{0000-0003-4689-3134}\inst{\ref{aff45}}
\and A.~Grazian\orcid{0000-0002-5688-0663}\inst{\ref{aff46}}
\and F.~Grupp\inst{\ref{aff45},\ref{aff47}}
\and S.~V.~H.~Haugan\orcid{0000-0001-9648-7260}\inst{\ref{aff48}}
\and S.~Hemmati\orcid{0000-0003-2226-5395}\inst{\ref{aff49}}
\and H.~Hoekstra\orcid{0000-0002-0641-3231}\inst{\ref{aff50}}
\and W.~Holmes\orcid{0009-0007-8554-4646}\inst{\ref{aff51}}
\and F.~Hormuth\inst{\ref{aff52}}
\and A.~Hornstrup\orcid{0000-0002-3363-0936}\inst{\ref{aff53},\ref{aff54}}
\and K.~Jahnke\orcid{0000-0003-3804-2137}\inst{\ref{aff55}}
\and M.~Jhabvala\inst{\ref{aff56}}
\and B.~Joachimi\orcid{0000-0001-7494-1303}\inst{\ref{aff57}}
\and S.~Kermiche\orcid{0000-0002-0302-5735}\inst{\ref{aff38}}
\and A.~Kiessling\orcid{0000-0002-2590-1273}\inst{\ref{aff51}}
\and B.~Kubik\orcid{0009-0006-5823-4880}\inst{\ref{aff32}}
\and H.~Kurki-Suonio\orcid{0000-0002-4618-3063}\inst{\ref{aff58},\ref{aff59}}
\and A.~M.~C.~Le~Brun\orcid{0000-0002-0936-4594}\inst{\ref{aff60}}
\and S.~Ligori\orcid{0000-0003-4172-4606}\inst{\ref{aff22}}
\and P.~B.~Lilje\orcid{0000-0003-4324-7794}\inst{\ref{aff48}}
\and V.~Lindholm\orcid{0000-0003-2317-5471}\inst{\ref{aff58},\ref{aff59}}
\and I.~Lloro\orcid{0000-0001-5966-1434}\inst{\ref{aff61}}
\and M.~Magliocchetti\orcid{0000-0001-9158-4838}\inst{\ref{aff39}}
\and G.~Mainetti\orcid{0000-0003-2384-2377}\inst{\ref{aff62}}
\and E.~Maiorano\orcid{0000-0003-2593-4355}\inst{\ref{aff12}}
\and O.~Mansutti\orcid{0000-0001-5758-4658}\inst{\ref{aff3}}
\and O.~Marggraf\orcid{0000-0001-7242-3852}\inst{\ref{aff63}}
\and M.~Martinelli\orcid{0000-0002-6943-7732}\inst{\ref{aff26},\ref{aff64}}
\and N.~Martinet\orcid{0000-0003-2786-7790}\inst{\ref{aff65}}
\and F.~Marulli\orcid{0000-0002-8850-0303}\inst{\ref{aff66},\ref{aff12},\ref{aff15}}
\and R.~J.~Massey\orcid{0000-0002-6085-3780}\inst{\ref{aff67}}
\and N.~Mauri\orcid{0000-0001-8196-1548}\inst{\ref{aff28},\ref{aff15}}
\and S.~Maurogordato\inst{\ref{aff8}}
\and E.~Medinaceli\orcid{0000-0002-4040-7783}\inst{\ref{aff12}}
\and S.~Mei\orcid{0000-0002-2849-559X}\inst{\ref{aff7},\ref{aff68}}
\and M.~Meneghetti\orcid{0000-0003-1225-7084}\inst{\ref{aff12},\ref{aff15}}
\and E.~Merlin\orcid{0000-0001-6870-8900}\inst{\ref{aff46}}
\and G.~Meylan\orcid{0000-0001-6503-0209}\inst{\ref{aff69}}
\and P.~Monaco\orcid{0000-0003-2083-7564}\inst{\ref{aff1},\ref{aff3},\ref{aff4},\ref{aff2}}
\and A.~Mora\orcid{0000-0002-1922-8529}\inst{\ref{aff70}}
\and M.~Moresco\orcid{0000-0002-7616-7136}\inst{\ref{aff66},\ref{aff12}}
\and C.~Moretti\orcid{0000-0003-3314-8936}\inst{\ref{aff3},\ref{aff2},\ref{aff4}}
\and L.~Moscardini\orcid{0000-0002-3473-6716}\inst{\ref{aff66},\ref{aff12},\ref{aff15}}
\and C.~T.~Mpetha\orcid{0000-0002-7805-2500}\inst{\ref{aff56}}
\and C.~Neissner\orcid{0000-0001-8524-4968}\inst{\ref{aff71},\ref{aff25}}
\and S.-M.~Niemi\orcid{0009-0005-0247-0086}\inst{\ref{aff72}}
\and C.~Padilla\orcid{0000-0001-7951-0166}\inst{\ref{aff71}}
\and S.~Paltani\orcid{0000-0002-8108-9179}\inst{\ref{aff37}}
\and F.~Pasian\orcid{0000-0002-4869-3227}\inst{\ref{aff3}}
\and K.~Pedersen\inst{\ref{aff73}}
\and V.~Pettorino\orcid{0000-0002-4203-9320}\inst{\ref{aff72}}
\and A.~Pezzotta\orcid{0000-0003-0726-2268}\inst{\ref{aff11}}
\and S.~Pires\orcid{0000-0002-0249-2104}\inst{\ref{aff40}}
\and G.~Polenta\orcid{0000-0003-4067-9196}\inst{\ref{aff74}}
\and M.~Poncet\inst{\ref{aff75}}
\and L.~A.~Popa\inst{\ref{aff76}}
\and F.~Raison\orcid{0000-0002-7819-6918}\inst{\ref{aff45}}
\and A.~Renzi\orcid{0000-0001-9856-1970}\inst{\ref{aff77},\ref{aff78},\ref{aff12}}
\and J.~Rhodes\orcid{0000-0002-4485-8549}\inst{\ref{aff51}}
\and G.~Riccio\inst{\ref{aff19}}
\and I.~Risso\orcid{0000-0003-2525-7761}\inst{\ref{aff16},\ref{aff17},\ref{aff11}}
\and E.~Romelli\orcid{0000-0003-3069-9222}\inst{\ref{aff3}}
\and M.~Roncarelli\orcid{0000-0001-9587-7822}\inst{\ref{aff12}}
\and R.~Saglia\orcid{0000-0003-0378-7032}\inst{\ref{aff47},\ref{aff45}}
\and Z.~Sakr\orcid{0000-0002-4823-3757}\inst{\ref{aff79},\ref{aff80},\ref{aff81}}
\and A.~G.~S\'anchez\orcid{0000-0003-1198-831X}\inst{\ref{aff45}}
\and D.~Sapone\orcid{0000-0001-7089-4503}\inst{\ref{aff82}}
\and B.~Sartoris\orcid{0000-0003-1337-5269}\inst{\ref{aff47},\ref{aff3}}
\and P.~Schneider\orcid{0000-0001-8561-2679}\inst{\ref{aff63}}
\and A.~Secroun\orcid{0000-0003-0505-3710}\inst{\ref{aff38}}
\and E.~Sefusatti\orcid{0000-0003-0473-1567}\inst{\ref{aff3},\ref{aff2},\ref{aff4}}
\and P.~Simon\inst{\ref{aff63}}
\and C.~Sirignano\orcid{0000-0002-0995-7146}\inst{\ref{aff77},\ref{aff78}}
\and G.~Sirri\orcid{0000-0003-2626-2853}\inst{\ref{aff15}}
\and L.~Stanco\orcid{0000-0002-9706-5104}\inst{\ref{aff78}}
\and P.~Tallada-Cresp\'{i}\orcid{0000-0002-1336-8328}\inst{\ref{aff24},\ref{aff25}}
\and A.~N.~Taylor\inst{\ref{aff30}}
\and I.~Tereno\orcid{0000-0002-4537-6218}\inst{\ref{aff83},\ref{aff84}}
\and N.~Tessore\orcid{0000-0002-9696-7931}\inst{\ref{aff85}}
\and S.~Toft\orcid{0000-0003-3631-7176}\inst{\ref{aff86},\ref{aff87}}
\and R.~Toledo-Moreo\orcid{0000-0002-2997-4859}\inst{\ref{aff88},\ref{aff89}}
\and F.~Torradeflot\orcid{0000-0003-1160-1517}\inst{\ref{aff25},\ref{aff24}}
\and I.~Tutusaus\orcid{0000-0002-3199-0399}\inst{\ref{aff43},\ref{aff42},\ref{aff80}}
\and J.~Valiviita\orcid{0000-0001-6225-3693}\inst{\ref{aff58},\ref{aff59}}
\and T.~Vassallo\orcid{0000-0001-6512-6358}\inst{\ref{aff3},\ref{aff44}}
\and Y.~Wang\orcid{0000-0002-4749-2984}\inst{\ref{aff49}}
\and J.~Weller\orcid{0000-0002-8282-2010}\inst{\ref{aff47},\ref{aff45}}
\and G.~Zamorani\orcid{0000-0002-2318-301X}\inst{\ref{aff12}}
\and F.~M.~Zerbi\orcid{0000-0002-9996-973X}\inst{\ref{aff11}}
\and E.~Zucca\orcid{0000-0002-5845-8132}\inst{\ref{aff12}}
\and M.~Ballardini\orcid{0000-0003-4481-3559}\inst{\ref{aff90},\ref{aff91},\ref{aff12}}
\and A.~Boucaud\orcid{0000-0001-7387-2633}\inst{\ref{aff7}}
\and E.~Bozzo\orcid{0000-0002-8201-1525}\inst{\ref{aff37}}
\and C.~Burigana\orcid{0000-0002-3005-5796}\inst{\ref{aff92},\ref{aff41}}
\and R.~Cabanac\orcid{0000-0001-6679-2600}\inst{\ref{aff80}}
\and M.~Calabrese\orcid{0000-0002-2637-2422}\inst{\ref{aff93},\ref{aff23}}
\and A.~Cappi\inst{\ref{aff8},\ref{aff12}}
\and J.~A.~Escartin~Vigo\inst{\ref{aff45}}
\and J.~Garc\'ia-Bellido\orcid{0000-0002-9370-8360}\inst{\ref{aff79}}
\and T.~Gasparetto\orcid{0000-0002-7913-4866}\inst{\ref{aff26}}
\and L.~Ingoglia\orcid{0000-0002-7587-0997}\inst{\ref{aff92}}
\and J.~Macias-Perez\orcid{0000-0002-5385-2763}\inst{\ref{aff94}}
\and R.~Maoli\orcid{0000-0002-6065-3025}\inst{\ref{aff95},\ref{aff26}}
\and J.~Mart\'{i}n-Fleitas\orcid{0000-0002-8594-569X}\inst{\ref{aff96}}
\and A.~Montoro\orcid{0000-0003-4730-8590}\inst{\ref{aff43},\ref{aff42}}
\and M.~P\"ontinen\orcid{0000-0001-5442-2530}\inst{\ref{aff58}}
\and E.~Sarpa\orcid{0000-0002-1256-655X}\inst{\ref{aff3}}
\and V.~Scottez\orcid{0009-0008-3864-940X}\inst{\ref{aff97},\ref{aff98}}
\and M.~Sereno\orcid{0000-0003-0302-0325}\inst{\ref{aff12},\ref{aff15}}
\and M.~Tenti\orcid{0000-0002-4254-5901}\inst{\ref{aff15}}
\and M.~Tucci\inst{\ref{aff37}}
\and M.~Viel\orcid{0000-0002-2642-5707}\inst{\ref{aff2},\ref{aff3},\ref{aff13},\ref{aff4},\ref{aff6}}
\and M.~Wiesmann\orcid{0009-0000-8199-5860}\inst{\ref{aff48}}
\and J.~A.~Acevedo~Barroso\orcid{0000-0002-9654-1711}\inst{\ref{aff51}}
\and Y.~Akrami\orcid{0000-0002-2407-7956}\inst{\ref{aff79},\ref{aff99}}
\and G.~Alguero\inst{\ref{aff94}}
\and I.~T.~Andika\orcid{0000-0001-6102-9526}\inst{\ref{aff47}}
\and S.~Anselmi\orcid{0000-0002-3579-9583}\inst{\ref{aff78},\ref{aff77},\ref{aff100}}
\and M.~Archidiacono\orcid{0000-0003-4952-9012}\inst{\ref{aff101},\ref{aff102}}
\and G.~Arico\orcid{0000-0002-2802-2928}\inst{\ref{aff15}}
\and F.~Atrio-Barandela\orcid{0000-0002-2130-2513}\inst{\ref{aff103}}
\and M.~Baes\orcid{0000-0002-3930-2757}\inst{\ref{aff104}}
\and L.~Bazzanini\orcid{0000-0003-0727-0137}\inst{\ref{aff90},\ref{aff12}}
\and P.~Bergamini\orcid{0000-0003-1383-9414}\inst{\ref{aff12}}
\and D.~Bertacca\orcid{0000-0002-2490-7139}\inst{\ref{aff77},\ref{aff46},\ref{aff78}}
\and M.~Bethermin\orcid{0000-0002-3915-2015}\inst{\ref{aff105}}
\and F.~Beutler\orcid{0000-0003-0467-5438}\inst{\ref{aff30}}
\and L.~Blot\orcid{0000-0002-9622-7167}\inst{\ref{aff106},\ref{aff60}}
\and M.~Bonici\orcid{0000-0002-8430-126X}\inst{\ref{aff107},\ref{aff23}}
\and M.~L.~Brown\orcid{0000-0002-0370-8077}\inst{\ref{aff108}}
\and S.~Bruton\orcid{0000-0002-6503-5218}\inst{\ref{aff109}}
\and A.~Calabro\orcid{0000-0003-2536-1614}\inst{\ref{aff26}}
\and B.~Camacho~Quevedo\orcid{0000-0002-8789-4232}\inst{\ref{aff2},\ref{aff13},\ref{aff3}}
\and F.~Caro\orcid{0009-0003-1053-0507}\inst{\ref{aff26}}
\and C.~S.~Carvalho\inst{\ref{aff84}}
\and F.~Cogato\orcid{0000-0003-4632-6113}\inst{\ref{aff66},\ref{aff12}}
\and A.~R.~Cooray\orcid{0000-0002-3892-0190}\inst{\ref{aff110}}
\and P.~Corcho-Caballero\orcid{0000-0001-6327-7080}\inst{\ref{aff111}}
\and B.~Csizi\orcid{0000-0003-3227-6581}\inst{\ref{aff9}}
\and O.~Cucciati\orcid{0000-0002-9336-7551}\inst{\ref{aff12}}
\and H.~Dannerbauer\orcid{0000-0001-7147-3575}\inst{\ref{aff29},\ref{aff112}}
\and T.~de~Boer\orcid{0000-0001-5486-2747}\inst{\ref{aff113}}
\and F.~De~Paolis\orcid{0000-0001-6460-7563}\inst{\ref{aff114},\ref{aff115},\ref{aff116}}
\and G.~Desprez\orcid{0000-0001-8325-1742}\inst{\ref{aff111}}
\and A.~D\'iaz-S\'anchez\orcid{0000-0003-0748-4768}\inst{\ref{aff117}}
\and S.~Di~Domizio\orcid{0000-0003-2863-5895}\inst{\ref{aff16},\ref{aff17}}
\and J.~M.~Diego\orcid{0000-0001-9065-3926}\inst{\ref{aff118}}
\and V.~Duret\orcid{0009-0009-0383-4960}\inst{\ref{aff38}}
\and M.~Y.~Elkhashab\orcid{0000-0001-9306-2603}\inst{\ref{aff3},\ref{aff4},\ref{aff1},\ref{aff2}}
\and Y.~Fang\orcid{0000-0002-0334-6950}\inst{\ref{aff47}}
\and A.~Farina\orcid{0009-0000-3420-929X}\inst{\ref{aff11},\ref{aff17}}
\and A.~Finoguenov\orcid{0000-0002-4606-5403}\inst{\ref{aff58}}
\and A.~Franco\orcid{0000-0002-4761-366X}\inst{\ref{aff114},\ref{aff115},\ref{aff116}}
\and K.~Ganga\orcid{0000-0001-8159-8208}\inst{\ref{aff7}}
\and R.~Gavazzi\orcid{0000-0002-5540-6935}\inst{\ref{aff65},\ref{aff119}}
\and E.~Gaztanaga\orcid{0000-0001-9632-0815}\inst{\ref{aff43},\ref{aff42},\ref{aff120}}
\and Z.~Ghaffari\orcid{0000-0002-6467-8078}\inst{\ref{aff3},\ref{aff2}}
\and F.~Giacomini\orcid{0000-0002-3129-2814}\inst{\ref{aff15}}
\and F.~Gianotti\orcid{0000-0003-4666-119X}\inst{\ref{aff12}}
\and E.~J.~Gonzalez\orcid{0000-0002-0226-9893}\inst{\ref{aff121},\ref{aff122}}
\and G.~Gozaliasl\orcid{0000-0002-0236-919X}\inst{\ref{aff123},\ref{aff58}}
\and A.~Gruppuso\orcid{0000-0001-9272-5292}\inst{\ref{aff12},\ref{aff15}}
\and M.~Guidi\orcid{0000-0001-9408-1101}\inst{\ref{aff14},\ref{aff12}}
\and C.~M.~Gutierrez\orcid{0000-0001-7854-783X}\inst{\ref{aff29},\ref{aff112}}
\and A.~Hall\orcid{0000-0002-3139-8651}\inst{\ref{aff30}}
\and N.~A.~Hatch\orcid{0000-0001-5600-0534}\inst{\ref{aff124}}
\and C.~Hern\'andez-Monteagudo\orcid{0000-0001-5471-9166}\inst{\ref{aff112},\ref{aff29}}
\and H.~Hildebrandt\orcid{0000-0002-9814-3338}\inst{\ref{aff125}}
\and J.~J.~E.~Kajava\orcid{0000-0002-3010-8333}\inst{\ref{aff126},\ref{aff127},\ref{aff128}}
\and Y.~Kang\orcid{0009-0000-8588-7250}\inst{\ref{aff37}}
\and V.~Kansal\orcid{0000-0002-4008-6078}\inst{\ref{aff129},\ref{aff130}}
\and D.~Karagiannis\orcid{0000-0002-4927-0816}\inst{\ref{aff90},\ref{aff131}}
\and K.~Kiiveri\orcid{0000-0002-3711-3346}\inst{\ref{aff132}}
\and J.~Kim\orcid{0000-0003-2776-2761}\inst{\ref{aff133}}
\and C.~C.~Kirkpatrick\inst{\ref{aff132}}
\and A.~Kov\'acs\orcid{0000-0002-5825-579X}\inst{\ref{aff134},\ref{aff135}}
\and I.~Kova{\v{c}}i{\'{c}}\orcid{0000-0001-6751-3263}\inst{\ref{aff104}}
\and K.~Koyama\orcid{0000-0001-6727-6915}\inst{\ref{aff120}}
\and S.~Kruk\orcid{0000-0001-8010-8879}\inst{\ref{aff10}}
\and M.~C.~Lam\orcid{0000-0002-9347-2298}\inst{\ref{aff30}}
\and F.~Lepori\orcid{0009-0000-5061-7138}\inst{\ref{aff136}}
\and G.~Leroy\orcid{0009-0004-2523-4425}\inst{\ref{aff137},\ref{aff67}}
\and G.~F.~Lesci\orcid{0000-0002-4607-2830}\inst{\ref{aff66},\ref{aff12}}
\and J.~Lesgourgues\orcid{0000-0001-7627-353X}\inst{\ref{aff138}}
\and T.~I.~Liaudat\orcid{0000-0002-9104-314X}\inst{\ref{aff139}}
\and L.~Linke\orcid{0000-0002-2622-8113}\inst{\ref{aff9}}
\and S.~J.~Liu\orcid{0000-0001-7680-2139}\inst{\ref{aff39}}
\and F.~Mannucci\orcid{0000-0002-4803-2381}\inst{\ref{aff140}}
\and C.~J.~A.~P.~Martins\orcid{0000-0002-4886-9261}\inst{\ref{aff141},\ref{aff142}}
\and M.~Migliaccio\inst{\ref{aff143},\ref{aff144}}
\and M.~Miluzio\inst{\ref{aff10},\ref{aff145}}
\and G.~Morgante\inst{\ref{aff12}}
\and S.~Nadathur\orcid{0000-0001-9070-3102}\inst{\ref{aff120}}
\and K.~Naidoo\orcid{0000-0002-9182-1802}\inst{\ref{aff120},\ref{aff55}}
\and A.~Navarro-Alsina\orcid{0000-0002-3173-2592}\inst{\ref{aff63}}
\and S.~Nesseris\orcid{0000-0002-0567-0324}\inst{\ref{aff79}}
\and F.~Oppizzi\orcid{0000-0003-3904-8370}\inst{\ref{aff17}}
\and F.~Pace\orcid{0000-0001-8039-0480}\inst{\ref{aff20},\ref{aff21},\ref{aff22}}
\and D.~Paoletti\orcid{0000-0003-4761-6147}\inst{\ref{aff12},\ref{aff41}}
\and F.~Passalacqua\orcid{0000-0002-8606-4093}\inst{\ref{aff77},\ref{aff78}}
\and K.~Paterson\orcid{0000-0001-8340-3486}\inst{\ref{aff55}}
\and C.~Pattison\orcid{0000-0003-3272-2617}\inst{\ref{aff120}}
\and R.~Paviot\orcid{0009-0002-8108-3460}\inst{\ref{aff40},\ref{aff75}}
\and D.~Potter\orcid{0000-0002-0757-5195}\inst{\ref{aff146}}
\and G.~W.~Pratt\inst{\ref{aff40}}
\and S.~Quai\orcid{0000-0002-0449-8163}\inst{\ref{aff66},\ref{aff12}}
\and M.~Radovich\orcid{0000-0002-3585-866X}\inst{\ref{aff46}}
\and W.~Roster\orcid{0000-0002-9149-6528}\inst{\ref{aff45}}
\and S.~Sacquegna\orcid{0000-0002-8433-6630}\inst{\ref{aff147}}
\and M.~Sahl\'en\orcid{0000-0003-0973-4804}\inst{\ref{aff148}}
\and D.~B.~Sanders\orcid{0000-0002-1233-9998}\inst{\ref{aff113}}
\and A.~Schneider\orcid{0000-0001-7055-8104}\inst{\ref{aff146}}
\and D.~Sciotti\orcid{0009-0008-4519-2620}\inst{\ref{aff26},\ref{aff64}}
\and E.~Sellentin\inst{\ref{aff149},\ref{aff50}}
\and S.~Serjeant\orcid{0000-0002-0517-7943}\inst{\ref{aff150}}
\and I.~Szapudi\orcid{0000-0003-2274-0301}\inst{\ref{aff113}}
\and K.~Tanidis\orcid{0000-0001-9843-5130}\inst{\ref{aff151}}
\and F.~Tarsitano\orcid{0000-0002-5919-0238}\inst{\ref{aff152},\ref{aff153},\ref{aff37}}
\and G.~Testera\orcid{0000-0003-2970-766X}\inst{\ref{aff17}}
\and R.~Teyssier\orcid{0000-0001-7689-0933}\inst{\ref{aff154}}
\and S.~Tosi\orcid{0000-0002-7275-9193}\inst{\ref{aff16},\ref{aff11},\ref{aff17}}
\and A.~Troja\orcid{0000-0003-0239-4595}\inst{\ref{aff3}}
\and C.~Uhlemann\orcid{0000-0001-7831-1579}\inst{\ref{aff155},\ref{aff156}}
\and C.~Valieri\inst{\ref{aff15}}
\and A.~Venhola\orcid{0000-0001-6071-4564}\inst{\ref{aff157}}
\and D.~Vergani\orcid{0000-0003-0898-2216}\inst{\ref{aff12}}
\and G.~Verza\orcid{0000-0002-1886-8348}\inst{\ref{aff158},\ref{aff159}}
\and S.~Vinciguerra\orcid{0009-0005-4018-3184}\inst{\ref{aff65}}
\and M.~von~Wietersheim-Kramsta\orcid{0000-0003-4986-5091}\inst{\ref{aff67},\ref{aff137}}
\and N.~A.~Walton\orcid{0000-0003-3983-8778}\inst{\ref{aff160}}
\and L.~Wang\orcid{0000-0002-6736-9158}\inst{\ref{aff161},\ref{aff111}}
\and A.~H.~Wright\orcid{0000-0001-7363-7932}\inst{\ref{aff125}}}

\institute{Dipartimento di Fisica - Sezione di Astronomia, Universit\`a di Trieste, Via Tiepolo 11, 34131 Trieste, Italy\label{aff1}
\and
IFPU, Institute for Fundamental Physics of the Universe, via Beirut 2, 34151 Trieste, Italy\label{aff2}
\and
INAF-Osservatorio Astronomico di Trieste, Via G. B. Tiepolo 11, 34143 Trieste, Italy\label{aff3}
\and
INFN, Sezione di Trieste, Via Valerio 2, 34127 Trieste TS, Italy\label{aff4}
\and
Department of Mathematical Physics, Institute of Physics, University of S\~ao Paulo, R. do Mat\~ao 1371, 05508-090, S\~ao Paulo, SP, Brazil\label{aff5}
\and
ICSC - Centro Nazionale di Ricerca in High Performance Computing, Big Data e Quantum Computing, Via Magnanelli 2, Bologna, Italy\label{aff6}
\and
Universit\'e Paris Cit\'e, CNRS, Astroparticule et Cosmologie, 75013 Paris, France\label{aff7}
\and
Universit\'e C\^{o}te d'Azur, Observatoire de la C\^{o}te d'Azur, CNRS, Laboratoire Lagrange, Bd de l'Observatoire, CS 34229, 06304 Nice cedex 4, France\label{aff8}
\and
Universit\"at Innsbruck, Institut f\"ur Astro- und Teilchenphysik, Technikerstr. 25/8, 6020 Innsbruck, Austria\label{aff9}
\and
ESAC/ESA, Camino Bajo del Castillo, s/n., Urb. Villafranca del Castillo, 28692 Villanueva de la Ca\~nada, Madrid, Spain\label{aff10}
\and
INAF-Osservatorio Astronomico di Brera, Via Brera 28, 20122 Milano, Italy\label{aff11}
\and
INAF-Osservatorio di Astrofisica e Scienza dello Spazio di Bologna, Via Piero Gobetti 93/3, 40129 Bologna, Italy\label{aff12}
\and
SISSA, International School for Advanced Studies, Via Bonomea 265, 34136 Trieste TS, Italy\label{aff13}
\and
Dipartimento di Fisica e Astronomia, Universit\`a di Bologna, Via Gobetti 93/2, 40129 Bologna, Italy\label{aff14}
\and
INFN-Sezione di Bologna, Viale Berti Pichat 6/2, 40127 Bologna, Italy\label{aff15}
\and
Dipartimento di Fisica, Universit\`a di Genova, Via Dodecaneso 33, 16146, Genova, Italy\label{aff16}
\and
INFN-Sezione di Genova, Via Dodecaneso 33, 16146, Genova, Italy\label{aff17}
\and
Department of Physics "E. Pancini", University Federico II, Via Cinthia 6, 80126, Napoli, Italy\label{aff18}
\and
INAF-Osservatorio Astronomico di Capodimonte, Via Moiariello 16, 80131 Napoli, Italy\label{aff19}
\and
Dipartimento di Fisica, Universit\`a degli Studi di Torino, Via P. Giuria 1, 10125 Torino, Italy\label{aff20}
\and
INFN-Sezione di Torino, Via P. Giuria 1, 10125 Torino, Italy\label{aff21}
\and
INAF-Osservatorio Astrofisico di Torino, Via Osservatorio 20, 10025 Pino Torinese (TO), Italy\label{aff22}
\and
INAF-IASF Milano, Via Alfonso Corti 12, 20133 Milano, Italy\label{aff23}
\and
Centro de Investigaciones Energ\'eticas, Medioambientales y Tecnol\'ogicas (CIEMAT), Avenida Complutense 40, 28040 Madrid, Spain\label{aff24}
\and
Port d'Informaci\'{o} Cient\'{i}fica, Campus UAB, C. Albareda s/n, 08193 Bellaterra (Barcelona), Spain\label{aff25}
\and
INAF-Osservatorio Astronomico di Roma, Via Frascati 33, 00078 Monteporzio Catone, Italy\label{aff26}
\and
INFN section of Naples, Via Cinthia 6, 80126, Napoli, Italy\label{aff27}
\and
Dipartimento di Fisica e Astronomia "Augusto Righi" - Alma Mater Studiorum Universit\`a di Bologna, Viale Berti Pichat 6/2, 40127 Bologna, Italy\label{aff28}
\and
Instituto de Astrof\'{\i}sica de Canarias, E-38205 La Laguna, Tenerife, Spain\label{aff29}
\and
Institute for Astronomy, University of Edinburgh, Royal Observatory, Blackford Hill, Edinburgh EH9 3HJ, UK\label{aff30}
\and
European Space Agency/ESRIN, Largo Galileo Galilei 1, 00044 Frascati, Roma, Italy\label{aff31}
\and
Universit\'e Claude Bernard Lyon 1, CNRS/IN2P3, IP2I Lyon, UMR 5822, Villeurbanne, F-69100, France\label{aff32}
\and
Institut de Ci\`{e}ncies del Cosmos (ICCUB), Universitat de Barcelona (IEEC-UB), Mart\'{i} i Franqu\`{e}s 1, 08028 Barcelona, Spain\label{aff33}
\and
Instituci\'o Catalana de Recerca i Estudis Avan\c{c}ats (ICREA), Passeig de Llu\'{\i}s Companys 23, 08010 Barcelona, Spain\label{aff34}
\and
Institut de Ciencies de l'Espai (IEEC-CSIC), Campus UAB, Carrer de Can Magrans, s/n Cerdanyola del Vall\'es, 08193 Barcelona, Spain\label{aff35}
\and
UCB Lyon 1, CNRS/IN2P3, IUF, IP2I Lyon, 4 rue Enrico Fermi, 69622 Villeurbanne, France\label{aff36}
\and
Department of Astronomy, University of Geneva, ch. d'Ecogia 16, 1290 Versoix, Switzerland\label{aff37}
\and
Aix-Marseille Universit\'e, CNRS/IN2P3, CPPM, Marseille, France\label{aff38}
\and
INAF-Istituto di Astrofisica e Planetologia Spaziali, via del Fosso del Cavaliere, 100, 00100 Roma, Italy\label{aff39}
\and
Universit\'e Paris-Saclay, Universit\'e Paris Cit\'e, CEA, CNRS, AIM, 91191, Gif-sur-Yvette, France\label{aff40}
\and
INFN-Bologna, Via Irnerio 46, 40126 Bologna, Italy\label{aff41}
\and
Institut d'Estudis Espacials de Catalunya (IEEC),  Edifici RDIT, Campus UPC, 08860 Castelldefels, Barcelona, Spain\label{aff42}
\and
Institute of Space Sciences (ICE, CSIC), Campus UAB, Carrer de Can Magrans, s/n, 08193 Barcelona, Spain\label{aff43}
\and
University Observatory, LMU Faculty of Physics, Scheinerstr.~1, 81679 Munich, Germany\label{aff44}
\and
Max Planck Institute for Extraterrestrial Physics, Giessenbachstr. 1, 85748 Garching, Germany\label{aff45}
\and
INAF-Osservatorio Astronomico di Padova, Via dell'Osservatorio 5, 35122 Padova, Italy\label{aff46}
\and
Universit\"ats-Sternwarte M\"unchen, Fakult\"at f\"ur Physik, Ludwig-Maximilians-Universit\"at M\"unchen, Scheinerstr.~1, 81679 M\"unchen, Germany\label{aff47}
\and
Institute of Theoretical Astrophysics, University of Oslo, P.O. Box 1029 Blindern, 0315 Oslo, Norway\label{aff48}
\and
Caltech/IPAC, 1200 E. California Blvd., Pasadena, CA 91125, USA\label{aff49}
\and
Leiden Observatory, Leiden University, Einsteinweg 55, 2333 CC Leiden, The Netherlands\label{aff50}
\and
Jet Propulsion Laboratory, California Institute of Technology, 4800 Oak Grove Drive, Pasadena, CA, 91109, USA\label{aff51}
\and
Felix Hormuth Engineering, Goethestr. 17, 69181 Leimen, Germany\label{aff52}
\and
Technical University of Denmark, Elektrovej 327, 2800 Kgs. Lyngby, Denmark\label{aff53}
\and
Cosmic Dawn Center (DAWN), Denmark\label{aff54}
\and
Max-Planck-Institut f\"ur Astronomie, K\"onigstuhl 17, 69117 Heidelberg, Germany\label{aff55}
\and
NASA Goddard Space Flight Center, Greenbelt, MD 20771, USA\label{aff56}
\and
Department of Physics and Astronomy, University College London, Gower Street, London WC1E 6BT, UK\label{aff57}
\and
Department of Physics, P.O. Box 64, University of Helsinki, 00014 Helsinki, Finland\label{aff58}
\and
Helsinki Institute of Physics, Gustaf H{\"a}llstr{\"o}min katu 2, University of Helsinki, 00014 Helsinki, Finland\label{aff59}
\and
Laboratoire d'etude de l'Univers et des phenomenes eXtremes, Observatoire de Paris, Universit\'e PSL, Sorbonne Universit\'e, CNRS, 92190 Meudon, France\label{aff60}
\and
SKAO, Jodrell Bank, Lower Withington, Macclesfield SK11 9FT, UK\label{aff61}
\and
Centre de Calcul de l'IN2P3/CNRS, 21 avenue Pierre de Coubertin 69627 Villeurbanne Cedex, France\label{aff62}
\and
Universit\"at Bonn, Argelander-Institut f\"ur Astronomie, Auf dem H\"ugel 71, 53121 Bonn, Germany\label{aff63}
\and
INFN-Sezione di Roma, Piazzale Aldo Moro, 2 - c/o Dipartimento di Fisica, Edificio G. Marconi, 00185 Roma, Italy\label{aff64}
\and
Aix-Marseille Universit\'e, CNRS, CNES, LAM, Marseille, France\label{aff65}
\and
Dipartimento di Fisica e Astronomia "Augusto Righi" - Alma Mater Studiorum Universit\`a di Bologna, via Piero Gobetti 93/2, 40129 Bologna, Italy\label{aff66}
\and
Department of Physics, Institute for Computational Cosmology, Durham University, South Road, Durham, DH1 3LE, UK\label{aff67}
\and
CNRS-UCB International Research Laboratory, Centre Pierre Bin\'etruy, IRL2007, CPB-IN2P3, Berkeley, USA\label{aff68}
\and
Institute of Physics, Laboratory of Astrophysics, Ecole Polytechnique F\'ed\'erale de Lausanne (EPFL), Observatoire de Sauverny, 1290 Versoix, Switzerland\label{aff69}
\and
Telespazio UK S.L. for European Space Agency (ESA), Camino bajo del Castillo, s/n, Urbanizacion Villafranca del Castillo, Villanueva de la Ca\~nada, 28692 Madrid, Spain\label{aff70}
\and
Institut de F\'{i}sica d'Altes Energies (IFAE), The Barcelona Institute of Science and Technology, Campus UAB, 08193 Bellaterra (Barcelona), Spain\label{aff71}
\and
European Space Agency/ESTEC, Keplerlaan 1, 2201 AZ Noordwijk, The Netherlands\label{aff72}
\and
DARK, Niels Bohr Institute, University of Copenhagen, Jagtvej 155, 2200 Copenhagen, Denmark\label{aff73}
\and
Space Science Data Center, Italian Space Agency, via del Politecnico snc, 00133 Roma, Italy\label{aff74}
\and
Centre National d'Etudes Spatiales -- Centre spatial de Toulouse, 18 avenue Edouard Belin, 31401 Toulouse Cedex 9, France\label{aff75}
\and
Institute of Space Science, Str. Atomistilor, nr. 409 M\u{a}gurele, Ilfov, 077125, Romania\label{aff76}
\and
Dipartimento di Fisica e Astronomia "G. Galilei", Universit\`a di Padova, Via Marzolo 8, 35131 Padova, Italy\label{aff77}
\and
INFN-Padova, Via Marzolo 8, 35131 Padova, Italy\label{aff78}
\and
Instituto de F\'isica Te\'orica UAM-CSIC, Campus de Cantoblanco, 28049 Madrid, Spain\label{aff79}
\and
Institut de Recherche en Astrophysique et Plan\'etologie (IRAP), Universit\'e de Toulouse, CNRS, UPS, CNES, 14 Av. Edouard Belin, 31400 Toulouse, France\label{aff80}
\and
Universit\'e St Joseph; Faculty of Sciences, Beirut, Lebanon\label{aff81}
\and
Departamento de F\'isica, FCFM, Universidad de Chile, Blanco Encalada 2008, Santiago, Chile\label{aff82}
\and
Departamento de F\'isica, Faculdade de Ci\^encias, Universidade de Lisboa, Edif\'icio C8, Campo Grande, PT1749-016 Lisboa, Portugal\label{aff83}
\and
Instituto de Astrof\'isica e Ci\^encias do Espa\c{c}o, Faculdade de Ci\^encias, Universidade de Lisboa, Tapada da Ajuda, 1349-018 Lisboa, Portugal\label{aff84}
\and
Mullard Space Science Laboratory, University College London, Holmbury St Mary, Dorking, Surrey RH5 6NT, UK\label{aff85}
\and
Cosmic Dawn Center (DAWN)\label{aff86}
\and
Niels Bohr Institute, University of Copenhagen, Jagtvej 128, 2200 Copenhagen, Denmark\label{aff87}
\and
Universidad Polit\'ecnica de Cartagena, Departamento de Electr\'onica y Tecnolog\'ia de Computadoras,  Plaza del Hospital 1, 30202 Cartagena, Spain\label{aff88}
\and
European University of Technology EUt+, European Union\label{aff89}
\and
Dipartimento di Fisica e Scienze della Terra, Universit\`a degli Studi di Ferrara, Via Giuseppe Saragat 1, 44122 Ferrara, Italy\label{aff90}
\and
Istituto Nazionale di Fisica Nucleare, Sezione di Ferrara, Via Giuseppe Saragat 1, 44122 Ferrara, Italy\label{aff91}
\and
INAF, Istituto di Radioastronomia, Via Piero Gobetti 101, 40129 Bologna, Italy\label{aff92}
\and
Astronomical Observatory of the Autonomous Region of the Aosta Valley (OAVdA), Loc. Lignan 39, I-11020, Nus (Aosta Valley), Italy\label{aff93}
\and
Univ. Grenoble Alpes, CNRS, Grenoble INP, LPSC-IN2P3, 53, Avenue des Martyrs, 38000, Grenoble, France\label{aff94}
\and
Dipartimento di Fisica, Sapienza Universit\`a di Roma, Piazzale Aldo Moro 2, 00185 Roma, Italy\label{aff95}
\and
Aurora Technology for European Space Agency (ESA), Camino bajo del Castillo, s/n, Urbanizacion Villafranca del Castillo, Villanueva de la Ca\~nada, 28692 Madrid, Spain\label{aff96}
\and
Institut d'Astrophysique de Paris, 98bis Boulevard Arago, 75014, Paris, France\label{aff97}
\and
ICL, Junia, Universit\'e Catholique de Lille, LITL, 59000 Lille, France\label{aff98}
\and
CERCA/ISO, Department of Physics, Case Western Reserve University, 10900 Euclid Avenue, Cleveland, OH 44106, USA\label{aff99}
\and
Laboratoire Univers et Th\'eorie, Observatoire de Paris, Universit\'e PSL, Universit\'e Paris Cit\'e, CNRS, 92190 Meudon, France\label{aff100}
\and
Dipartimento di Fisica "Aldo Pontremoli", Universit\`a degli Studi di Milano, Via Celoria 16, 20133 Milano, Italy\label{aff101}
\and
INFN-Sezione di Milano, Via Celoria 16, 20133 Milano, Italy\label{aff102}
\and
Departamento de F{\'\i}sica Fundamental. Universidad de Salamanca. Plaza de la Merced s/n. 37008 Salamanca, Spain\label{aff103}
\and
Universiteit Gent, Department of Physics and Astronomy, Proeftuinstraat 86 N3, 9000 Ghent, Belgium
\label{aff104}
\and
Universit\'e de Strasbourg, CNRS, Observatoire astronomique de Strasbourg, UMR 7550, 67000 Strasbourg, France\label{aff105}
\and
Center for Data-Driven Discovery, Kavli IPMU (WPI), UTIAS, The University of Tokyo, Kashiwa, Chiba 277-8583, Japan\label{aff106}
\and
Waterloo Centre for Astrophysics, University of Waterloo, Waterloo, Ontario N2L 3G1, Canada\label{aff107}
\and
Jodrell Bank Centre for Astrophysics, Department of Physics and Astronomy, University of Manchester, Oxford Road, Manchester M13 9PL, UK\label{aff108}
\and
California Institute of Technology, 1200 E California Blvd, Pasadena, CA 91125, USA\label{aff109}
\and
Department of Physics \& Astronomy, University of California Irvine, Irvine CA 92697, USA\label{aff110}
\and
Kapteyn Astronomical Institute, University of Groningen, PO Box 800, 9700 AV Groningen, The Netherlands\label{aff111}
\and
Universidad de La Laguna, Dpto. Astrof\'\i sica, E-38206 La Laguna, Tenerife, Spain\label{aff112}
\and
Institute for Astronomy, University of Hawaii, 2680 Woodlawn Drive, Honolulu, HI 96822, USA\label{aff113}
\and
Department of Mathematics and Physics E. De Giorgi, University of Salento, Via per Arnesano, CP-I93, 73100, Lecce, Italy\label{aff114}
\and
INFN, Sezione di Lecce, Via per Arnesano, CP-193, 73100, Lecce, Italy\label{aff115}
\and
INAF-Sezione di Lecce, c/o Dipartimento Matematica e Fisica, Via per Arnesano, 73100, Lecce, Italy\label{aff116}
\and
Departamento F\'isica Aplicada, Universidad Polit\'ecnica de Cartagena, Campus Muralla del Mar, 30202 Cartagena, Murcia, Spain\label{aff117}
\and
Instituto de F\'isica de Cantabria, Edificio Juan Jord\'a, Avenida de los Castros, 39005 Santander, Spain\label{aff118}
\and
Institut d'Astrophysique de Paris, UMR 7095, CNRS, and Sorbonne Universit\'e, 98 bis boulevard Arago, 75014 Paris, France\label{aff119}
\and
Institute of Cosmology and Gravitation, University of Portsmouth, Portsmouth PO1 3FX, UK\label{aff120}
\and
Departament de F\'{\i}sica, Universitat Aut\`onoma de Barcelona, 08193 Bellaterra (Barcelona), Spain\label{aff121}
\and
Instituto de Astronomia Teorica y Experimental (IATE-CONICET), Laprida 854, X5000BGR, C\'ordoba, Argentina\label{aff122}
\and
Department of Computer Science, Aalto University, PO Box 15400, Espoo, FI-00 076, Finland\label{aff123}
\and
School of Physics and Astronomy, University of Nottingham, University Park, Nottingham NG7 2RD, UK\label{aff124}
\and
Ruhr University Bochum, Faculty of Physics and Astronomy, Astronomical Institute (AIRUB), German Centre for Cosmological Lensing (GCCL), 44780 Bochum, Germany\label{aff125}
\and
Department of Physics and Astronomy, Vesilinnantie 5, University of Turku, 20014 Turku, Finland\label{aff126}
\and
Finnish Centre for Astronomy with ESO (FINCA), Quantum, Vesilinnantie 5, University of Turku, 20014 Turku, Finland\label{aff127}
\and
Serco for European Space Agency (ESA), Camino bajo del Castillo, s/n, Urbanizacion Villafranca del Castillo, Villanueva de la Ca\~nada, 28692 Madrid, Spain\label{aff128}
\and
ARC Centre of Excellence for Dark Matter Particle Physics, Melbourne, Australia\label{aff129}
\and
Centre for Astrophysics \& Supercomputing, Swinburne University of Technology,  Hawthorn, Victoria 3122, Australia\label{aff130}
\and
Department of Physics and Astronomy, University of the Western Cape, Bellville, Cape Town, 7535, South Africa\label{aff131}
\and
Department of Physics and Helsinki Institute of Physics, Gustaf H\"allstr\"omin katu 2, University of Helsinki, 00014 Helsinki, Finland\label{aff132}
\and
Department of Physics, Oxford University, Keble Road, Oxford OX1 3RH, UK\label{aff133}
\and
MTA-CSFK Lend\"ulet Large-Scale Structure Research Group, Konkoly-Thege Mikl\'os \'ut 15-17, H-1121 Budapest, Hungary\label{aff134}
\and
Konkoly Observatory, HUN-REN CSFK, MTA Centre of Excellence, Budapest, Konkoly Thege Mikl\'os {\'u}t 15-17. H-1121, Hungary\label{aff135}
\and
Departement of Theoretical Physics, University of Geneva, Switzerland\label{aff136}
\and
Department of Physics, Centre for Extragalactic Astronomy, Durham University, South Road, Durham, DH1 3LE, UK\label{aff137}
\and
Institute for Theoretical Particle Physics and Cosmology (TTK), RWTH Aachen University, 52056 Aachen, Germany\label{aff138}
\and
IRFU, CEA, Universit\'e Paris-Saclay 91191 Gif-sur-Yvette Cedex, France\label{aff139}
\and
INAF-Osservatorio Astrofisico di Arcetri, Largo E. Fermi 5, 50125, Firenze, Italy\label{aff140}
\and
Centro de Astrof\'{\i}sica da Universidade do Porto, Rua das Estrelas, 4150-762 Porto, Portugal\label{aff141}
\and
Instituto de Astrof\'isica e Ci\^encias do Espa\c{c}o, Universidade do Porto, CAUP, Rua das Estrelas, PT4150-762 Porto, Portugal\label{aff142}
\and
Dipartimento di Fisica, Universit\`a di Roma Tor Vergata, Via della Ricerca Scientifica 1, Roma, Italy\label{aff143}
\and
INFN, Sezione di Roma 2, Via della Ricerca Scientifica 1, Roma, Italy\label{aff144}
\and
HE Space for European Space Agency (ESA), Camino bajo del Castillo, s/n, Urbanizacion Villafranca del Castillo, Villanueva de la Ca\~nada, 28692 Madrid, Spain\label{aff145}
\and
Department of Astrophysics, University of Zurich, Winterthurerstrasse 190, 8057 Zurich, Switzerland\label{aff146}
\and
INAF - Osservatorio Astronomico d'Abruzzo, Via Maggini, 64100, Teramo, Italy\label{aff147}
\and
Theoretical astrophysics, Department of Physics and Astronomy, Uppsala University, Box 516, 751 37 Uppsala, Sweden\label{aff148}
\and
Mathematical Institute, University of Leiden, Einsteinweg 55, 2333 CA Leiden, The Netherlands\label{aff149}
\and
School of Physical Sciences, The Open University, Milton Keynes, MK7 6AA, UK\label{aff150}
\and
Center for Astrophysics and Cosmology, University of Nova Gorica, Nova Gorica, Slovenia\label{aff151}
\and
Kobayashi-Maskawa Institute for the Origin of Particles and the Universe, Nagoya University, Chikusa-ku, Nagoya, 464-8602, Japan\label{aff152}
\and
Institute for Particle Physics and Astrophysics, Dept. of Physics, ETH Zurich, Wolfgang-Pauli-Strasse 27, 8093 Zurich, Switzerland\label{aff153}
\and
Department of Astrophysical Sciences, Peyton Hall, Princeton University, Princeton, NJ 08544, USA\label{aff154}
\and
Fakult\"at f\"ur Physik, Universit\"at Bielefeld, Postfach 100131, 33501 Bielefeld, Germany\label{aff155}
\and
School of Mathematics, Statistics and Physics, Newcastle University, Herschel Building, Newcastle-upon-Tyne, NE1 7RU, UK\label{aff156}
\and
Space physics and astronomy research unit, University of Oulu, Pentti Kaiteran katu 1, FI-90014 Oulu, Finland\label{aff157}
\and
International Centre for Theoretical Physics (ICTP), Strada Costiera 11, 34151 Trieste, Italy\label{aff158}
\and
Center for Computational Astrophysics, Flatiron Institute, 162 5th Avenue, 10010, New York, NY, USA\label{aff159}
\and
Institute of Astronomy, University of Cambridge, Madingley Road, Cambridge CB3 0HA, UK\label{aff160}
\and
SRON Netherlands Institute for Space Research, Landleven 12, 9747 AD, Groningen, The Netherlands\label{aff161}}

\abstract{We present \texttt{CosmoPostProcess}, a simulation-based forward model algorithm calibrated to reproduce optical cluster observables in line with measurements taken with the \Euclid space telescope. The main deliverable of \texttt{CosmoPostProcess} is a correction for stacked surface-density profiles, binned in terms of richness and redshift, that accounts for selection-related systematic effects. These corrections take into account the modification to the stacked weak lensing signal from richness-selected samples of clusters identified in the photometric \Euclid survey compared with an unbiased reference sample. In this work, we focus on the \Euclid richness definition currently foreseen for the cosmological analysis, which does not apply a colour selection. \Euclid also provides an alternative richness estimate based on a red-sequence galaxy selection, which is not considered here. The algorithm processes $N$-body simulations by painting galaxies with a halo-occupation model and emulating the survey’s detection and richness-assignment algorithms. We implemented a novel algorithm to estimate the optical cluster centres from galaxy-projected densities and validated it against the official \Euclid pipelines. The effect of baryonic physics on the halo density profiles is incorporated through a correction calibrated on hydrodynamical simulations, yielding total-matter profiles consistent with those measured in the reference hydrodynamical simulations. In validation against hydrodynamical simulations, the baryon-corrected excess surface density is in agreement to within \( 2\,\%\) for cluster-centric radii, \(r\in[0.1,\,5]\,h^{-1}\,\mathrm{Mpc}\). To assess the impact of the different contributions to the selection bias, we performed dedicated tests that included variations of both cosmological parameters and the parameters of the mass-richness relation. Across all these tests, the selection bias induced by projection alone follows a robust pattern: a large-scale structure that is physically correlated with the main halo and projected along the line of sight enhances the stacked surface density profile near the transition from one-halo to two-halo dominance, with a peak radius of about \(1\,h^{-1}\,\mathrm{Mpc}\) and an amplitude of about \(20\!-\!40\,\%\), with a dependence on richness and redshift. This behaviour is mild at low and intermediate redshifts ($z\lesssim0.7$), where the impact remains at the level of a few per cent, but it becomes increasingly relevant at higher redshift ($z\gtrsim0.7$), with a more pronounced enhancement of the transition-scale peak.
 Finally, baryonic modifications remain sub-dominant outside the core, with an impact of about \( 2\,\%\) beyond \(r\gtrsim 0.3\,h^{-1}\,\mathrm{Mpc}\). As an outcome of this analysis, the \texttt{CosmoPostProcess} framework delivers radial profile corrections with associated uncertainties, combining the selection bias from projection effects with the impact of baryonic physics and miscentring. These corrections will be a key ingredient to ensure controlled systematics in the \Euclid DR1 galaxy cluster cosmological analysis.}


\titlerunning{{Galaxy clusters selection bias}}
\authorrunning{{Euclid Collaboration: R. Ingrao et al.}}

\maketitle
\nolinenumbers

\section{\label{sc:intro}Introduction}

Galaxy clusters represent the most massive gravitationally bound systems formed through hierarchical gravitational collapse. They anchor the nodes of the cosmic web and trace the growth of structure, providing a powerful probe of cosmology \citep[e.g.][]{2001Natur.409...39B,2011ARA&A..49..409A,2012ARA&A..50..353K}. The comoving number density of clusters as a function of mass and redshift is sensitive to the total matter density parameter, $\Omega_\mathrm{m}$, and to $\sigma_8$, which sets the normalisation of the linear matter power spectrum at the scale of $8\,h^{-1}\,\mathrm{Mpc}$  \citep[e.g.][]{2001MNRAS.323....1S,2008ApJ...688..709T,2013MNRAS.433.1230W,2016MNRAS.456.2486D,2023A&A...671A.100E}.
Furthermore, modern surveys demonstrated the capability of clusters to yield competitive constraints on dark energy and structure growth by comparing observed counts with halo mass function predictions \citep[e.g.][]{2015MNRAS.449..199M,2019MNRAS.488.4779C,2021A&A...653A..19G,2022A&A...665A.100L,2024A&A...689A.298G,2025PhRvD.111f3533B}.

To exploit this sensitivity, cluster cosmology requires robust selection functions and precise mass calibration \citep{2013PhR...530...87W,2019SSRv..215...25P}. Therefore, cluster samples ought to be detected efficiently over wide areas, with well-characterised selection observables (e.g. optical richness, X-ray flux, or a Sunyaev--Zeldovich signal) and an accurately calibrated mapping between these observables and halo mass. Forthcoming wide-field surveys will sharpen these requirements.

\Euclid \citep{Laureijs11,EuclidSkyOverview} will survey the extragalactic sky with two complementary instruments: the Visible Imager (VIS), providing high-resolution optical imaging for shape measurements \citep{Cropper16,EuclidSkyVIS}, and the Near-Infrared Spectrometer and Photometer (NISP), delivering photometry in the \YE, \JE, and \HE bands \citep{EuclidSkyNISP,EuclidSkyNISPCU}. Together, they will cover about $14\,000\,\mathrm{deg}^2$. The wide survey is expected to detect of order $\mathcal{O}(10^{5})$ galaxy clusters and proto-clusters out to $z \gtrsim 2$ \citep{2016MNRAS.459.1764S}. These data will provide large homogeneous cluster samples with high-quality lensing and photometric information. Over such an area, statistical uncertainties become sub-dominant and the constraining power is set increasingly by the control of measurement and selection effects, particularly in the cluster mass calibration.

Achieving this level of precision requires an accurate calibration of the relation between a survey mass proxy, such as the optical richness, $\lambda$, and the true halo mass. Weak lensing measurements of cluster density profiles are the primary tool because they probe the total matter distribution in this context \citep[e.g.][]{2014MNRAS.439....2V,2019MNRAS.482.1352M,2024A&A...681A..66E,2025A&A...697A.184G}, without having to rely on assumptions of hydrostatic or dynamical equilibrium, and they are only weakly sensitive to the dynamical state of individual clusters. They can also be obtained homogeneously over wide areas and broad redshift ranges, and combined in stacked analyses to reduce the statistical noise for low-mass or high-redshift systems. Results from the Dark Energy Survey \citep[DES,][]{2005astro.ph.10346T} illustrate both the power of this approach and its sensitivity to analysis choices.\footnote{See also \url{https://www.darkenergysurvey.org/} for an updated overview of the survey.} The DES Year 1 analysis of optically selected clusters found a strong tension with {\it Planck} cosmic microwave background cosmology \citep{2019MNRAS.488.4779C,2020MNRAS.496.4468S,2020A&A...641A...6P}, later traced to biases in the weak lensing mass calibration induced by selection effects. A forward-modelling treatment of selection, consistently accounting for line-of-sight (LoS) projections, brought DES Year 1 into agreement with {\it Planck} \citep{2024PhRvL.133v1002S,2025PhRvD.112h3535A}. These results highlight the need to control three main sources of systematic uncertainty in weak lensing cluster mass calibration: LoS projections, miscentring, and baryonic physics.

Projection effects originate from structures located along the same LoS that contaminate the observables of photometrically selected galaxy clusters; this confusion is primarily induced by the uncertainty in photometric redshift estimates. As a result, both richness measurements and weak lensing signals can be biased \citep{2014ApJ...797...34M,2024A&A...681A..67E,EP-Ragagnin}. These superposed structures may artificially increase stacked density profiles and enhance the inferred clustering amplitude \citep{Wu2022,2023MNRAS.521.5064S,2024PhRvD.110j3508Z}. The effect is strongly scale dependent, with the largest impact generally found around the cluster boundary \citep{2026PhRvD.113f3559N,2025PhRvD.111f3502L}.

Miscentring constitutes a second major source of uncertainty. In optical catalogues, the cluster centre is typically assigned to the brightest cluster galaxy (BCG) or to the peak of the galaxy density map. However, these proxies can be offset from the assumed halo centre, defined in simulations as the minimum of the gravitational potential or the maximum of the mass density field. Furthermore other observational tracers such as X-ray or Sunyaev--Zeldovich (SZ) peaks can provide a pathway to calibrate this systematic \citep{2015MNRAS.454.2305S,2019ApJS..244...22H,2021A&A...653A..19G,2025MNRAS.536..572D}. Offsets between assigned and true centres bias the inner lensing profile and, if uncorrected, lead to underestimated halo masses. From observations, miscentring affects $20\,\%$ to $40\,\%$ of clusters, depending on redshift and selection method. This is commonly accounted by assuming a radial offset distribution often calibrated with ICM data \citep{2007arXiv0709.1159J,2014MNRAS.438...49R,2022A&A...661A..11C,2024MNRAS.532.3359S}.

Finally, baryonic effects from gas cooling, star formation, and active galactic nucleus (AGN) feedback modify halo density profiles and therefore the corresponding lensing signal. Hydrodynamical simulations provide detailed predictions for these effects, but are computationally expensive. As an alternative, semi-analytic baryonification methods \citep{2014MNRAS.441.1769C,2015JCAP...12..049S,Schneider19,2019OJAp....2E...4C,2021MNRAS.500.2316C,2024A&A...690A.188A,EP-Castro} allow us to model the redistribution of stars, cold gas, and hot gas, while displacing dark matter particles accordingly, with parameters calibrated on hydrodynamical simulations. This provides a realistic correction to dark matter-only predictions while avoiding the cost of full hydrodynamical runs.

As we demonstrate in this work, these systematics are not independent, as projection, miscentring, and baryonic effects can couple in nontrivial ways, impacting both selection and mass calibration. Therefore, a coherent forward modeling approach is essential to meeting the requirements for the control on systematic errors in \Euclid.

In this paper, we describe \texttt{CosmoPostProcess}, a forward-modelling pipeline developed to generate realistic mock observables for \Euclid optical cluster analyses. The code takes as input large-volume $N$-body simulations and produces cluster catalogues with galaxy-based richness and stacked weak lensing signals, designed to mirror as closely as possible the measurement choices adopted within the \Euclid cluster cosmology framework.

The pipeline paints galaxies onto the simulated matter distribution with calibrated halo-occupation prescriptions and assigns richness with a neural-network emulator that reproduces the behaviour of the official \Euclid richness algorithm. Weak lensing observables are then constructed from the same realisation, including two key effects that influence both richness and lensing: 1) offsets between the adopted and true centres; and 2) baryonic modifications of the matter profiles. These ingredients are implemented directly at the particle level, using flexible prescriptions adapted to \Euclid.

With these mocks, we have the ability to quantify the lensing-selection bias by comparing a richness-selected sample with a mass-matched reference sample. Here, we study how the effect varies with scale, redshift, and richness. The same framework also allows us to test the stability of the inferred trends against changes in the forward-model assumptions. The results presented here provide a baseline for the treatment of these systematics in stacked-lensing measurements for \Euclid Data Release~1 (DR1).

The paper is organised as follows: \cref{sc:PF} summarises the \Euclid cluster finders and the richness estimator, which define the behaviour we emulate in \texttt{CosmoPostProcess}. \Cref{sc:Simulations} describes the simulations used as input to the mock generation and for calibration. \Cref{sc:Methodology} details the forward model, including galaxy painting and richness assignment, the computation of projected density and lensing profiles, miscentring, and the implementation of baryonic effects, which are discussed in depth in \cref{sc:Baryon}. Next, \cref{sec:results} presents the selection bias, separating the contributions from projections, miscentring, and baryons, while quantifying their variation with halo-occupation assumptions and cosmology. In \cref{sec:conclusions}, we discuss the results and draw our conclusions.

\section{\label{sc:PF}Galaxy clusters detection in \Euclid}

In this section, we introduce the main \Euclid algorithms used for the optical cluster detection and probabilistic richness assignment. These are the components of the official \Euclid pipeline that \texttt{CosmoPostProcess} emulates to construct realistic mock cluster catalogues (see \cref{sc:membership}). \Euclid adopts two complementary cluster finders \citep{2019A&A...627A..23E,2019MNRAS.485..498M}: \texttt{AMICO} is an optimal matched-filter algorithm that produces minimum-variance estimates of cluster amplitude and assigns probabilistic memberships through an iterative cleaning of the filtered maps \citep{2018MNRAS.473.5221B}; \texttt{PZWav} \citep{Gonzalez2014Sexten,2023MNRAS.519.2630W,2024A&A...685A..98D} is an adaptive tomographic wavelet method that constructs probability-weighted density maps in redshift slices and identifies significant peaks with minimal model assumptions \citep{2019A&A...627A..23E}. These two cluster-finding algorithms have demonstrated robust performance on \Euclid data, as shown by \citet{2025arXiv250319196E}, where they successfully recovered hundreds of high-significance systems with richness estimates that trace external mass proxies using the first \Euclid Quick Data Release (Q1). Besides  \Euclid, \texttt{AMICO} has been validated in wide weak lensing imaging \citep{2025A&A...701A.201M,2025A&A...703A..25L} and in deep fields such as COSMOS, a two square degree multi-wavelength survey, and COSMOS-Web, a deep JWST NIRCam survey that probes structure at earlier cosmic times \citep{2024A&A...687A..56T,2025A&A...697A.197T,Thongkham_2026}. As for \texttt{PZWav}, it has been tested in realistic multi-band datasets and in J-PAS, a wide photometric survey that observes galaxies through many narrow optical filters to obtain accurate information on photometric redshifts, where it displays a close agreement with \texttt{AMICO} as shown in \cite{2024A&A...685A..98D}. Taken together, these results support the adoption of \texttt{AMICO} and \texttt{PZWav} as the baseline cluster finders for \Euclid. Although \texttt{AMICO} also provides a richness estimate, the richness definition adopted for the \Euclid cosmological analysis can be obtained with the dedicated \texttt{RICH-CL} algorithm; further details are given in \cref{sc:RICH-CL}.
Their application to real data demonstrates also well-quantified selection functions, competitive purity and completeness, and consistent agreement with external mass proxies \citep{2019A&A...627A..23E,2024A&A...687A..56T,2025arXiv250319196E}.
We employed runs using both finders on official \Euclid galaxy mock light cones based on the \Euclid \texttt{Flagship2} simulation \citep{EuclidSkyFlagship} to keep the emulator tied to pipeline observables and to the operational definitions of the algorithms (see \cref{sc:Simulations,sc:galaxy-painting,sc:membership}).

As a case study, in the following we present results based on the \texttt{PZWav} cluster catalogue. However, given the similar performance of the two cluster-finding algorithms, we expect that the main results presented in this paper can be easily extended to \texttt{AMICO}.

\section{\label{sc:Simulations}Simulations}

This section describes the numerical simulations used to calibrate and apply \texttt{CosmoPostProcess}. The halo occupation model and the probabilistic richness emulator are calibrated on the forward-modelled \Euclid mock galaxy catalogue \texttt{Flagship2} (\cref{sc:Flagship-sim}). The baryonic modification of halo density profiles is calibrated using the hydrodynamical \texttt{Magneticum} simulations (\cref{sc:Magneticum}). The fully calibrated pipeline is then applied to large-volume dark matter simulations from the \texttt{PICCOLO} suite (\cref{sc:PICCOLO}), which provide the underlying matter distribution used to quantify selection biases.

\subsection{\label{sc:PICCOLO}\texttt{PICCOLO}}

The dark matter large-scale structure we applied the forward model to was based on simulations from the \texttt{PICCOLO} suite, originally presented by \citet{2023A&A...671A.100E}. That work introduced a set of 15 $\Lambda$CDM cosmologies designed to sample parameter combinations consistent with cluster abundance constraints from the joint DES and South Pole Telescope analysis \citep{2021PhRvD.103d3522C}.

In the present analysis, we extended that original set by including six additional cosmologies and by re-running both the original and the newly added models at higher mass resolution. These additional simulations were produced in view of the future \Euclid\ DR1 selection bias analysis, for which a broader sampling of cosmological parameter space will be required. The simulations retain the same box size and numerical framework as in \citet{2023A&A...671A.100E}, while increasing the particle number to improve the modelling of small-scale density profiles and projected observables relevant for cluster lensing.

Each simulation follows the evolution of $4 \times 2560^3$ particles in a periodic box of side length $1290\,h^{-1}\,\mathrm{Mpc}$. For the fiducial cosmology this corresponds to a particle mass of $2.8 \times 10^9\,h^{-1}\,M_\odot$. Initial conditions are generated at $z_{\mathrm{init}} = 24$ using third-order Lagrangian perturbation theory with the \texttt{monofonIC} code \citep{2020ascl.soft08024H}, and the evolution is performed with the TreePM gravity solver implemented in \texttt{OpenGadget3} \citep{2023MNRAS.526..616G,2024A&A...692A..81D,2025arXiv250401061D}. For each cosmology, we stored 16 snapshots spanning the redshift range $0 \le z \le 2$.

Among the available cosmologies, we focus here on two representative models, labelled C0 and C1. The reference cosmology C0 is consistent with {\it Planck} results \citep{2020A&A...641A...6P} and corresponds to the fiducial model of the original \texttt{PICCOLO} suite. It adopts $\Omega_{\mathrm m}=0.3158$ and $\sigma_8=0.8102$, yielding $S_8\equiv\sigma_8\sqrt{\Omega_{\mathrm m}/0.3}=0.831$ (\cref{tab:piccoloC0C1}).

The second cosmology C1 is characterised by a lower matter density, $\Omega_{\mathrm m}=0.1986$, and a higher fluctuation amplitude, $\sigma_8=0.8590$, corresponding to $S_8=0.699$. The contrast between C0 and C1 probes variations both along and across the $\Omega_{\mathrm m}$--$\sigma_8$ degeneracy direction that affects many large-scale structure observables. Differences between these models primarily impact the amplitude of the two-halo contribution in cluster lensing profiles and therefore provide a useful test of the sensitivity of the selection bias signal to cosmological parameters. For the scope of this paper, we restricted the analysis to the fiducial C0 run and to C1, which together serve to illustrate the capability of the pipeline to capture the cosmological dependence of the bias. A comprehensive exploration of the full cosmological grid, motivated by the future DR1 analysis, is deferred to future work, for the currently used sets of cosmological parameters are summarised in \cref{tab:piccoloC0C1}.

\begin{table}[ht!]
\centering
\small
\setlength{\tabcolsep}{5pt}
\caption{\texttt{PICCOLO} high-resolution simulations.}
\label{tab:piccoloC0C1}
\begin{tabularx}{\columnwidth}{@{} l *{4}{>{\centering\arraybackslash}X} @{}}
\toprule
Run & $\Omega_{\mathrm m}$ & $h$ & $n_{\mathrm s}$ & $\sigma_8$ \\
\midrule
C0 & 0.3158 & 0.6732 & 0.9661 & 0.8102 \\
C1 & 0.1986 & 0.7267 & 0.9775 & 0.8590 \\
\bottomrule
\end{tabularx}
\tablefoot{Columns list the present-day matter density, $\Omega_{\mathrm m}$, reduced Hubble parameter, $h\equiv H_0/(100\,\mathrm{km\,s^{-1}\,Mpc^{-1}})$, scalar spectral index, $n_{\mathrm s}$, and linear-theory fluctuation amplitude, $\sigma_8$.}
\end{table}

\subsection{\label{sc:Flagship-sim}\Euclid \texttt{Flagship2}}

\texttt{Flagship2} is the reference mock galaxy catalogue for \Euclid, built from a $(3600\,h^{-1}\,\mathrm{Mpc})^3$ dark matter simulation evolved with $16\,000^3$ particles using the \texttt{PKDGRAV3} code \citep{2017ComAC...4....2P} and adopting a flat $\Lambda$CDM cosmology consistent with {\it Planck}.\footnote{For reference, \texttt{Flagship2} assumes $\Omega_{\mathrm m}=0.319$, $\Omega_{\mathrm b}=0.049$, $h=0.67$, $n_{\mathrm s}=0.96$, and $\sigma_8=0.83$ \citep{EuclidSkyFlagship}.} A light cone covering the full sky out to $z=3$ is produced on the fly; an octant of approximately $5200\,\mathrm{deg}^2$ is released with galaxies painted using a hybrid halo occupation and abundance-matching model to reproduce the \Euclid photometric and spectroscopic selection functions \citep{EuclidSkyFlagship}. In this work, we used the \texttt{Flagship2} catalogues retrieved through \texttt{CosmoHub} \citep{2017ehep.confE.488C,2020A&C....3200391T} and restricted the analysis to an eight-tile $400\,\mathrm{deg}^2$ portion of the light cone, corresponding to the same subset used in \Cref{app:flagship} to calibrate the halo occupation model and to train the neural-network emulator of $P_{\mathrm{mem}}$. This choice ensures that the catalogue used in the main analysis is fully consistent with the reference sample underlying both the mass--richness calibration and the richness-emulation framework. It also reflects a pragmatic compromise: for the specific calibration target considered here, the adopted $400\,\mathrm{deg}^2$ subset already provides sufficient statistics, while retaining a calibration scheme that is computationally light and can be readily rerun and retuned as the modelling choices are updated. Although using the full octant would further reduce the statistical noise, \Cref{app:flagship} shows that the emulator trained on these eight tiles preserves the mass--richness relation with residuals consistent with zero within the uncertainty band, and that the resulting selection bias signal is consistent with that measured directly on \texttt{Flagship2}. Optical cluster catalogues are generated using the \texttt{AMICO} and \texttt{PZWav} detection pipelines, and cluster richness is assigned with \texttt{RICH-CL} using the probabilistic scheme described in \cref{sc:RICH-CL}. The resulting catalogue provides the reference membership probabilities and mass--richness relation used to calibrate and validate the halo occupation model and the membership emulator implemented in \texttt{CosmoPostProcess}. For this purpose, detected clusters are matched to \texttt{Flagship2} halos using a galaxy-membership-based scheme, associating each cluster with the halo contributing the largest fraction of its observed richness. We refer to quantities obtained with this procedure as membership-matched. This matching strategy is intended to anchor the calibration to the per-galaxy membership information extracted from \texttt{Flagship2}; namely, the relation between the membership probability and the observables entering the richness assignment. In this sense, the \texttt{Flagship2}-based calibration is not meant to define the final DR1 galaxy population model, but rather to demonstrate that \texttt{CosmoPostProcess} can reproduce the relevant cluster observables in a controlled way for a given HOD prescription. The residual sensitivity to the assumed galaxy population is expected to enter primarily through the HOD modelling, and will be assessed through the suite of HOD variations adopted in the production of the final DR1 mocks.

\subsection{\label{sc:Magneticum}\texttt{Magneticum}}

\texttt{Magneticum} is a suite of cosmological hydrodynamical simulations run with the \texttt{P-Gadget3} code, an evolution of the public \texttt{P-Gadget2} code \citep{2005MNRAS.364.1105S}. It includes sub-resolution models for radiative cooling, star formation, chemical enrichment, supernova-driven winds, black-hole growth, and AGN feedback \citep{2014MNRAS.442.2304H}. The background cosmology corresponds to WMAP7 \citep{2011ApJS..192...18K}.

We used two configurations: the high-resolution Box 3 run and a multi-cosmology variant spanning 15 cosmologies obtained by varying $\Omega_{\mathrm m}$, $\Omega_{\mathrm b}$, $\sigma_8$, and $h$ around the WMAP7 fiducial values. Subgrid physics is held fixed across all runs \citep{2023A&A...675A..77R,2020MNRAS.494.3728S}. These simulations were used to calibrate the baryonic modification of halo density profiles implemented in the forward model. In \cref{tab:magneticum-wide}, we summarise the relevant configurations.

\begin{table*}[t]
\caption{\texttt{Magneticum} configurations used in this work.}
\label{tab:magneticum-wide}
\centering
\setlength{\tabcolsep}{6pt}
\renewcommand{\arraystretch}{1.15}

\begin{tabular*}{\textwidth}{@{\extracolsep{\fill}} l c c c @{}}
\toprule
Run &
\parbox[c]{2.6cm}{\centering $L_{\rm box}$\\$[h^{-1}\,\mathrm{Mpc}]$} &
\parbox[c]{3.2cm}{\centering Number of\\particles} &
\parbox[c]{3.0cm}{\centering $m_{\rm part}$\\$[h^{-1}\,M_\odot]$} \\
\midrule
Box 3/uhr & 128 & $2 \times 1536^{3}$ & $3.6 \times 10^{7}$ \\
Multi-cosmology (M1--M15) & 896 & $2 \times 1536^{3}$ & $1.3 \times 10^{10}$ \\
\bottomrule
\end{tabular*}
\tablefoot{Cosmologies M1--M15 vary $\Omega_{\mathrm{m}}$, $\Omega_{\mathrm{b}}$, $\sigma_8$, and $h$ around the WMAP7 fiducial cosmology \citep{2023A&A...675A..77R,2020MNRAS.494.3728S}.}
\end{table*}

\section{\label{sc:Methodology}Methodology}

Here, we describe the methodology used to self-consistently simulate projected density profiles (\cref{sc:density-profiles}) and richness values, which are two key observables of the \Euclid cluster catalogue. The assignment of richness proceeds by first painting galaxies  on the simulations (\cref{sc:galaxy-painting}) and then assigning the membership to each cluster (\cref{sc:membership}). Finally, we assess the effect of miscentring on the observables (\cref{sc:miscentering}).

\subsection{\label{sc:density-profiles}Density profiles}

We computed 3D and 2D density profiles selecting and projecting dark matter particles by applying a tree-search based on the \texttt{KDTree} implementation in \texttt{SciPy} \citep{2020SciPy-NMeth}. This choice provides good scaling with the number $N$ of particles, $\mathcal{O}(N\log{N})$, which is convenient given the large number of particles per box (see \cref{sc:Simulations}).

For the 3D profiles, we searched for dark matter particles inside spheres centred on \texttt{SubFind} halos \citep{Springel2001,2009MNRAS.399..497D}. We used thirty log-equispaced radial bins from $0.01\,h^{-1}\,{\rm Mpc}$ to $5\,h^{-1}\,{\rm Mpc}$. By summing all particles enclosed in a sphere, we can compute the cumulative mass as
\begin{equation}
M(<r_{\rm i}) = N_i^{\mathrm{part}} m_{\rm part}\,,
\end{equation}
where $m_{\rm part}$ is the mass of a DM particle in our simulations, and $N_{i}^{\mathrm{part}}$ is the number of particles in the $i$-th sphere around the target halo centre. We apply this to all halos with $\Mvir \geq 10^{13}\,h^{-1}M_{\odot}$. These profiles serve as input for the baryonic correction model (see \cref{sc:Baryon}).

To compute surface density profiles, we counted particles inside a cylinder \citep[see e.g.][]{Wu2022,2024PhRvL.133v1002S}. The cylinder has a depth of $\pm 50\,h^{-1}\,{\rm Mpc}$ around the target cluster. The default value of the base radius is $R_{\text{max}} = 10\, h^{-1}\,{\rm Mpc}$ and can be customised. The cylinder is then split using $20$ log-spaced annuli from $R_{\rm min}=0.01\,h^{-1}{\rm Mpc}$ to $R_{\text{max}}$. We projected particles along the chosen axis and computed
\begin{equation}
\Sigma(r_{i}^{\mathrm{cen}}) \;=\; \frac{\left(N^{\mathrm{part}}_{i}-{N_{i-1}^{\mathrm{part}}}\right)\,m_{\rm part}}{\pi\,(R_{i}^2-R_{i-1}^2)}\,,
\label{eq:profiles}
\end{equation}
where $R_{i-1}$ and $R_{i}$ are the bin edges and $r_{i}^{\mathrm{cen}} = \sqrt{R_{i}\,R_{i-1}}$.

\begin{figure}
    \centering
    \includegraphics[scale=0.55]{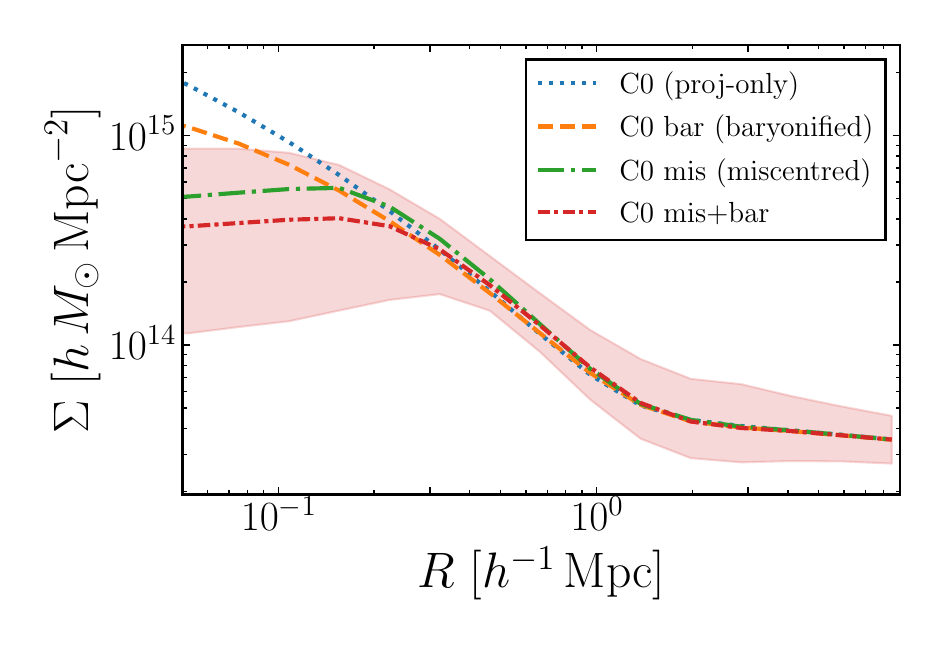}
    \caption{Median surface density profiles, $\Sigma$, and relative scatter for about $3000$ objects within a narrow mass bin centred on $M=10^{14}\,h^{-1}\,M_{\odot}$. The blue dotted curve is for the original average profile, while dashed orange, dot-long dashed green, and dot-short dashed red curves refer to the effect of including baryonic correction, miscentring, and both effects together, respectively. For the latter, the shaded area indicates the $1\,\sigma$ statistical uncertainty.}
    \label{fig:profiles}
\end{figure}

 We then applied baryonic corrections (see \cref{sc:Baryon}) inside a spherical region covering $5\,h^{-1}\,{\rm Mpc}$ in radius. This region encompasses the full one-halo term and extends slightly beyond the transition to the two-halo regime, allowing us to trace the full baryonic density profile measured in \texttt{Magneticum} (see \Cref{app:calib}). An actual split in one-halo and two-halo component of the profiles is applied afterward, during cylinder projection, to disentangle the effects of correlated structures from that of the main halo. More specifically, this decomposition is defined geometrically in configuration space rather than through a separate halo-model fit. During the projection step, particles inside the cylinder are assigned to the one-halo component if their three-dimensional (3D) distance from the halo centre satisfies $R^2+z^2 \leq r_{\rm 1h}^2$, where $R$ is the projected separation, $z$ is the LoS separation, and $r_{\rm 1h}=r_{\rm vir}$ is taken as the reference halo radius used in the simulation catalogue, optionally rescaled by a user-provided factor. Particles outside this radius, but still within the projection cylinder, contribute to the two-halo component. The two terms are then accumulated separately in the same projected radial bins. For illustrative purposes, in \cref{fig:profiles} we show the median projected profile (blue dashed line) measured in the \texttt{PICCOLO} C0 simulation, measured in a narrow mass bin (using the virial mass definition) around $M=10^{14}\,h^{-1}\,M_{\odot}$ together with the modifications introduced by the various systematics (see \cref{sc:miscentering,sc:Baryon}). To increase the statistics of our sample, we extract three profile catalogues by projecting along the three Cartesian axes for each box analysed. As an additional feature for studying orientation biases, we also measured the triaxial shape parameters and the orientation angle between the main axis of the inertia momentum tensor of the halo and the LoS (see \Cref{app:triaxiality} for further details).

\subsection{\label{sc:galaxy-painting}Galaxy painting}

Selection bias arises from the correlation, at fixed halo mass, between the up-scatter of the observed richness used for sample selection and the up-scatter of the weak lensing signal. To model this correlation consistently, halos must be populated with galaxies using the same dark matter structures employed to measure the projected density profiles. We therefore implemented a halo occupation distribution (HOD) model calibrated on the \texttt{Flagship2} mock galaxy catalogue and subsequently applied to the \texttt{PICCOLO} simulations.

The HOD parameters were determined by fitting the occupation statistics to the \texttt{Flagship2} cluster catalogue, ensuring that the simulated galaxy population reproduced the reference mass–richness relation described in \Cref{app:flagship}. Once calibrated, the same parametrisation was used to populate halos in \texttt{PICCOLO}, thereby isolating selection and projection effects from changes in the underlying galaxy model.

The model assigns central galaxies at the \texttt{SubFind} centre of each halo by drawing from a Bernoulli distribution with a probability of
\begin{equation}
P_{\mathrm{cen}}(\Mvir) = \frac{1}{2} \left\{ 1 + \mathrm{erf}\left[ \frac{1}{\sigma_{\log M, \mathrm{cen}}} \logten\left( \frac{\Mvir}{M_{\min, \mathrm{cen}}} \right) \right] \right\}\,.
\label{eq:cen-mass}
\end{equation}
The parameter \(M_{\min,\mathrm{cen}}\) sets the characteristic minimum halo mass required to host a central galaxy, while \(\sigma_{\log M,\mathrm{cen}}\) controls the width of the transition between halos with negligible central probability and those almost certainly hosting one. The function increases smoothly with halo mass, capturing the intrinsic scatter in the central–halo mass relation.

For halos hosting a central galaxy, the number, \(N_{\mathrm{sat}}\), of satellite galaxies is drawn from a Poisson distribution with an expectation value of
\begin{equation}
\label{eq:sat-mass}
N_{\mathrm{sat}}(\Mvir, z_{\mathrm{cl}}) =
N_{\mathrm{cen}}(\Mvir)
\left( \frac{\Mvir - M_{\min,\mathrm{cen}}}{M_{1,\mathrm{sat}} - M_{\min,\mathrm{cen}}} \right)^{\alpha}
\left( \frac{1 + z_{\mathrm{cl}}}{1 + z_0} \right)^{\epsilon}\,.
\end{equation}
Here, \(M_{1,\mathrm{sat}}\) defines the mass scale at which halos host on average one satellite galaxy, and \(\alpha\) governs the growth of satellite occupation with halo mass. The parameter \(\epsilon\) introduces a redshift dependence to account for evolutionary effects. Variations in \(M_{1,\mathrm{sat}}\) and \(\alpha\) modify the satellite abundance at fixed \(\Mvir\), thereby affecting both the richness distribution and its correlation with projected mass density.

The HOD described by \cref{eq:cen-mass,eq:sat-mass} was calibrated using the mass--richness relation measured from a set of optical cluster detections produced with \texttt{PZWav} on the \texttt{Flagship2} light cone. Detected clusters were associated to \texttt{Flagship2} halos through a membership-based matching scheme, whereby each cluster was matched to the halo contributing the largest fraction of its assigned richness. This choice suppresses, by construction, the contribution of chance projections to the calibration sample, yielding a reference relation that traces the intrinsic halo occupation at fixed mass.

The HOD parameters can then be obtained by fitting this matched mass--richness relation. Calibrating on a projection-minimised reference ensures that when \texttt{CosmoPostProcess} is applied to \texttt{PICCOLO}, the resulting richness boosts and scatter are driven by the \texttt{PICCOLO} density field. In this way, the projection-induced corrections inferred from \texttt{PICCOLO} quantify the selection effects native to those simulations, rather than residual contamination inherited from the calibration step. To apply the painting in the \texttt{PICCOLO} simulation, we passed  the virial mass derived with \texttt{SubFind} and the redshift for all halos with $M_{\rm vir} \geq 10^{13}\,h^{-1}\,M_{\odot}$ to the model. The redshift of the target halos are derived from the simulation box redshift \(z_{\mathrm{box}}\) and the LoS comoving coordinate, \(x_{\mathrm{proj}}\) in $h^{-1}\,\mathrm{Mpc}$, via 
\begin{equation}
\label{eq:redshift-comoving}
z_{\mathrm{cl}} = z_{\mathrm{box}} + D_{\rm c}^{-1}(x_{\mathrm{proj}})\,,
\end{equation}
where \(D_{\rm c}^{-1}\) denotes the inverse comoving distance–redshift relation.

\begin{figure}[ht]
    \centering
    \includegraphics[scale=0.5]{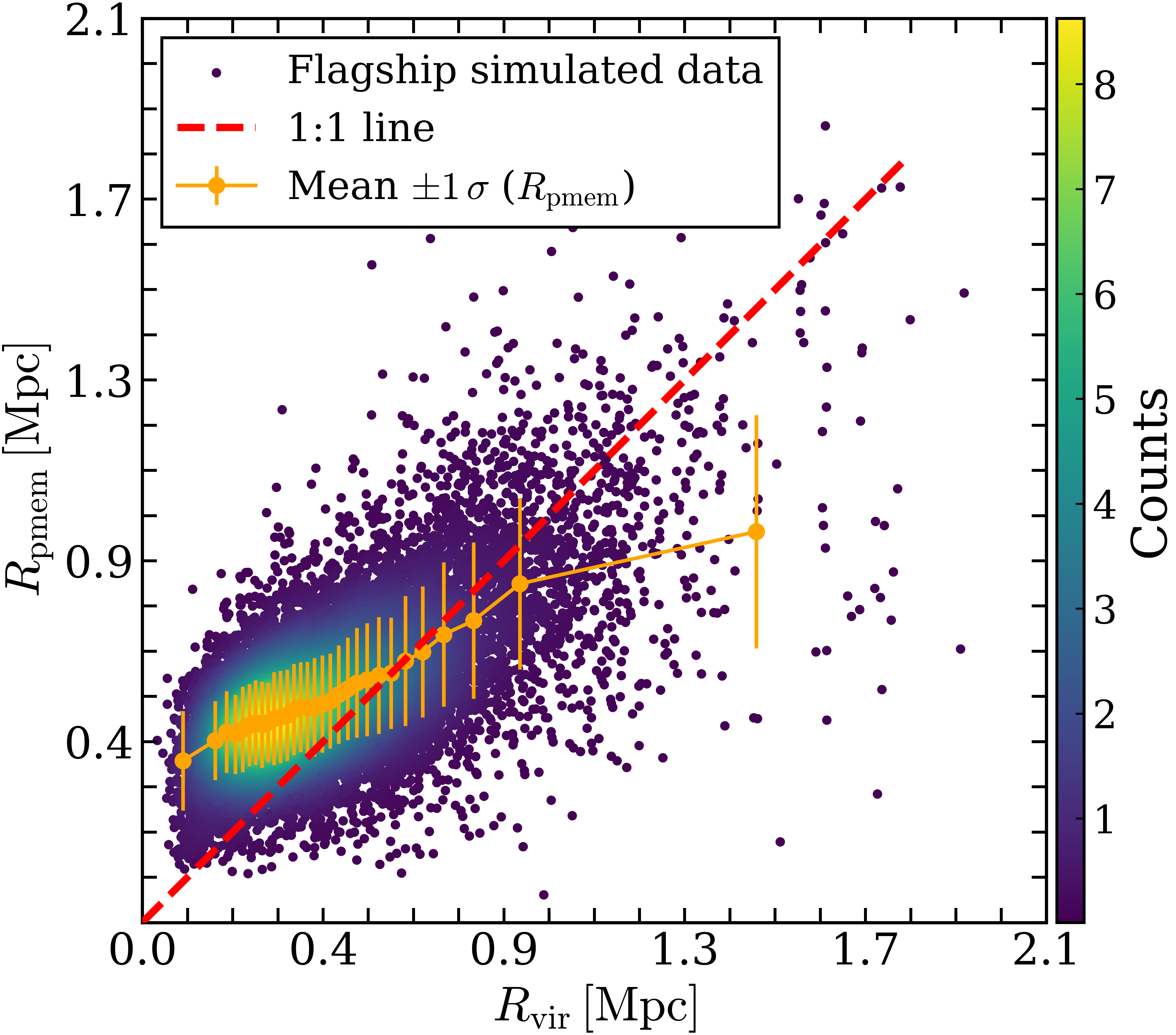}
    \caption{Relation between $R_{\rm vir}$ and $R_{\rm pmem}$ in proper $\mathrm{Mpc}$ units from \texttt{Flagship2}. Orange points show the binned mean and the standard deviation as a function of $R_{\rm vir}$.}
    \label{fig:rvir-vs-rpmem}
\end{figure}

Satellite galaxies were placed by uniformly selecting \(N_{\mathrm{sat}}\) dark matter particles from the host halo. For the painted galaxy catalogue to be consistent with the richness definition adopted later in the pipeline, this selection was restricted to the projected aperture entering the probabilistic \texttt{RICH-CL} estimator, namely, \(R_{\rm pmem}\). Since the \texttt{PICCOLO} halo catalogues provide virial radii \(R_{\rm vir}\) rather than \(R_{\rm pmem}\), we constructed a proxy mapping between the two scales. This mapping can be obtained by interpolating the empirical relation between \(R_{\rm vir}\) and \(R_{\rm pmem}\) measured in the membership-matched \texttt{Flagship2} cluster catalogue, including its intrinsic scatter. The resulting correspondence is shown in \cref{fig:rvir-vs-rpmem}.

This choice ensures that the spatial extent of the painted satellite population is tied to the same effective aperture used to estimate richness in the probabilistic framework described in \cref{sc:RICH-CL}. The uniform selection of dark matter particles preserves the triaxial structure of halos and maintains a direct connection between galaxy positions and the underlying mass distribution. The resulting galaxy catalogue provides the input for the membership assignment described in the following section.

\subsection{Membership assignation}

Optical richness is an observational mass proxy designed to trace the underlying cluster galaxy population. In the \Euclid cluster cosmology pipeline, cluster masses are inferred from stacked tangential shear profiles using the \texttt{COMB-CL} pipeline, which constructs individual and stacked lensing profiles under homogeneous source selection and lensing modelling. This approach has been validated on Stage-III data and simulations \citep{2024A&A...689A.252E, 2025A&A...695A.280E}. The optical richness entering the mass--richness calibration is estimated by the \texttt{RICH-CL} algorithm.

The \texttt{RICH-CL} code includes two alternative richness estimators: a red-sequence-based implementation and a probabilistic membership implementation. In this work, we adopted the latter as reference, since it is the estimator currently foreseen for the official \Euclid DR1 cosmological analysis. In contrast to the red-sequence-based definition, the probabilistic estimator does not apply a colour selection and is therefore expected to be more affected by projection effects from structures aligned along the LoS. Throughout this section, the \texttt{RICH-CL} richness refers to this probabilistic definition. Within \texttt{CosmoPostProcess}, we reproduce it through a neural-network emulator trained to match the \texttt{RICH-CL} output on \texttt{Flagship2}. We first summarise the theoretical framework underlying this estimator and then describe its implementation in the forward model.

\subsubsection{\label{sc:RICH-CL}Theoretical framework}

In optical cluster cosmology, richness estimators do not correspond to a direct count of member galaxies; rather, they are defined as weighted sums over galaxies in the cluster region, where the weights encode the probability of membership and account for projection effects and background contamination \citep[e.g.][]{2009ApJ...699..768R, 2014ApJ...785..104R, 2015MNRAS.453...38R, 2015A&A...582A.100A}. Our description follows the methodology presented by \citet[][hereafter \citetalias{2016A&A...595A.111C}]{2016A&A...595A.111C}, which provides a procedure closely related to that implemented in \texttt{RICH-CL} and serves as a practical reference for defining the inputs and validation steps of a \Euclid-like richness estimator. We focus, in particular, on the no-threshold variant of the probabilistic richness introduced by \citetalias{2016A&A...595A.111C}. This discussion is intended as a guide to the structure of a probabilistic membership estimator and to the meaning of the relevant ingredients. It should not be interpreted as a one-to-one description of the implementation adopted in \texttt{CosmoPostProcess}, for which the operative photometric-redshift treatment is described below in \cref{sc:membership}.

The richness $\lambda$ is computed as a weighted sum over galaxies within a projected aperture, \(R_{\rm pmem}\),
\begin{equation}
\lambda = \sum_{R \le R_{\rm pmem}}
\frac{P_{\mathrm{mem}}}{(1-f_{\rm bkg})\,C(m,z)}\,,
\label{eq:lambda_sum_CB16}
\end{equation}
where the sum runs over all galaxies whose projected separation from the cluster centre satisfies \(R \le R_{\rm pmem}\). In practice, \(R_{\rm pmem}\) defines the characteristic aperture within which the probabilistic membership assignment, and hence the richness estimate, is evaluated. This is the same aperture scale that we use in the galaxy-painting step to set the projected region from which satellite tracers are drawn. The factor \(1-f_{\rm bkg}\) accounts for the correction for survey geometry and masking, where \(f_{\mathrm{bkg}}\) is the fraction of the aperture lost to survey masking. The completeness correction \(C(m, z)\) accounts for the fraction of galaxies missing from the catalogue due to magnitude limits. In \citetalias{2016A&A...595A.111C}, richness is evaluated within a cluster-size aperture, taken as \(r_{\rm vir}\) in their tests.\footnote{The virial radius \(r_{\rm vir}\) is defined following the spherical-overdensity criterion adopted in the \texttt{SubFind} group catalogues, that is as the radius enclosing a mean interior density \(\bar{\rho}(<r_{\rm vir}) = \Delta_{\rm vir}(z)\,\rho_{\rm c}(z)\), where \(\rho_{\rm c}(z)=3H^2(z)/(8\pi G)\) is the critical density of the Universe at redshift \(z\), and \(\Delta_{\rm vir}(z)\) is the redshift-dependent overdensity predicted by the spherical-collapse model \citep{1998ApJ...495...80B}. The corresponding virial mass is \(M_{\rm vir}\equiv M(<r_{\rm vir})\).}

The membership probability, $P_{\mathrm{mem}}$, combines redshift information from both the galaxy and the cluster,
\begin{equation}
P_{\mathrm{mem}} =
\frac{(1 - \beta)\sum_{i} P_{\mathrm{gal}}(z_i) P_{\mathrm{cl}}(z_i)}
{\mathcal{N}(z_{\mathrm{cl}}, \sigma_0, \delta z)}\,,
\label{eq:Pmem_final_CB16}
\end{equation}
where $P_{\mathrm{gal}}(z)$ and $P_{\mathrm{cl}}(z)$ are the redshift probability distributions of the galaxy and the cluster. The sum runs over a redshift grid of spacing $\delta z$, and $\mathcal{N}$ ensures proper normalisation of the discrete redshift-overlap term. The galaxy redshift probability distribution is modelled as
\begin{equation}
P_{\mathrm{gal}}(z) =
\frac{1}{\sqrt{2\pi}\,\sigma_z}
\exp\left[-\frac{(z - z_{\rm gal}^{\rm photo})^2}{2\,\sigma_z^2}\right]\,,
\,\,\,\,\,
\sigma_z = \sigma_0(1 + z)\,,
\label{eq:Pg_CB16}
\end{equation}
where $\sigma_0$ sets the width of the illustrative Gaussian photometric-redshift kernel in this reference formulation. In the mock implementation described below, the effective photometric-redshift scatter is instead modelled with \cref{eq:z-dispersion}.

The factor $\beta$ quantifies contamination from background galaxies and is defined as
\begin{equation}
\beta =
\frac{\langle N_{\mathrm{bkg}}(m_{\mathrm{gal}}, z_{\mathrm{cl}}) \rangle}
{\langle N_{\mathrm{tot}}(m_{\mathrm{gal}}, z_{\mathrm{cl}}, R) \rangle}\,,
\label{eq:beta_CB16}
\end{equation}
where $N_{\mathrm{bkg}}$ and $N_{\mathrm{tot}}$ denote the background and total galaxy counts at magnitude $m_{\mathrm{gal}}$ and redshift $z_{\mathrm{cl}}$. In the \citetalias{2016A&A...595A.111C} framework, $N_{\mathrm{tot}}$ is evaluated locally at the projected cluster-centric separation $R \equiv R_{\rm c,g}$, while $N_{\mathrm{bkg}}$ is estimated from an outer annulus around the cluster. Consequently, $1-\beta$ carries an explicit radial dependence: near the cluster centre, $N_{\mathrm{tot}}\gg N_{\mathrm{bkg}}$ and $\beta\ll 1$, while at large projected radii $\beta\rightarrow 1$. As a simple approximation,
\begin{equation}
1-\beta(R)\approx
\frac{\Sigma_{\rm mod}(R)-\Sigma_{\rm bkg}}{\Sigma_{\rm mod}(R)}\,,
\end{equation}
where $\Sigma_{\rm mod}(R)$ is the projected surface density predicted by a reference model profile \citep{1997ApJ...490..493N}, and $\Sigma_{\rm bkg}$ is the projected background surface density estimated at large radii. In \citetalias{2016A&A...595A.111C}, no specific reference surface density profile is assumed, whereas more recent versions of \texttt{RICH-CL} constrain \(R_{\rm pmem}\) iteratively using a power-law profile. This probabilistic formulation defines richness as a sum over membership probabilities and provides the observable that we reproduce within the forward model. Further details on the \texttt{RICH-CL} probabilistic membership scheme in its definitive configuration for the Euclid DR1 cosmological analysis will be presented in Euclid Collaboration: Benoist et al.\ (in prep.).

\subsubsection{\label{sc:membership}Membership emulator}

The probabilistic richness defined in \cref{eq:lambda_sum_CB16} depends on the membership probability, $P_{\mathrm{mem}}$, given by \cref{eq:Pmem_final_CB16,eq:beta_CB16}. In \texttt{CosmoPostProcess}, we do not emulate the richness directly as a black-box mass--richness relation. Instead, for each galaxy-cluster pair entering the richness aperture, we emulate the corresponding membership probability and denote it by $P_{\mathrm{mem}}^{\mathrm{emu}}$, see \Cref{app:flagship} for extra details. This quantity is intended as a fast surrogate for the probabilistic \texttt{RICH-CL} membership weight, so that the observed richness can still be constructed as a sum over per-galaxy membership probabilities, consistently with \cref{eq:lambda_sum_CB16}. This preserves the structure and functional dependencies of the \texttt{RICH-CL} estimator while allowing an efficient evaluation for large simulated catalogues.

We emulate $P_{\mathrm{mem}}^{\mathrm{emu}}$ with a fully connected neural network comprising three hidden layers with Leaky ReLU activations and a sigmoid output, ensuring $0 \le P^{\mathrm{emu}}_{\mathrm{mem}} \le 1$. The implementation uses \texttt{PyTorch} \citep{PyTorch}. Training and validation are performed on the \texttt{Flagship2} simulation, where the reference \texttt{RICH-CL} membership probabilities are available. The trained network is subsequently applied to the painted \texttt{PICCOLO} catalogues. Besides reproducing the operational \texttt{RICH-CL} probabilistic estimator, this emulation provides a computationally efficient evaluation of $P_{\mathrm{mem}}^{\mathrm{emu}}$ for large simulated catalogues. This is particularly important for \texttt{CosmoPostProcess}, whose main goal is to trace how cosmology affects the amount of projection-induced contamination entering the richness selection, and therefore the resulting weak lensing selection bias, by processing many independent realisations and multiple cosmological parameter choices. In this framework, the $P_{\mathrm{mem}}$ emulator is not intended to encode a direct dependence on cosmological parameters. Rather, the same calibrated membership prescription is applied to cosmology-dependent galaxy and halo populations, so that the cosmology dependence of the bias arises from the projected structure entering the richness estimate. A direct evaluation of the full membership model for all simulated catalogues would be prohibitively expensive. The physical information is nevertheless preserved, since the input variables reflect directly the quantities entering the probabilistic formulation: cluster redshift, projected galaxy-cluster separation normalised by $R_{\rm pmem}$, background-area fraction, $f_{\mathrm{bkg}}$, and galaxy-cluster redshift separation, $\Delta z$.

To compute $\Delta z$, we simulate photometric redshift estimates for both galaxies and clusters. Starting from the true galaxy redshift, we added a Gaussian scatter with standard deviation,
\begin{equation}
\sigma_{\mathrm{photo}\mbox{-}z}(z)=z_0 + b\,z + a\,z^2\,,
\label{eq:z-dispersion}
\end{equation}
with default values $a=0.041$, $b=-0.046$, and $z_0=0.034$, derived from a \texttt{Phosphoros} DR1-like run on the same \texttt{Flagship2} sample. An analogous scatter is applied to the cluster redshift, reduced by a factor $1/\sqrt{N_{\mathrm{mem}}}$, consistently with the \texttt{RICH-CL} probabilistic definition. This choice should be regarded as an educated baseline configuration adopted to demonstrate the capabilities of the mock-generation framework, rather than as a final description of the DR1 photometric-redshift uncertainties. As the understanding of the DR1 data improves, both the galaxy and cluster photometric-redshift uncertainties will likely require a broader effective description than the one assumed here. The observed richness is then computed by summing the emulated per-galaxy membership probabilities within $R_{\rm pmem}$,

\begin{equation}
\lambda_{\rm obs} =
\sum_{R \le R_{\rm pmem}}
P^{\mathrm{emu}}_{\mathrm{mem}}\,.
\end{equation}

\begin{figure}[ht]
    \centering
    \includegraphics[scale=0.3]{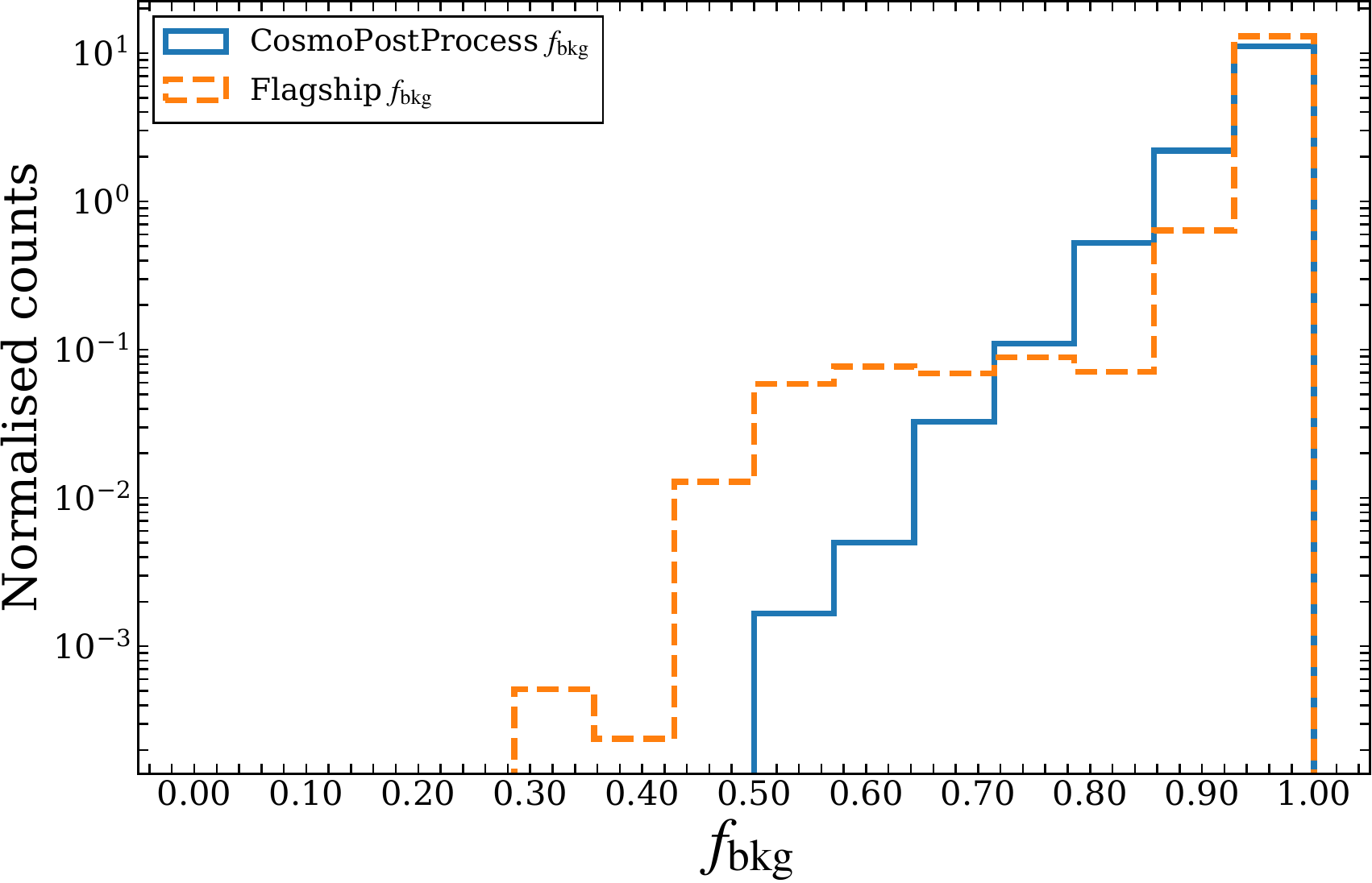}
    \caption{Normalised counts as a function of $f_{\mathrm{bkg}}$, as predicted by the \texttt{CosmoPostProcess} model (solid blue line) and by the \texttt{Flagship2} model (orange dashed line).}
    \label{fig:f_bkg}
\end{figure}

\begin{figure}[ht]
    \centering
    \includegraphics[scale=0.48]{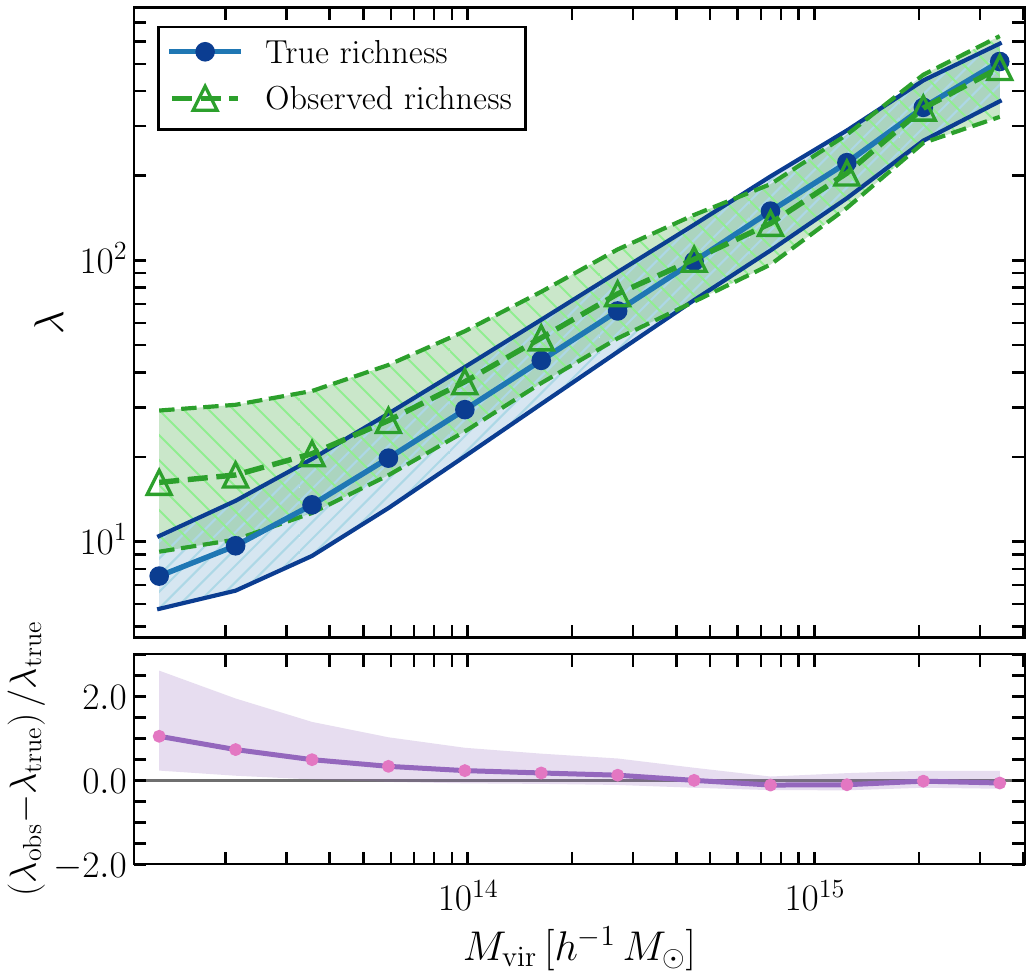}
    \caption{\textit{Upper panel:} Median mass--richness relation and $1\sigma$ scatter envelope. The blue curve shows the relation for the true richness calibrated on \texttt{Flagship2}, while the green curve shows the observed richness obtained by recomputing the richness with the \texttt{CosmoPostProcess} membership-probability emulator. Here, observed richness denotes the final richness returned by the mock pipeline after the emulation and post-processing steps, in contrast to the true richness defined at the HOD level. \textit{Lower panel:} Relative variation of the observed richness with respect to the true richness, $(\lambda_{\rm obs}-\lambda_{\rm true})/\lambda_{\rm true}$. The offset is not introduced through a separate correction term, but arises because projected galaxies and galaxies overlapping in photometric redshift can receive non-zero emulated membership probabilities, $P_{\mathrm{mem}}^{\mathrm{emu}}$, and therefore contribute to $\lambda_{\rm obs}$. This naturally produces an upward shift and an increased scatter in the reconstructed richness, especially at low halo masses, with projection effects acting on top of the usual scatter induced by Eddington bias.}
    \label{fig:SR}
\end{figure}

As previously described for $R_{\rm pmem}$, we implemented a prescription to model an $f_{\mathrm{bkg}}$ proxy in \texttt{CosmoPostProcess}. For each cluster included in the richness catalogue, by default all galaxy-painted-halos with $\Mvir>10^{13}\,h^{-1}\,M_{\odot}$, we consider an annulus centred on the projected cluster position spanning $R_{\min}^{\mathrm{area}}=3\,h^{-1}\,{\rm Mpc}$ to $R_{\max}^{\mathrm{area}}=5\,h^{-1}\,{\rm Mpc}$, and compute the fraction of this annular area obscured by neighbouring projected clusters. In practice, we evaluate the area covered by circles of radius $R_{\rm pmem}$ centred on the projected positions of clusters overlapping the annulus; the remaining unobscured fraction defines our estimate of $f_{\mathrm{bkg}}$. \Cref{fig:f_bkg} compares the overall distribution of $f_{\mathrm{bkg}}$ for objects above $10^{13}\,h^{-1}\,M_{\odot}$ measured in \texttt{Flagship2} and in \texttt{CosmoPostProcess} applied to \texttt{PICCOLO}. The agreement is only qualitative, but this is sufficient for the present purpose. In particular, the main role of the $f_{\mathrm{bkg}}$ model is to capture the contribution from neighbouring foreground clusters, while we do not attempt here to reproduce the full survey masking pattern, which is more complex and not yet in a form that can be injected straightforwardly into the simulations for DR1. As shown in \Cref{app:flagship}, $f_{\mathrm{bkg}}$ is the least informative input for $P_{\mathrm{mem}}$, and tests in which it is omitted altogether produce no appreciable change in the calibration. For this reason, we regard the simplified treatment adopted here as sufficient for the current implementation.

\begin{figure}[h]
    \begin{adjustwidth}{-0.005\linewidth}{-0.005\linewidth}
    \centering
    \includegraphics[width=1.0\linewidth]{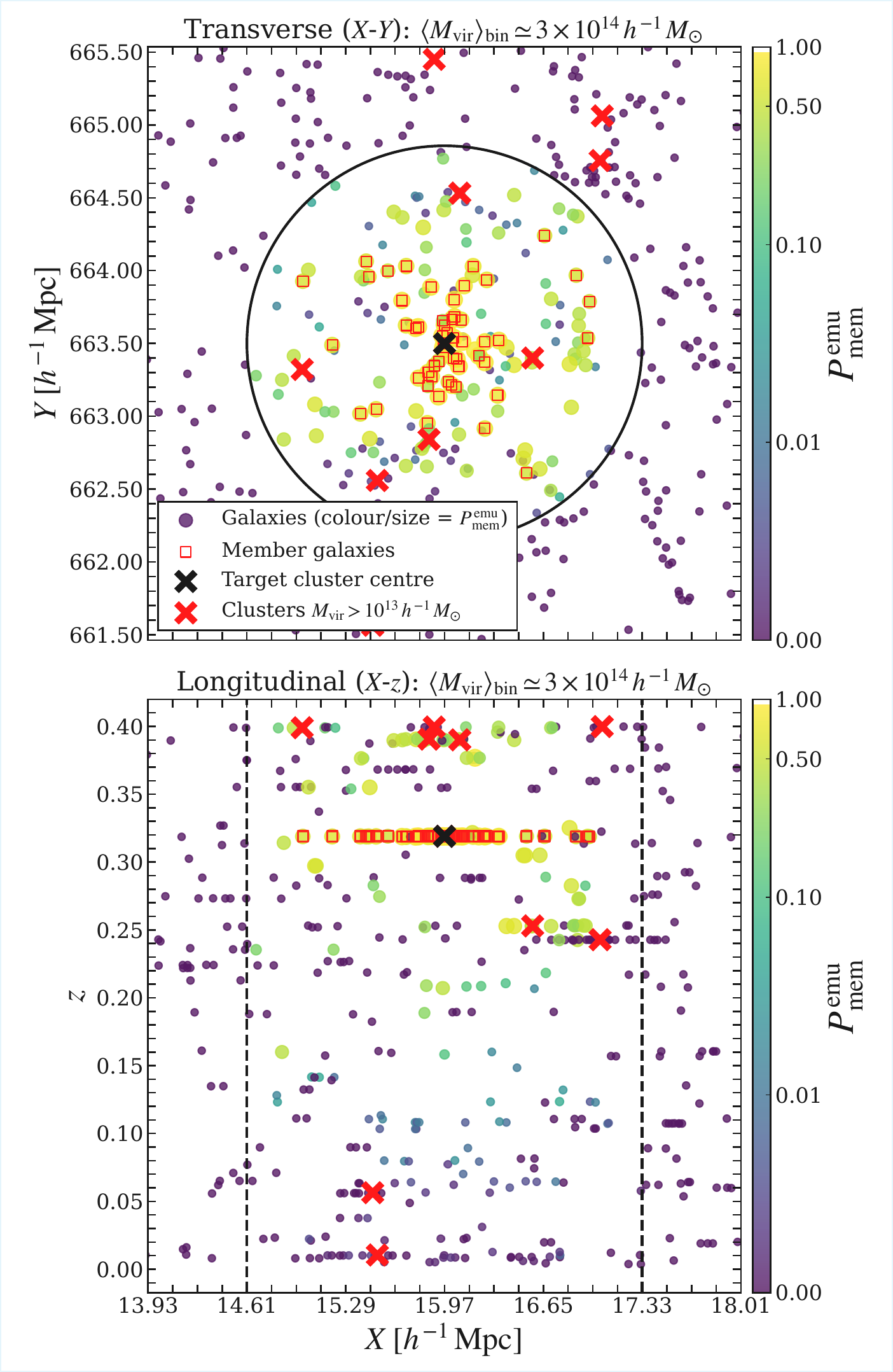}
    \end{adjustwidth}
    \caption{\textit{Top panel:} Transverse section of a cluster with $M_{\rm vir}\simeq 3 \times 10^{14}\,h^{-1}\,M_{\odot}$ from the \texttt{PICCOLO} C0 simulation processed with \texttt{CosmoPostProcess}. The black circle marks the $R_{\rm pmem}$ projected radial cut, with galaxies colour-coded according to their $P^{\mathrm{emu}}_{\mathrm{mem}}$ values. True member galaxies are highlighted by red boxes, while red crosses indicate all objects above $M_{\rm vir} = 10^{14}\,h^{-1}\,M_{\odot}$. \textit{Bottom panel:} LoS view of the same object restricted to the redshift interval covered by the first \texttt{PICCOLO} snapshot.}
    \label{fig:membership}
\end{figure}

In \cref{fig:SR}, we show the true and observed mass--richness relations measured in \texttt{PICCOLO} via \texttt{CosmoPostProcess} using the emulator calibrated as described in \Cref{app:flagship}, where the observed richness is constructed by summing the emulated per-galaxy membership probabilities $P_{\mathrm{mem}}^{\mathrm{emu}}$. Projection effects are already visible at this stage, since they boost the observed richness and increase the scatter at fixed mass compared to the intrinsic relation. This effect has been empirically confirmed with SZ \citep{2025A&A...700A..15G} and spectroscopic \citep{2021MNRAS.505...33M,2025MNRAS.544.2080M} follow-up observation of optically selected clusters. Furthermore, to illustrate the emulator behaviour in a representative case, we show the predicted membership probabilities $P_{\mathrm{mem}}^{\mathrm{emu}}$ for a cluster of mass close to $3 \times 10^{14}\,h^{-1}\,M_\odot$ in \cref{fig:membership}. The top panel displays the transverse distribution of galaxies within $R_{\rm pmem}$, colour-coded by $P_{\mathrm{mem}}^{\mathrm{emu}}$, while the bottom panel shows the same system along the LoS within a redshift slice of width $\sigma_{\mathrm{photo}\mbox{-}z}(z_{\mathrm{cl}})$, as defined in \cref{eq:z-dispersion}. This visualisation highlights how the emulator combines projected separation and redshift proximity to down-weight background structures and to assign higher membership probabilities to galaxies that are both spatially concentrated and compatible in photometric-redshift space.

\subsection{Miscentring\label{sc:miscentering}}

\texttt{CosmoPostProcess} injects miscentring by shifting the projected friends-of-friends (FoF) centre, identified via the halo-finding algorithm \texttt{SubFind} \citep{2009MNRAS.399..497D}, in polar coordinates within the plane perpendicular to the LoS. Miscentring is a known systematic effect, affecting both cluster lensing and richness estimates.

We implemented a substructure-based method to inject miscentring on top of the FoF centres of our halos. Around each target halo, we selected a substructures inside a cylinder centred on the FoF position with an aperture, $n \, R_{\rm pmem}$. The factor $n=1.5$ is tuned to match a chosen radial miscentring reference, here the one measured from the \texttt{Flagship2} cluster catalogue. More specifically, the reference distribution is the radial offset between the centre of the membership-matched halo, taken as the true centre, and the centre returned by the cluster finder in the \texttt{Flagship2} catalogue. In the present case study, the calibration is performed against the \texttt{PZWav} centres, consistently with the calibration catalogue used in this work. The cylinder depth follows the photometric redshift dispersion in \cref{eq:z-dispersion}, so we included structures satisfying $z_{\mathrm{cl}}-\sigma_z \leq z \leq z_{\mathrm{cl}}+\sigma_z$. We convolved the projected substructure positions with a Gaussian kernel with standard deviation of $\sigma_{\mathrm{mc}}=4\,\mathrm{px}$. We discretised the map on a grid with 100 by 100 pixels. The new centre is the weighted mean of the pixels with counts above the median.

\begin{figure}[h!]
    \centering
    \includegraphics[scale=0.4]{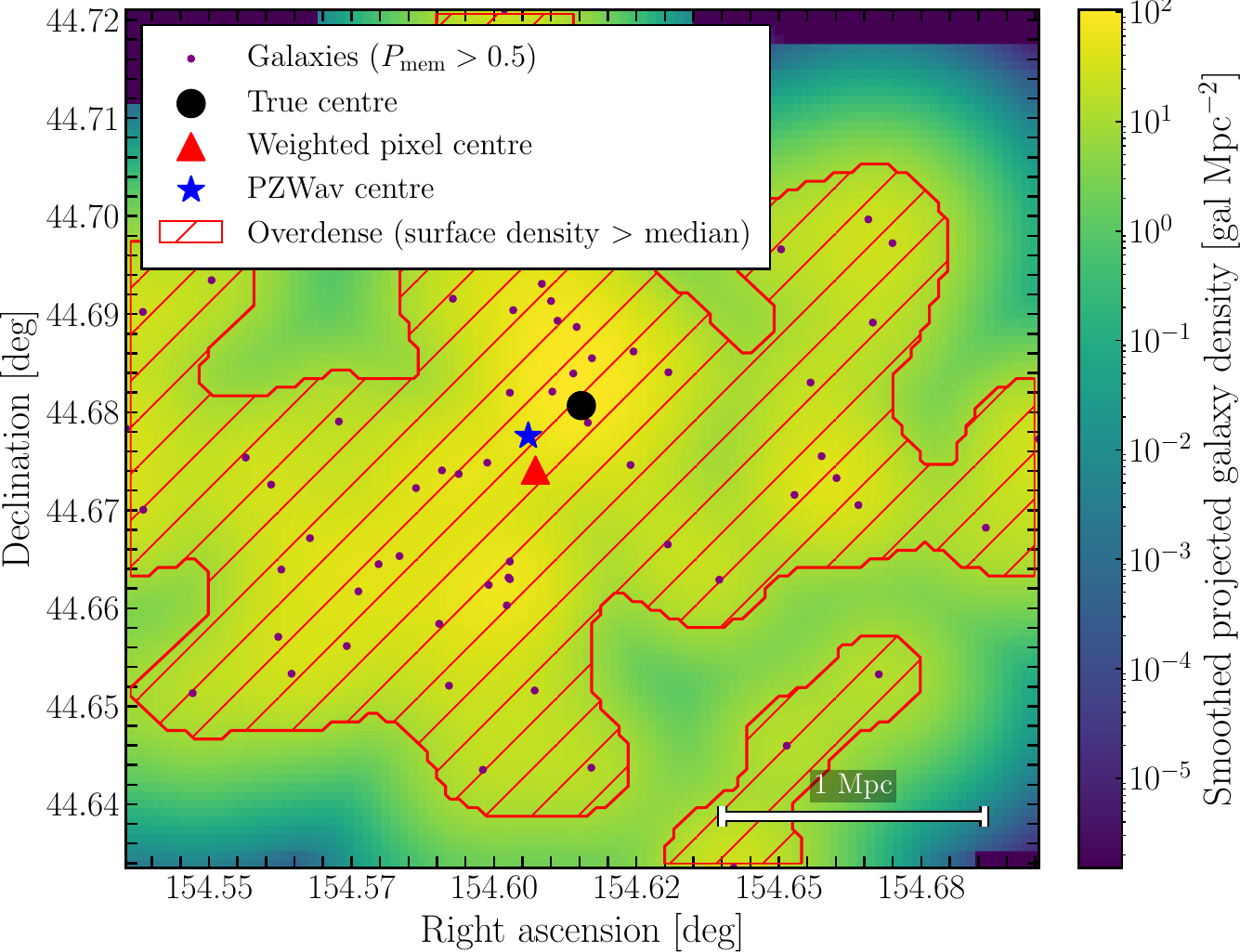}
    \caption{Illustration of our miscentring scheme, applied to a
$M_{\rm vir}=1.62\times10^{14}\,h^{-1}\,M_{\odot}$ cluster in the
\texttt{Flagship2} simulation. Purple points show galaxies with
$P_{\rm mem}>0.5$ and
$z_{\rm cl}-\sigma_z \leq z_{\rm gal} \leq z_{\rm cl}+\sigma_z$.
The colour map shows the Gaussian-smoothed projected galaxy number
density of the same galaxies, in units of galaxies per
$\mathrm{Mpc}^{-2}$. The red hatched region marks pixels
with surface density above the median value of the map. The red
triangle, blue star, and black dot mark the weighted pixel centre, the
\texttt{PZWav} centre, and the true centre, respectively. The scale bar
shown in the panel corresponds to $1\,\mathrm{Mpc}$ at the cluster
redshift. For a full validation of the miscentring model, including the
agreement at the level of the miscentring distribution, we refer the
reader to \cref{fig:mis-distribution}.}
    \label{fig:massive-1}
\end{figure}

\begin{figure*}
    \centering
    \includegraphics[scale=0.515]{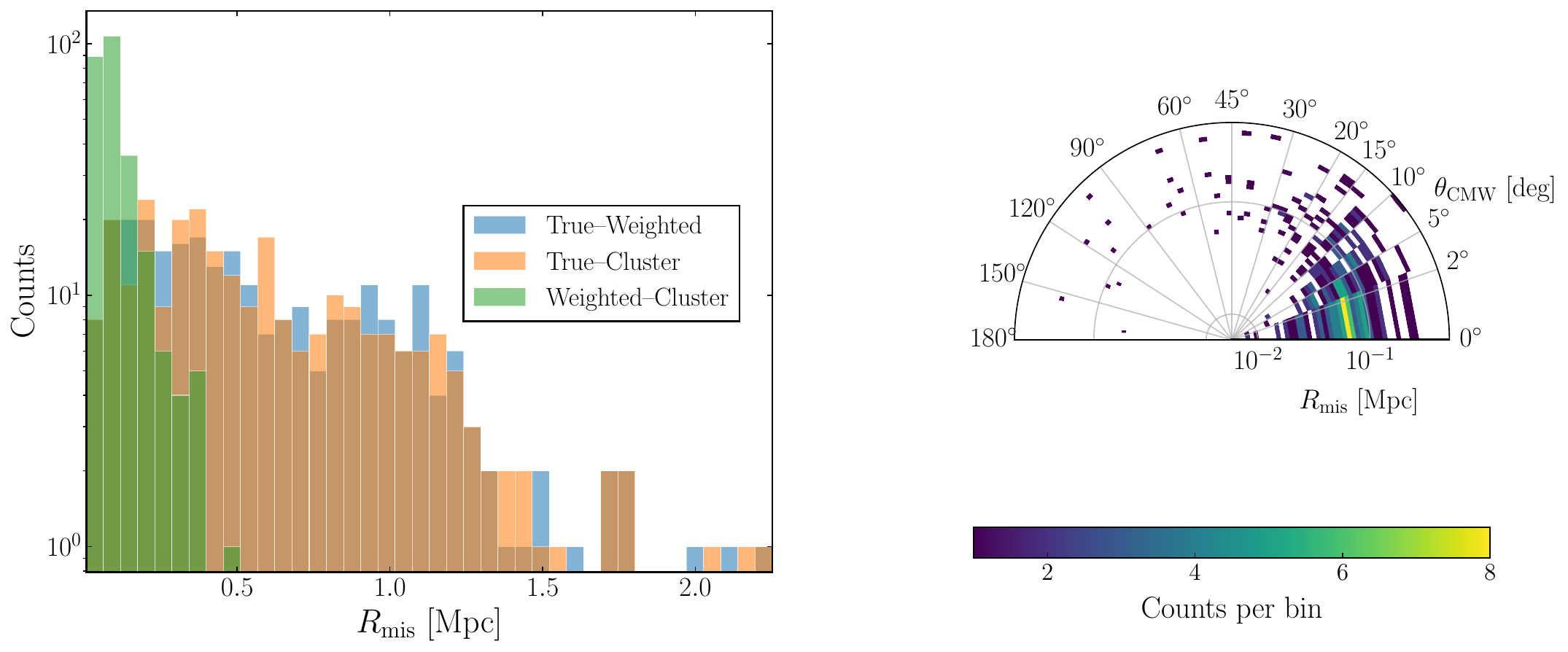}
    \caption{\textit{Left panel:} Radial miscentring distributions. Blue compares the membership centre, which is the truth, with the pixel weighted centre. Orange compares the membership centre with the cluster finder centre. Green compares the cluster finder centre with the pixel weighted centre. The overlap of the blue and orange histograms and the peak at low $R_{\mathrm{mis}}$ (in proper $\mathrm{Mpc}$) in the green histogram indicate that the weighted pixel centre tracks the cluster finder centre closely.
\textit{Right panel:} 2D histogram of the projected separation, $R_{\mathrm{mis}}$, and angular aperture with respect to the true centre, $\theta_{\widehat{\mathrm{CMW}}}$, between the pixel-weighted and cluster-finder centres. The pixel colour encodes the counts in log-scale. The distribution concentrates near zero degrees at small $R_{\mathrm{mis}}$ demonstrating the ability of our miscentring scheme to reproduce the actual direction of the centre displacement.}
    \label{fig:mis-distribution}
\end{figure*}

We validated this substructure method on the \texttt{Flagship2} galaxy catalogue adopting as true centre of the detection the centre of the matched halo. We used galaxies with $z$ values within $\sigma_z$ of the cluster redshift (\cref{fig:massive-1}). We also applied a cut on the membership probability of $P_{\mathrm{mem}}>0.5$ that empirically reproduces the coarser spatial distribution of substructures used in the \texttt{PICCOLO} miscentring calculation. We selected the 1000 most massive clusters and then restrict the validation sample to systems with $z_{\mathrm{cl}}<0.45$, yielding 358 objects. This cut is adopted conservatively to validate the miscentring prescription on the same calibration sample used in the rest of the analysis while restricting the comparison to a statistically robust \texttt{Flagship2} regime, where the detection behaviour is more stable. It also matches the depth of the first \texttt{PICCOLO} C0 snapshot used in the calibration. With this choice, we avoid over-interpreting the higher-redshift \texttt{Flagship2} behaviour in the validation step, while allowing the redshift dependence in the mock prescription to be driven by the adopted photometric-redshift uncertainties. \Cref{fig:mis-distribution} shows the recovered radial miscentring distribution (left panel) and the angular offset distribution between our pixel weighted centre and the \texttt{PZWav} centre (right panel). Both distributions are displayed as function of the physical separation of the projected centres, $R_{\rm mis}$. These results demonstrate that our miscentring scheme successfully reproduces both the magnitude and the direction of the centre displacement. In particular, a Kolmogorov--Smirnov (KS) test comparing our radial separation distribution to that measured relative to the \texttt{PZWav} centres shows excellent agreement with KS statistic $D=0.046$ and $\text{$p$-value}=0.95$. By contrast, a KS test of the measured angular separations against a uniform distribution strongly rejects uniformity producing $D=0.73$ which translates in $\text{$p$-value}\simeq 0$.

\subsection{\label{sc:Baryon}Baryonification}

Our aim is to map the dark matter-only (DMO) density profiles measured from the \texttt{PICCOLO} suite into a baryonified dark matter profile that reproduces hydrodynamical simulation results, while retaining speed and flexibility. While we based our baryonification on the model originally proposed by \cite{Schneider19}, we introduced two modifications to their method to trace the effect of baryons out to the cluster edge and to capture its redshift dependence. First, the hot gas was described by a unified core-tail model whose parameters are calibrated snapshot by snapshot, with the introduction of a tail component to trace the profile up to $5\,h^{-1}\,{\rm Mpc}$. Second, we applied a single, redshift-dependent adiabatic contraction (AC) mapping directly on the DMO profiles, following \citet{2025JCAP...12..043S}. The parameter values used in this work are calibrated against the \texttt{Magneticum} Box 3 data (see \cref{sc:Magneticum}). Here, we provide only a summary of the model, while the full specification of the profiles and mappings, including all relevant equations, is collected in \Cref{app:calib}.

\subsubsection{DMO profiles and baryonified reconstruction}

We started from the DMO profiles measured in \cref{sc:density-profiles} for the C0 and C1 \texttt{PICCOLO} runs. The code reads cumulative DMO shell masses built on the same radial edges $\{r_{i}\}$ used in the 3D density profiles, converts them to shell densities $\rho_{\rm DMO}(r)$, and applies a Savitzky--Golay smoothing in $\logten\rho_{\rm DMO}$ to suppress high-frequency particle noise while keeping the large-scale slope intact. Onto these smoothed DMO profiles we add the baryonic response associated with AC, namely the increase of the inner dark matter density induced by the slow condensation of baryons, which deepens the potential and approximately preserves orbital adiabatic invariants. In the standard spherical/circular-orbit picture this is often expressed as $r\,M(<r)\approx \mathrm{const}$ for mass shells, with improved prescriptions accounting for eccentric orbits via orbit-averaged radii \citep{1984Natur.311..517B,2004ApJ...616...16G}. We model AC in two steps: (i) we predict the hot-gas and central-stellar components as a function of mass and redshift; (ii) we displace the dark matter particle position with a single global mapping $\xi_{\rm ac}(r\,|\,z)$ evaluated at the density level. This procedure yielded a contracted dark matter density profile, $\rho_{\rm dm}^{\rm ac}(r)$, which was then converted, under the assumption of mass conservation, into the baryonified density profile $\rho_{\rm dmb}(r)$ used in our analysis.

\begin{figure*}[t]
    \centering
    \includegraphics[scale=0.575]{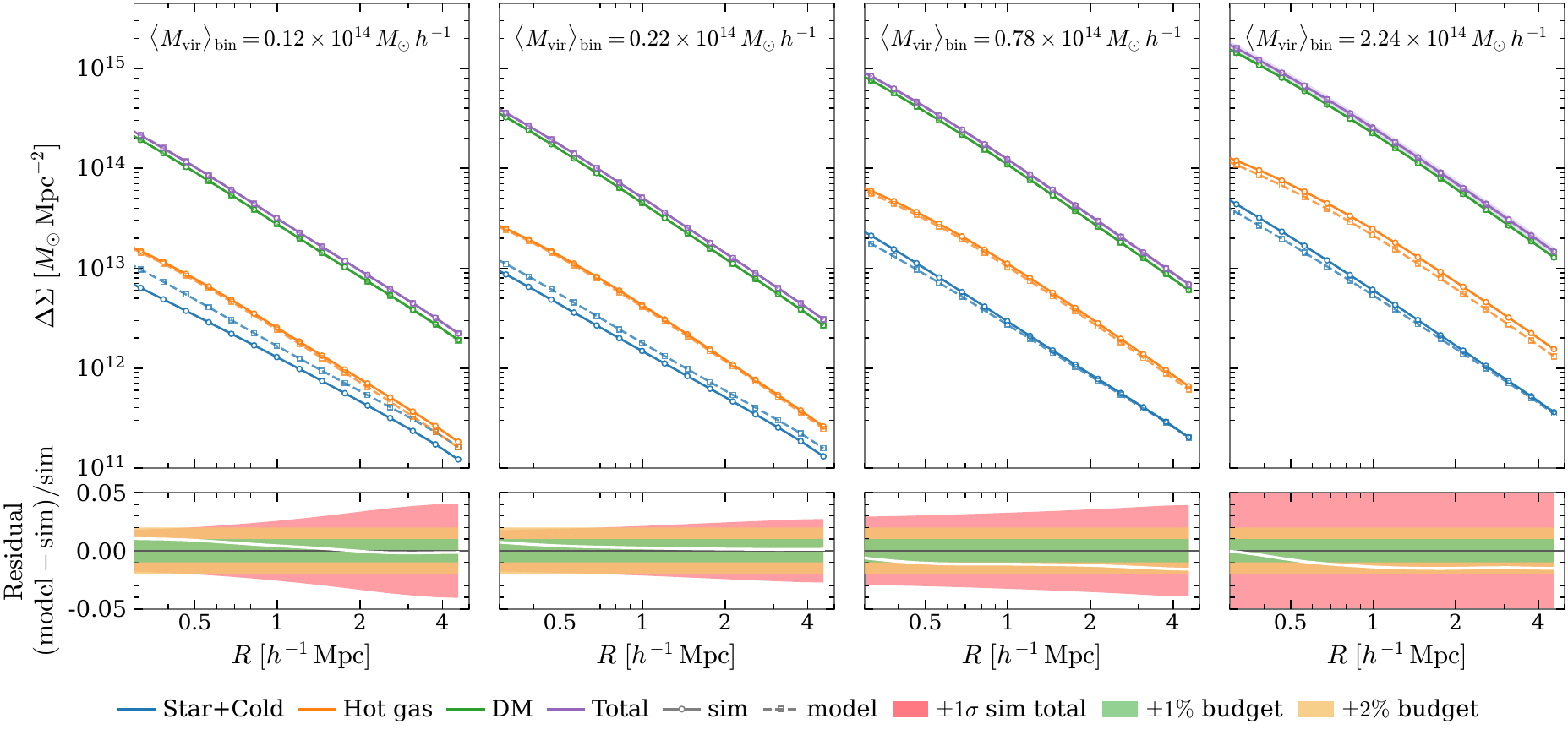}
    \caption{Excess surface density $\Delta \Sigma$ in four mass bins for the halos identified in \texttt{Magneticum} Box 3 at high resolution. The colour code indicates the species of the components with stars and cold gas in light blue, hot gas in orange, and the dark matter density after the quasi-AC in green. Solid lines denote the simulation data used for calibration, while the dashed lines are used for the model. In the \textit{bottom panels}, the coloured bands in the residuals panel indicate: the jackknife $1\,\sigma$ uncertainty on the total density profile (red), the $1\,\%$ relative error on the total (green), and the $2\,\%$ relative error (orange).}
    \label{fig:deltasigma-vs-magneticum}
\end{figure*}

The validation was performed by first establishing a one-to-one (biunivocal) halo correspondence between the hydrodynamical and DMO simulations. For each matched halo pair, profiles are computed and subsequently stacked within mass bins. With this approach, the model reproduces the stacked $\Delta\Sigma$ profiles of the hydrodynamical simulation to within $1$--$2\,\%$ over the range $0.1$--$5\,h^{-1}{\rm Mpc}$ across four mass bins (see \cref{fig:deltasigma-vs-magneticum}).

\subsubsection{Hot-gas model and redshift evolution}

To capture the redshift evolution of the hot component, we calibrated the parameters of our model at four simulation snapshots, $z_{\rm snap}\in\{0.00, 0.25, 0.47, 1.17\}$, and describe its evolution by interpolating the best-fit parameters between these epochs. To apply the baryonic correction, we selected the snapshot with redshift nearest to that of the target halo. The parameter vector governs the mass dependence of the hot-gas model through: a core pivot parameter and mixing index controlling the inner-slope steepening, an ejection scale, a tail-onset location, and a broken-slope description for the outer-tail decay. Parameter calibrated values are given in \Cref{tab:gas-snapshot-fits}.

The hot gas was modelled as the sum of a pressure-supported, normalised one-halo core and a diffuse tail that transitions smoothly to the cosmic mean density, thereby avoiding artificial truncations. The resulting model combines a tunable $\beta$-model \citep{1976A&A....49..137C} for the core region and a power-law tail, with a sigmoid convolution to ensure a smooth transition between the two (see \crefnp{eq:hot-gas-1,eq:hot-gas-2,eq:hot-gas-3}). The core region is tuned via a characteristic core size and the slope of the $\beta$-model and must obey a normalisation bound to enforce consistency with the gas fraction measured in \texttt{Magneticum}. The tail is controlled by an onset radius and an outer slope that vary with halo mass and redshift. Numerically, the core normalisation uses cached integrals to improve computational speed, while the tail employs a smooth logistic-to-power-law transition; stabilisation by parameter floors ensures robust behaviour for all halos.

\subsubsection{Stellar and gas mass components}

Stellar mass was split into a compact central component and a diffuse satellite term. The central galaxy (CGA) was modelled with a fixed scale radius $\rh=0.075\,\rvir$. The CGA contributes in the AC mapping, where the ratio between the cumulative mass of the central galaxy and the dark matter modulates the feedback of the latter. It is then used to compute the final total density in the validation tests. The satellite galaxy (SGA) term follows the local dark matter slope and contributes only implicitly to the quasi-AC, while hot gas and CGA components are regulated by a feedback strength parameter, which determines the modification they induce on the DMO profile.

Stellar and gas mass fractions were modelled as smooth functions of halo mass, with parameters that evolve with redshift and depend on the cosmic baryon fraction $\fb$. These fractions were used to normalise the corresponding stellar and gas density profiles that enter the baryonification scheme. At each redshift, we enforce global baryon conservation by requiring
$f_{\star}+f_{\rm gas}=\fb$, while the stellar mass associated with the central galaxy is constrained to satisfy, $f_{\rm cga}\le f_{\star}$. Unless otherwise specified, the baryon fraction is fixed to the cosmological value of the simulation, $f_{\rm b} = \Omega_{\rm b}/\Omega_{\rm m}$. The functional forms of the stellar and gas components, together with the normalisation prescriptions based on $f_{\star}$, $f_{\rm cga}$, and $f_{\rm gas}$, are defined in \Cref{app:calib} (see \crefnp{eq:fractions_star,eq:fractions_cga,eq:hot-gas-1,eq:hot-gas-3}).

\subsubsection{Quasi-AC mapping and particle remapping}

The impact of baryons on the dark matter distribution is incorporated through a single, global quasi-AC mapping applied at the level of the density profiles. Starting from the DMO profile $\rho_{\rm DMO}(r)$, we computed the enclosed mass profiles of the baryonic components; namely, hot gas, CGA, and SGA, using the density models introduced in \Cref{app:calib}. These enclosed masses enter the definition of a redshift-dependent contraction response function $\xi_{\rm ac}(r)$, whose amplitude depends on: (i) a set of normalisation and slope parameters that encode its redshift evolution; (ii) the enclosed mass fractions of central stars and hot gas; and (iii) the halo peak height $\nu(M,z)$. The contraction mapping $\xi_{\rm ac}(r)$ is applied to the cumulative dark matter mass profile, which is subsequently differentiated with respect to radius to obtain the contracted dark matter density profile $\rho_{\rm dm}^{\rm ac}(r)$ -- see \Cref{app:calib,sc:Methodology}.

To implement the contracted dark matter distribution consistently at the particle level, we apply a quantile-based remapping of particle radii. In practice, we construct the cumulative mass profile corresponding to $\rho_{\rm dm}^{\rm ac}(r)$ and define a monotonic mapping that associates each quantile of the original particle-radius distribution to the corresponding quantile of the contracted model. Particle positions are then rescaled radially with respect to the halo centre, while preserving their angular coordinates, and wrapped back into the simulation box when necessary. Since selection effects and projected density profiles are computed using particle masses, we also rescale the mass of each dark matter particle by a factor $f_{\rm CDM}=1-\fb$ (where CDM stands for cold dark matter). This ensures that the total projected mass associated with the contracted particle distribution is consistent with the baryonified dark matter profile and that correlations between richness and lensing are preserved throughout the forward model.

\subsubsection{Validation in \texttt{Magneticum}}

The baryonification scheme was validated by applying the quasi-AC to halos in the DMO counterpart of the \texttt{Magneticum} Box 3 simulation. For each halo, the contracted dark matter profile is combined with the associated baryonic components, and the corresponding 2D surface density profile is computed via an Abel transform. Finally, the surface density, $\Sigma(R)$, is converted into excess surface density, $\Delta\Sigma(R)$, according to
\begin{align}
\Delta\Sigma(R) &= \bar\Sigma(<R)-\Sigma(R)\,,\,\,\,\,\,
\bar\Sigma(<R)=\frac{2}{R^2}\int_{0}^{R}\Sigma(R')\,R'\,{\rm d}R'\,,
\label{eq:sigma_conv}
\end{align}
which is the conversion we adopted throughout this work to convert from surface to excess surface density. It is worth noticing that in \cref{sec:results}, we present the results of the selection bias in terms of $\Sigma(R)$, since in this case the features in the bias radial profile are easier to visualise.

We show in \cref{fig:deltasigma-vs-magneticum} a comparison between the $\Delta \Sigma(R)$ profiles predicted by our calibrated model of baryonification and those actually measured from the \texttt{Magneticum} simulations. Curves of different colours refer to the different components that make the total density profiles. With our baryonic correction model, we reproduced the excess surface density profiles from \texttt{Magneticum} with an accuracy of $1$--$2\,\%$. We point out that, to correctly propagate baryonic effects through our pipeline, the baryonic correction was applied to the particles before painting the galaxies. The characteristic suppression effects of baryonic correction (orange line) and miscentring (green line) can be seen in \cref{fig:profiles}, together with their combined effect (red line).

\section{\label{sec:results}Results}

\begin{figure*}[ht!]
    \centering
    \includegraphics[scale=0.38]{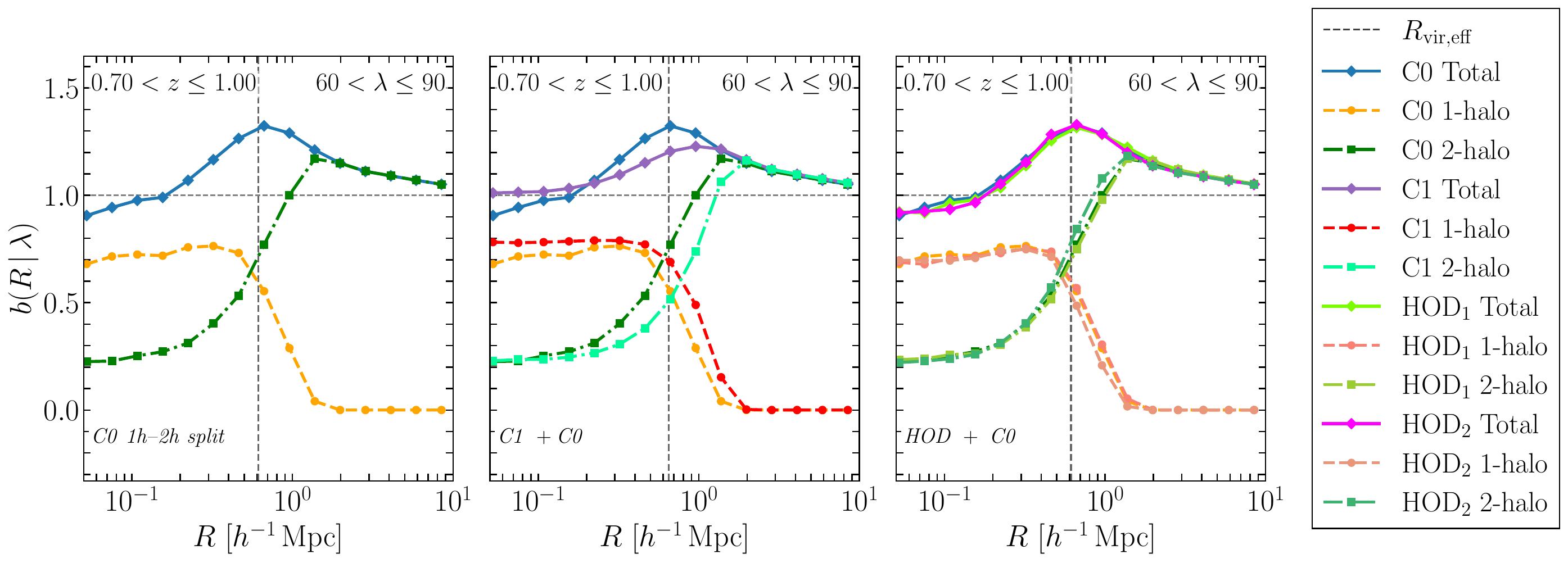}
    \caption{\label{fig:onebin_1h2h}Contributions of the one-halo and two-halo terms to the selection bias of the projected density profile in the $60<\lambda\leq90,\ 0.70<z\leq1.00$ bin. The bias is defined as $b(R\,|\,\lambda)=\Sigma^{\mathrm{sel}}_{\lambda}(R)/\Sigma^{\mathrm{ref}}_{\lambda}(R)$. In each panel, solid lines show the total bias, dashed lines the one-halo contribution, and dash-dotted lines the two-halo contribution; the vertical dashed line marks the median effective virial radius $R_{\rm vir,eff}$. \textit{Left:} C0 with miscentring and baryonification (total in blue; one-halo in orange; two-halo in green). \textit{Middle:} C1 (total in purple; one-halo in red; two-halo in turquoise), with the C0 curves overplotted for reference. \textit{Right:} two HOD variants overlaid on C0 mis+bar, with HOD$_1$ (total in lawngreen) defined by $(\alpha,\logten{[M_{1,\mathrm{sat}}\,h/M_{\odot}]})=(0.839,\,12.205)$ and HOD$_2$ (total in fuchsia) by $(\alpha,\logten{[M_{1,\mathrm{sat}}\,h/M_{\odot}]})=(0.918,\,12.176)$; for each HOD, the corresponding one-halo and two-halo terms are shown with the same line styles and lighter companion colours. The dashed vertical line marks the median $R_{\rm vir}$ of the mass-selected sample for the C0 fiducial case.}
\end{figure*}

Having established the structure and validation of the pipeline, we now quantify the systematics affecting the weak lensing signal of the optical cluster catalogue. Throughout this section we characterise these effects through the selection bias  estimator,
\begin{equation}
\label{eq:def_bias_results}
b(R\,|\,\lambda)\;\equiv\;\frac{\Sigma^{\rm sel}_{\lambda}(R)}{\Sigma^{\rm ref}_{\lambda}(R)}\,,
\end{equation}
which measures the scale-dependent modification of the projected density profile at fixed observed richness $\lambda$. The numerator $\Sigma^{\rm sel}_{\lambda}(R)$ denotes the stacked surface-density profile of richness-selected clusters, while the denominator $\Sigma^{\rm ref}_{\lambda}(R)$ is the corresponding mass-selected reference constructed as described in \cref{sc:Methodology}. Here, the individual object profiles that enter the stacking are computed using \cref{eq:profiles} in both the numerator and the denominator. The ratio therefore isolates the impact of richness selection on the inferred lensing signal.

Unless stated otherwise, we adopt the C0 cosmology and the HOD calibrated on the \texttt{Flagship2} cluster catalogue as our fiducial configuration. Uncertainties correspond to 68 per cent bootstrap intervals obtained by resampling within the $(\ln M, z)$ cells used to build the reference stacks.

\subsection{One-halo and two-halo contributions}

We first examined a representative bin, $60<\lambda\leq90,\ 0.70<z\leq1.00$, shown in \cref{fig:onebin_1h2h}. In the fiducial configuration including miscentring and baryonification, the bias exhibits a clear scale dependence, with a pronounced enhancement near the transition between the one-halo and two-halo regimes at $R \simeq 1\,h^{-1}\mathrm{Mpc}$. The peak reaches $\sim 20$–$40$ per cent relative to the mass-selected reference.

The decomposition demonstrates that the two-halo contribution dominates the bias beyond $R \simeq 0.5\,h^{-1}\mathrm{Mpc}$, indicating that the transition-scale enhancement is primarily sourced by correlated large-scale structure projected along the LoS. This behaviour is consistent with previous studies of projection-induced biases in optically selected cluster samples and forward-modelling analyses of richness–lensing coupling \citep[e.g.][]{2020MNRAS.496.4468S,Wu2022,2024PhRvD.110j3508Z,2023MNRAS.521.5064S,2024A&A...681A..67E,EP-Ragagnin}.

\subsection{Redshift and richness dependence}

\begin{figure*}[ht!]
    \centering
    \includegraphics[width=\textwidth]{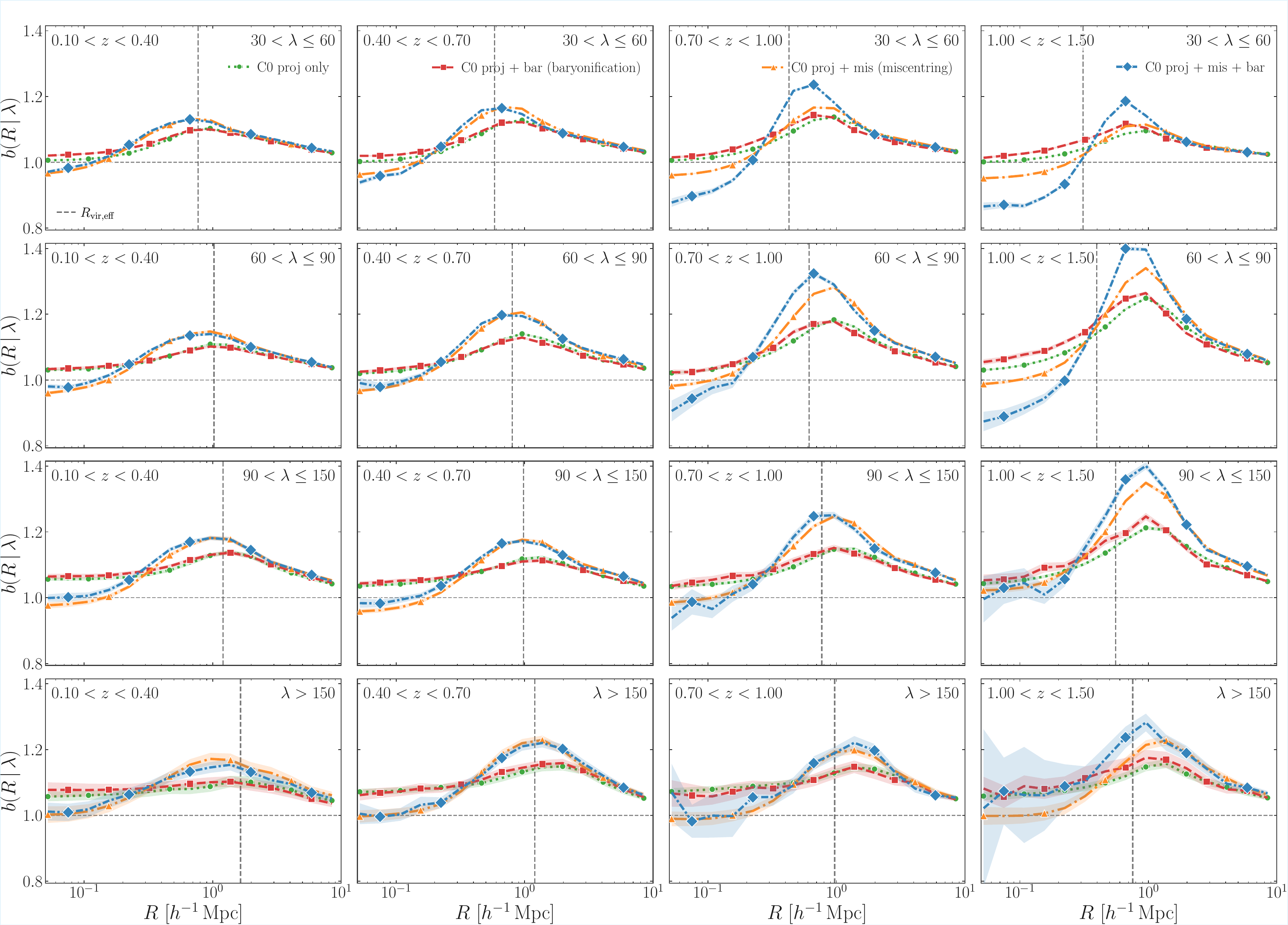}
    \caption{\label{fig:C0_grid}Bias, estimated through \cref{eq:def_bias_results}, in stacked richness bins for cosmology C0. Curves show the full case (blue), projections with miscentring (orange), projections with baryons (red), and projection-only (green). Shaded regions indicate 68 per cent bootstrap intervals; vertical dashed lines show an indicative separation scale for the one-halo and two-halo terms, at the median virial radius $R_{\rm vir,eff}$ in the bin. Baryons produce only a few-percent change in the inner profile at $z>0.7$, while miscentring dominates the bias by suppressing the central signal and enhancing the transition scale. When combined, baryons further amplify this inner suppression. The vertical dashed line display the median $R_{\rm vir}$ in the richness-selected bin.}
\end{figure*}

The full redshift and richness dependence of the bias in the C0 cosmology is shown in \cref{fig:C0_grid}. In the projection-only configuration, the bias is positive and peaks near the one-halo to two-halo transition. The amplitude increases with redshift, reaching its largest values in the $0.70<z\leq1.50$ bins.

The bias amplitude exhibits a non-monotonic dependence on observed richness. At intermediate richness, $60<\lambda\leq150$, the transition-scale enhancement is most pronounced. At the highest richness, $\lambda>150$, the peak amplitude decreases and the curves become flatter.

The reduction of projection contamination at high $\lambda_{\rm obs}$ is physically expected. At large observed richness, the selected sample is increasingly dominated by intrinsically massive halos whose internal density profile overwhelms typical LoS fluctuations, reducing the fractional contribution of projected structures. In our framework this behaviour is further modulated by halo geometry. Orientation and triaxiality can enhance projected densities and weak lensing signals at fixed mass, an effect that we discuss in \Cref{app:triaxiality} and that has been investigated in the context of halo structure studies (e.g. \citealt{2012MNRAS.426.1558G}). For the specific trend with observed richness, \citet{2025A&A...697A.184G} and \citet{2024A&A...681A..67E} find in hydrodynamical simulations that projection-induced richness contamination decreases toward high $\lambda_{\rm obs}$ as intrinsically massive systems dominate the selected sample, qualitatively corroborating the behaviour seen in \cref{fig:C0_grid}.

\subsection{Miscentring and baryonic effects}

Miscentring produces the expected suppression of the inner profile and enhancement around the typical offset scale. The suppression is strongest at small radii and becomes more pronounced at low richness. Because the displaced centre is often shifted toward locally overdense regions, richness can increase while the central surface density is suppressed. This redshift dependence arises naturally in our implementation because the miscentring prescription is coupled to the richness reconstruction through the substructure distribution used to define the displaced centre. In turn, the substructures entering this procedure are selected within a redshift slice whose width is set by the adopted photometric-redshift uncertainty. As this uncertainty broadens with redshift, the set of structures contributing to the effective centre assignment changes, which induces a corresponding redshift dependence in the impact of miscentring on the weak lensing bias.

Baryonic modifications alone produce comparatively modest changes in $b(R\,|\,\lambda)$. Although the 3D density profile can be altered at the $\sim 20$ per cent level in the inner regions, projection dilutes these modifications and confines their effect to small radii \citep[e.g.][]{Schneider19,EP-Ragagnin}. When miscentring and baryonic effects are combined, their impact is amplified at high redshift, where baryon-induced contraction steepens the inner density profile and deepens the central deficit relative to the reference stack.

\subsection{Cosmological dependence}

\begin{figure}[ht!]
    \centering
    \includegraphics[width=0.415\textwidth]{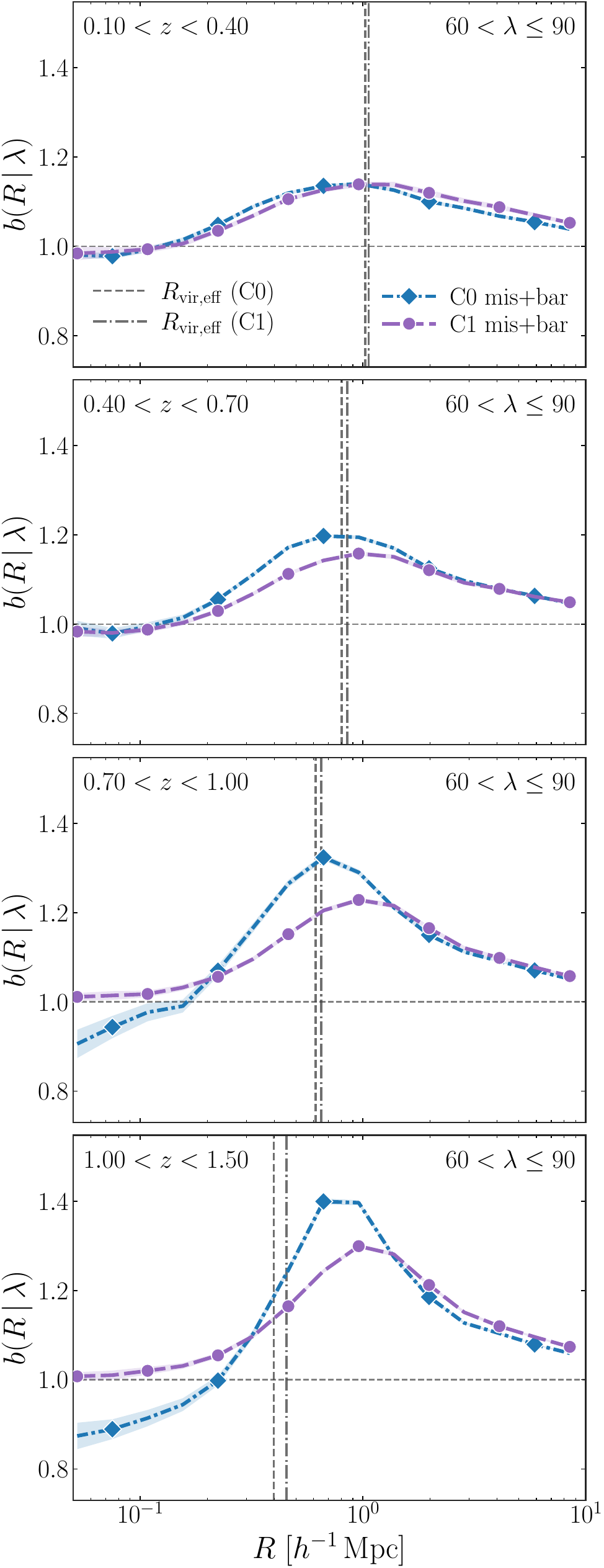}
    \caption{\label{fig:C1_full}Cosmological dependence of the selection bias, shown for the C1 cosmological parameter set including all configurations (baryonification and miscentring). The comparison is performed at fixed richness ($60 < \lambda \leq 90$). Differences with respect to the fiducial C0 case reflect the reduced matter density in C1, which leads to a lower bias amplitude and a modest outward shift of the transition between the one-halo and two-halo regimes. The dashed–dotted line indicates the median virial radius $R_{\rm vir}$ for the C1 configuration, highlighting the displacement of the peak relative to the characteristic halo scale at which the transition occurs.}
\end{figure}

To assess the robustness of the inferred selection bias to cosmological assumptions, we compared the fiducial simulation (C0) with the alternative cosmological configuration C1. The two runs share a similar fluctuation amplitude but differ primarily in the matter density parameter, with C1 characterised by a lower value of $\Omega_{\rm m}$. This change modifies both the abundance of dark matter halos and their large-scale clustering, which in turn affects the amount of correlated structure projected along the LoS.

\Cref{fig:C1_full} shows the resulting bias profile for clusters in the richness interval $60 < \lambda \leq 90$. The overall radial structure of the bias remains qualitatively unchanged between the two cosmologies, confirming that the characteristic peak at the transition between the one-halo and two-halo regimes is a robust feature of richness-selected samples. However, the amplitude of the effect decreases in the lower-density cosmology. In particular, the peak of the selection bias is reduced by up to $\sim 9$ per cent relative to the fiducial configuration.

The location of the peak also shifts slightly towards larger radii, by approximately $0.1$–$0.8\,h^{-1}\,\mathrm{Mpc}$. This behaviour reflects the change in the halo population associated with a fixed richness selection. In a cosmology with lower $\Omega_{\rm m}$ the abundance of massive halos is reduced, so clusters populating a given richness bin correspond on average to slightly lower masses and therefore somewhat larger relative virial radii when expressed in physical units. The dashed–dotted line in \cref{fig:C1_full} marks the median virial radius in the C1 configuration, illustrating the small displacement between the peak of the bias profile and the typical halo scale.

These results indicate that the detailed amplitude of the selection bias depends on the cosmological model through its impact on halo abundance and clustering. Nevertheless, the qualitative behaviour and radial structure of the effect remain stable across the explored cosmologies. At the same time, the non-negligible variation in the bias amplitude indicates that this ingredient should ultimately be treated self-consistently in the DR1 cosmological analysis. In practice, although the mock-generation framework is sufficiently modular to be retuned efficiently, it is not designed to be recomputed on the fly within the likelihood analysis. The current strategy is therefore to use the set of \texttt{PICCOLO} cosmologies to build an emulator that interpolates the weak lensing bias across cosmological parameter space. This would allow the correction to be matched to the cosmological region explored by the inference, while retaining the flexibility of the present framework and avoiding the need for a full rerun of the mocks at each likelihood step.

\begin{figure}[ht!]
    \centering
    \includegraphics[width=0.45\textwidth]{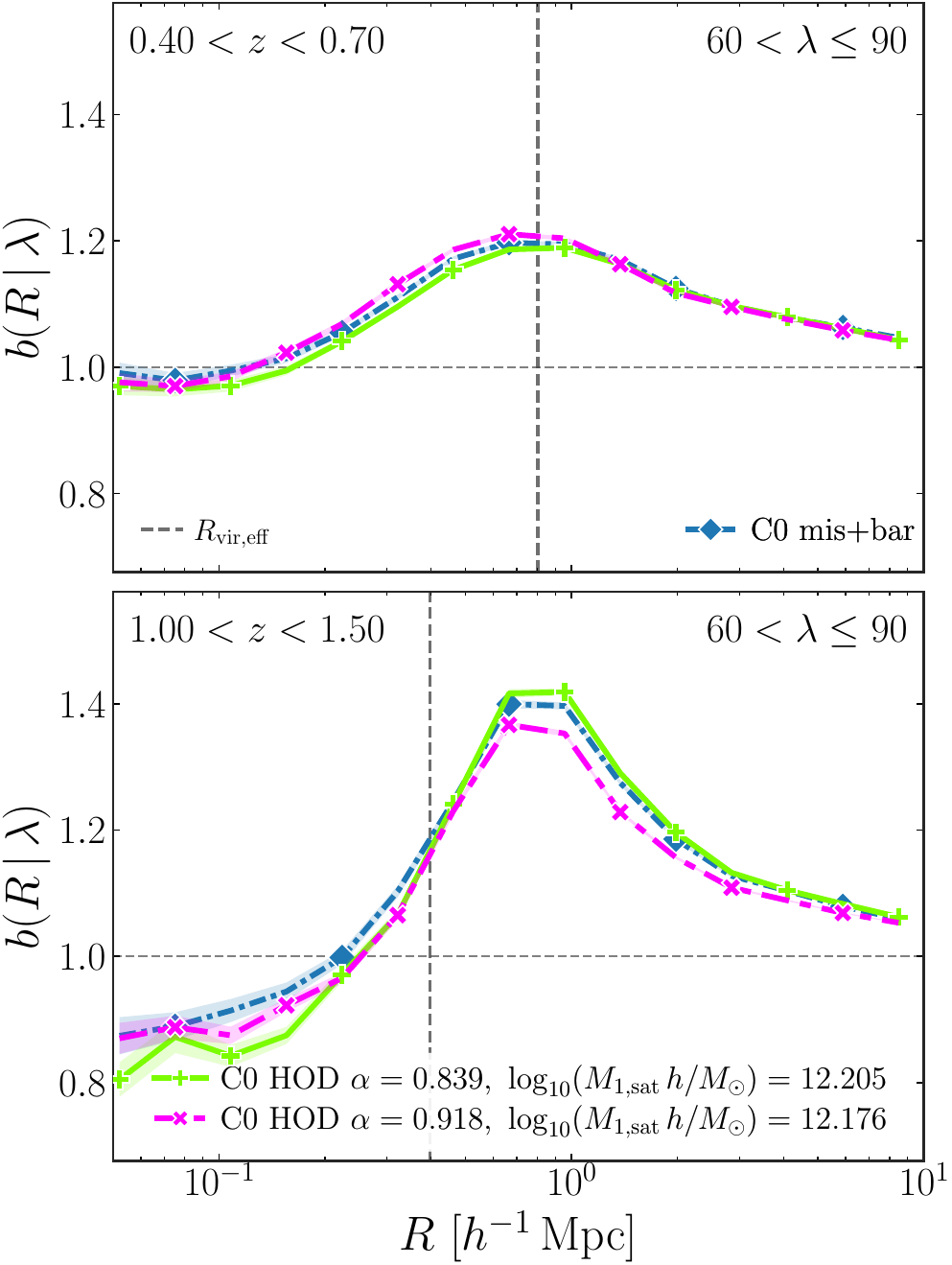}
    \caption{\label{fig:C0_HODtails_bias}Response of the selection bias to variations in the halo occupation distribution (HOD) parameters for the $60 < \lambda \leq 90$ richness bin. We vary the satellite pivot mass $\logten{(M_{1,\mathrm{sat}}\,h/M_{\odot})}$ and the slope $\alpha$ of the satellite occupation relation. Lower $\logten{(M_{1,\mathrm{sat}}\,h/M_{\odot})}$ and higher $\alpha$ increase the satellite contribution and the associated projection contamination, producing a bias that is a few per cent larger (magenta curve) than the fiducial case (blue curve). Conversely, higher $\logten{(M_{1,\mathrm{sat}}\,h/M_{\odot})}$ and shallower slopes result in a slightly reduced bias (green curve). Since the redshift dependence of the HOD-induced variations is generally mild, we show two representative redshift bins. At the highest redshift the trend partially reverses near the transition scale and at larger radii, while the inner region remains suppressed relative to the fiducial configuration for both HOD variants.}
\end{figure}

\subsection{HOD variations}

We also investigated the sensitivity of the selection bias to the galaxy population model used to populate dark matter halos. This can be achieved by varying the parameters of the halo occupation distribution (HOD), which determines the abundance of satellite galaxies at fixed halo mass. In particular, we vary the satellite normalisation scale $\logten{(M_{1,\mathrm{sat}}\,h/M_{\odot})}$ and the slope $\alpha$ of the satellite occupation relation. The explored variations follow the direction orthogonal to the main $\logten{(M_{1,\mathrm{sat}}\,h/M_{\odot})}$–$\alpha$ degeneracy and correspond approximately to a $2\sigma$ excursion around the fiducial HOD calibration.

\Cref{fig:C0_HODtails_bias} shows the resulting bias profiles for clusters in the richness range $60 < \lambda \leq 90$. Increasing the satellite abundance, obtained by lowering $\logten{(M_{1,\mathrm{sat}}\,h/M_{\odot})}$ and steepening $\alpha$, enhances the probability that galaxies associated with neighbouring halos are counted as members of the target cluster. This increases the level of projection contamination and produces a modest amplification of the bias profile, at the level of a few per cent relative to the fiducial case. Conversely, decreasing the satellite fraction reduces the probability of such projections and slightly suppresses the bias amplitude.

Therefore, the main effect of these HOD variations is a change in the overall amplitude of the bias, while the radial structure of the profile remains largely unchanged. The redshift dependence of this response is generally weak, which is why only two representative redshift bins are shown in \cref{fig:C0_HODtails_bias}. An exception occurs near the peak scale at the highest redshift, where the trend partially reverses. In this regime the larger intrinsic scatter in the mass–richness relation modifies how clusters scatter across richness thresholds, producing a mild inversion of the HOD dependence near the transition between the one-halo and two-halo regimes. Overall, the sensitivity of the selection bias to realistic HOD variations remains modest compared to the dominant contribution from projection effects.

\subsection{Mass calibration implications}

The scale-dependent distortions of the stacked density profiles discussed above translate directly into biases in weak lensing mass calibration. To quantify the net impact on inferred cluster masses, we fit the stacked excess surface density profiles with a standard Navarro--Frenk--White (NFW) model \citep{1997ApJ...490..493N} and compare the recovered masses to the true halo masses in the simulations. This allows us to separate the effects of projection, miscentring, and baryonic physics on the mass calibration, as well as to isolate the contribution arising purely from the richness-based selection,

\begin{equation}
b_{\rm mass} \equiv \left\langle \frac{M_{\rm fit}}{M_{\rm true}}\right\rangle\,.
\end{equation}

Projection-only stacks yield $b_{\rm mass}=1.22$. Including miscentring and baryons shifts masses to $b_{\rm mass}=0.86$. To isolate the bias induced purely by the richness selection, we compare the masses obtained from the richness-selected stacks to those derived from the mass-selected reference sample,

\begin{equation}
b_{\rm sel} \equiv \frac{M_{\rm fit}^{\rm (rich)}}{M_{\rm fit}^{\rm (mass\mbox{-}sel)}}\,.
\end{equation}

We find $\langle b_{\rm sel}\rangle = 1.17$ across the 16 $(z,\lambda)$ bins. This amplitude is consistent and the corresponding $\sim\,5$ per cent mitigation is consistent with recent forward-modelling and observational analyses of optical cluster samples, which assumed miscentred NFW profile \citep{2021MNRAS.507.5671G,2024A&A...687A.178G}. In simulations, \citet{2025A&A...697A.184G} and \citet{EP-Ragagnin} report richness-dependent lensing mass biases at the $\sim 10$–$20$ per cent level once projection and centring systematics are included. Although the precise value depends on modelling choices, our inferred $\langle b_{\rm sel}\rangle = 1.17$ lies well within the range reported in these studies, supporting the interpretation that correlated large-scale structure is the dominant driver of positive selection bias in richness-selected samples.

\subsection{Richness-density coupling}

\begin{figure}[ht!]
  \centering
  \includegraphics[scale=0.45]{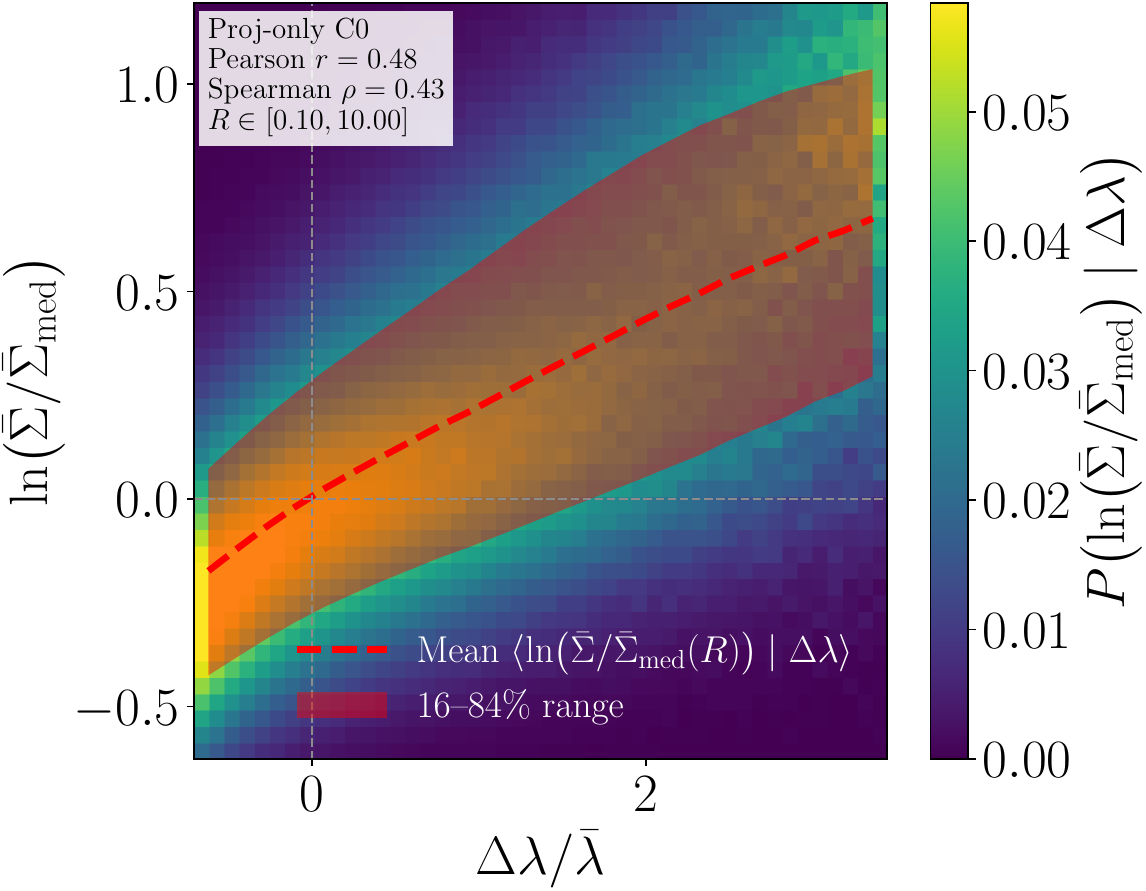}
  \caption{\label{fig:C0_van_lambda_sigma_correlation}
    Correlation between fluctuations in projected galaxy density and richness for the projection-only C0 sample.
    The colour map shows the conditional probability density
    $P\!\left[\ln(\bar{\Sigma}/\bar{\Sigma}_{\rm med})\,\middle|\,\Delta\lambda\right]$,
    where $\Delta\lambda = \lambda-\lambda_{\rm med}$ is the deviation from the median richness, and
    $\ln(\bar{\Sigma}/\bar{\Sigma}_{\rm med})$ is the logarithmic deviation of the projected surface-density profile from its median.
    The overbar denotes an average over radii,
    $R\in[0.1,\,10]\,h^{-1}\mathrm{Mpc}$.
    The red dashed line indicates the mean relation
    $\left\langle \ln(\bar{\Sigma}/\bar{\Sigma}_{\rm med}) \mid \Delta\lambda \right\rangle$,
    while the shaded band encloses the central $16$--$84$ percentile range.
    A clear positive correlation is observed, with Pearson $r=0.40$ and Spearman $\rho=0.34$,
    indicating that clusters with upward richness fluctuations systematically exhibit enhanced projected density profiles.
  }
\end{figure}

\Cref{fig:C0_van_lambda_sigma_correlation} shows a clear positive correlation between richness and projected density fluctuations, with Pearson $r=0.40$ and Spearman $\rho=0.34$. Upward fluctuations in $\lambda_{\rm obs}$ are systematically associated with enhanced projected densities, confirming that the transition-scale peak in $b(R\,|\,\lambda)$ originates from correlated projection effects.

\section{\label{sec:conclusions}Conclusions and outlook}

We introduce \texttt{CosmoPostProcess}, a fast and modular forward model that ingests large-volume $N$-body simulations, populates halos with an HOD calibrated on the \texttt{Flagship2} \Euclid simulation, emulates the \texttt{RICH-CL} procedure developed for \Euclid to obtain a survey-like richness, applies a substructure-informed miscentring prescription validated against membership-based centres, and incorporates baryonic physics
through a mass- and redshift-dependent correction calibrated on hydrodynamical simulations.
In this work, the \texttt{RICH-CL} richness refers to the probabilistic membership definition, which is currently foreseen for the official \Euclid DR1 cosmological analysis. \Euclid also provides an alternative red-sequence-based richness definition, which is not considered here and is expected to be less affected by projection effects owing to its colour selection. The miscentring prescription adopted here improves upon the isotropic models commonly employed in the literature by capturing the richness-lensing correlation induced by miscentring in a physically motivated manner.
Together, these components provide per-bin, selection-aware corrections to
stacked cluster-lensing profiles that are compatible with \Euclid
pipeline definitions.

Our results reveal a consistent physical picture for the optical selection bias on galaxy cluster lensing profiles across richness and redshift. The correlated LoS structure enhances the lensing
signal near the one-two halo transition, producing a peak of
\(20\!-\!40\,\%\) around \(R \simeq 1\,h^{-1}\mathrm{Mpc}\). The anti-correlation between richness and small-scale density profile induced by miscentring
suppresses the inner signal at \(r \lesssim 0.1\!-\!0.5\,h^{-1}\,\mathrm{Mpc}\), while leaving the transition-scale enhancement largely unchanged.

Baryonic physics modifies the projected density profile primarily at small radii and when considered in isolation, it is shown to affect the selection bias of a richness-selected sample at the percent level. However, baryons play an important role once miscentring is included. While the anti-correlation in the inner bias profile is driven primarily by miscentring, baryonic effects further enhance this suppression at the smallest scales and simultaneously amplify the bias peak at the transition between the one-halo and two-halo regimes. This behaviour demonstrates that baryonic physics non-negligibly contributes to shaping the radial dependence of the selection bias through its interplay with miscentring.

Variations in the underlying cosmological parameters primarily impact the selection bias by altering the abundance and large-scale clustering of massive halos. In particular, a lower matter density leads to a reduction in the overall bias amplitude and to an outward shift of the bias peak towards larger radii, reflecting the displacement of the transition between the one-halo and two-halo regimes. These effects are most pronounced at high redshift and high richness, where the bias is largest. This sensitivity motivates, for DR1, an emulator-based treatment of the correction, where the weak lensing bias is interpolated across the set of cosmologies sampled by the \texttt{PICCOLO} simulations, rather than recomputed on the fly within the likelihood analysis. Our tests with alternative HOD parameter choices show that reasonable variations of the mass--richness relation induce only modest changes in the bias,
typically around \(1\!-\!2\,\%\), with a mild inversion of the trend at high
redshift due to the interplay between redshift evolution in the mass--richness
relation and the increased scatter in the observed richness.

These findings highlight that a single set of corrections calibrated at fixed
cosmology and HOD cannot capture the full response of the bias across parameter
space. For \Euclid DR1, a practical approach is to provide
corrections derived at the fiducial cosmology and HOD, accompanied by an
uncertainty estimate reflecting the residual sensitivity to model variations.
For \Euclid DR1, our baseline strategy is therefore to combine corrections derived from the fiducial calibration with an emulator-based description of their cosmological dependence, trained on the set of \texttt{PICCOLO} runs. Beyond DR1, we plan to extend the framework by generating increasingly survey-realistic mocks and by developing a more flexible emulator-based description of the selection bias that can accommodate a broader range of cosmological and halo-occupation variations. In the present implementation, we therefore treat the uncertainty budget primarily through variations of the physical ingredients that dominate the bias prediction; namely, the HOD, baryonic, and miscentring prescriptions, while the residual statistical uncertainty of the individual $P_{\rm mem}$ emulator is assumed to be subdominant.

A remaining ingredient for a fully end-to-end correction at the catalogue level is the survey selection function. In the present work, the calibration and validation were performed on the basis of simulations where the richness mapping can be sampled down to arbitrarily low intrinsic and observed values. Thus, the forward model can effectively capture the full $P(\lambda_{\rm obs}\,|\,\lambda_{\rm true})$ without the additional truncations that (in real data) would translate to purity and completeness factors. For \Euclid DR1, the same forward-modelling approach will be combined with the empirically inferred selection function of the cluster finder to propagate these catalogue-level effects consistently in the cosmological analysis. A closely related strategy, in which the cluster selection enters explicitly in the inference from optically selected samples, has been adopted in recent cosmological applications of AMICO cluster catalogues \citep{2025A&A...703A..25L}.

The code architecture is naturally suited for simulation-based inference (SBI),
enabling future analyses in which cosmological and HOD parameters can be
constrained directly from forward-simulated catalogues and lensing observables.
A key advantage of the framework is that richness, lensing, miscentring, and baryonic corrections are derived consistently from the same simulated realisation, with the relevant modifications implemented directly at the particle level. This makes it possible to track how changes in the forward-modelling ingredients propagate jointly to the galaxy distribution and to the resulting observables.
Planned extensions of this work include expanding the training set of cosmological hydrodynamical simulations used for
the baryonic correction, sampling multiple feedback models to ensure a robust and
agnostic calibration, and incorporating more complete treatments of
photometric-redshift and source-selection systematics. Our long-term goal is to
lower modelling systematics below the statistical precision at the transition
scales most sensitive to selection effects, thereby enabling high-fidelity mass
calibration for cluster cosmology in forthcoming \Euclid data.

\begin{acknowledgements}
This research is supported by: the grant ASI n. 2024-10-HH.0 “Attività scientifiche per la missione \Euclid – fase E”; the Fondazione ICSC, Spoke 3 Astrophysics and Cosmos Observations. National Recovery and Resilience Plan (Piano Nazionale di Ripresa e Resilienza, PNRR) Project ID CN\_00000013 "Italian Research centre on High-Performance Computing, Big Data and Quantum Computing" funded by MUR Missione 4 Componente 2 Investimento 1.4: Potenziamento strutture di ricerca e creazione di "campioni nazionali di R\&S (M4C2-19)" - Next Generation EU (NGEU); the National Recovery and Resilience Plan (NRRP), Mission 4, Component 2, Investment 1.1, Call for tender No. 1409 published on 14.9.2022 by the Italian Ministry of University and Research (MUR), funded by the European Union – NextGenerationEU– Project Title "Space-based cosmology with \Euclid: the role of High-Performance Computing" – CUP J53D23019100001 - Grant Assignment Decree No. 962 adopted on 30/06/2023 by the Italian Ministry of Ministry of University and Research (MUR); in part by the INFN InDark Grant; in part by grant NSF PHY-2309135 to the Kavli Institute for Theoretical Physics (KITP). Michel Aguena, Lucie Baumont, Matteo Costanzi and Emiliano Munari are supported by the PRIN 2022 project EMC2 -\Euclid Mission Cluster Cosmology: unlock the full cosmological utility of the \Euclid photometric cluster catalogue (code no. J53D23001620006). We acknowledge the CINECA award under the ISCRA initiative, for the availability of high performance computing resources and support and the EuroHPC Joint Undertaking for awarding this project access to the EuroHPC supercomputer LEONARDO, hosted by CINECA (Italy) and the LEONARDO consortium through an EuroHPC [Development] Access call. Roberto Ingrao and Matteo Costanzi gratefully acknowledge support from the CNRS/IN2P3
Computing Center (Lyon - France) for providing computing and data-processing
resources needed for this work. \AckCosmoHub

 \AckEC
\end{acknowledgements}

\bibliography{aa60776-26}

\begin{appendix}

\captionsetup{font=small, skip=6pt}
\newcommand{\apptabletune}{
  \renewcommand{\arraystretch}{1.22}
  \setlength{\tabcolsep}{7pt}
}

\section{\label{app:triaxiality}Triaxiality and orientation }

\begin{figure*}[t]
    \centering
    \includegraphics[width=0.95\textwidth]{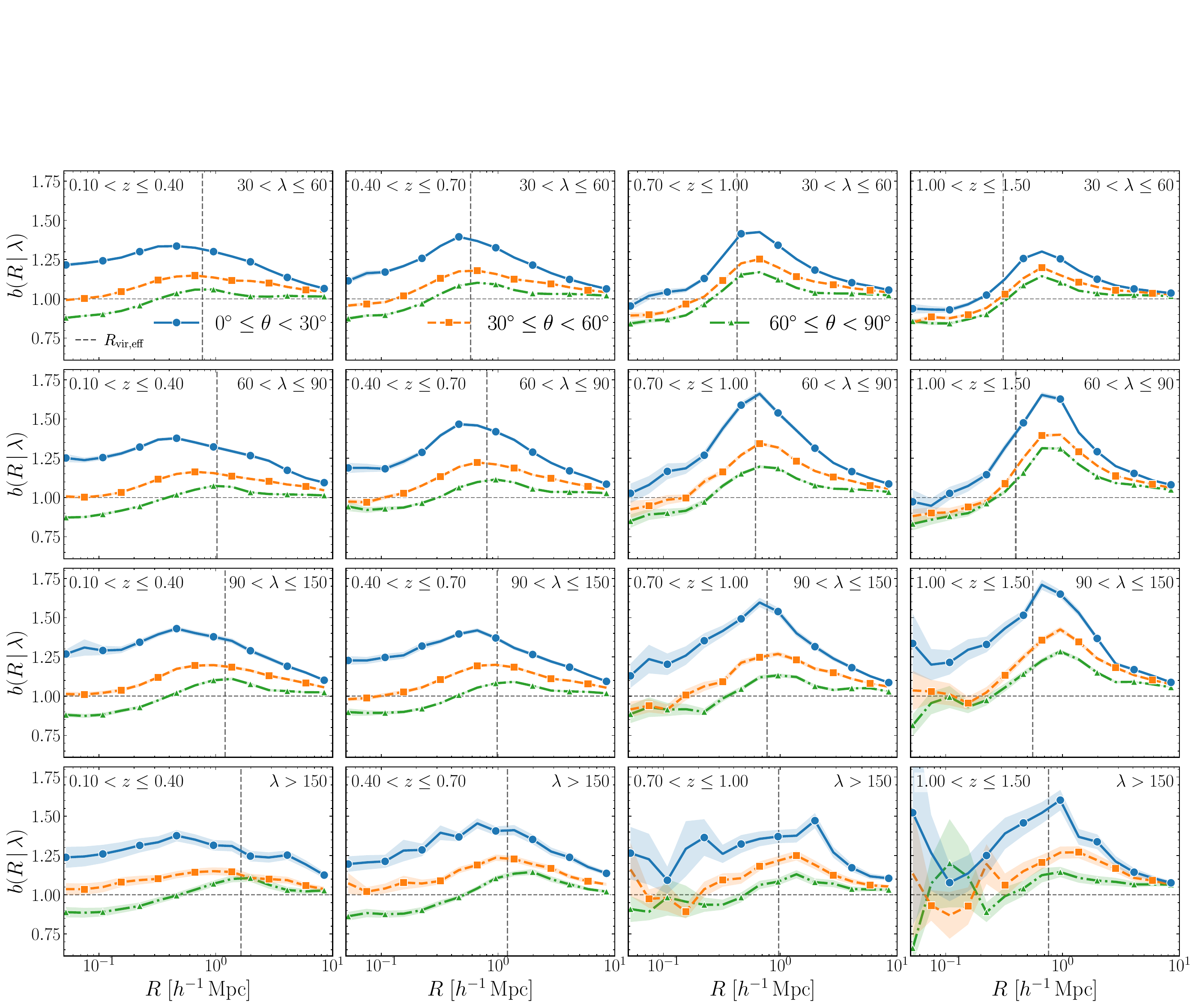}
    \caption{\label{fig:orientation-bias}Selection bias in $\Sigma$ profiles, for the miscentring and baryonified case. In each panel, the three curves with different colours and line types corresponds to different intervals of values of the orientation angle relative to the LoS: $\ang{0} \leq \theta < \ang{30}$ in blue, $\ang{30} \leq \theta < \ang{60}$ in orange, and $\ang{60} \leq \theta < \ang{90}$ green. For lower values of $\theta$ the projection bias is boosted due to the objects being more likely oblate and aligned with contaminants in projection. At the same time, these objects have a larger probability of being well-collimated, thus reducing the miscentring anti-correlation trend in the inner region of the profile.}
\end{figure*}

The code features a module to measure halo triaxiality inside $r_{\rm vir}$ and reports the inclination angle $\theta$ with respect to a chosen LoS, along with the axis ratios
$q=b/a$ and $s=c/a$, where $a\ge b\ge c$ are the semi–axes of the best-fitting ellipsoid.

By default we adopt principal component analysis (PCA) scheme, PCA identifies the orthogonal axes of maximum variance in the particle distribution and is used to infer the halo shape and orientation. Starting from particle positions relative to the halo centre, we first apply a spherical cut at $r_{\rm vir}$ to avoid box-alignment artefacts and then, if necessary, draw an unbiased subsample capped at $4096$ particles. Shapes are not attempted when fewer than $100$ particles survive this preselection and in any case to avoid bottleneck in the galaxy painting loop we limit this part of the analysis to halos above the mass threshold of $10^{13}\,h^{-1}M_{\odot}$, used also for the surface density profiles.

Initialisation comes from a 3D PCA of the centred positions \citep[implemented with \texttt{scikit-learn;}][]{scikit-learn}. The principal directions define a rotation matrix $R$; the associated spreads along those directions set provisional semi-axis lengths $(a,b,c)$, which immediately give the $q$ and $s$ shape parameters.

Refinement proceeds iteratively. At each pass we rotate into the current principal frame, keeping only particles inside the triaxial envelope
\begin{equation}
r_{\rm tri}=\sqrt{d_1^{2}+\left(\frac{d_2}{q}\right)^2+\left(\frac{d_3}{s}\right)^2}\;<\;r_{\rm vir}\,,
\end{equation}
where $d_i$ are the particle Cartesian coordinates in the reference frame of the halo. We then recompute the PCA on the retained set to update our set of geometrical variables: the rotation matrix and axes $(R,a,b,c)$ and therefore $(q,s)$. The process stops once both ratios stabilise,
\begin{equation}
\left|1-\frac{q_{\rm new}}{q}\right|<10^{-6}\quad\text{and}\quad \left|1-\frac{s_{\rm new}}{s}\right|<10^{-6}\,,
\end{equation}
or after $100$ iterations, in which case the halo is flagged as not converged. Finally, letting $\hat{\mathbf{A}}$ be the unit vector along the major axis (first row of $R$) and $\mathbf{e}_n$ the unit vector of the selected Cartesian LoS, the inclination angle can be written as
\begin{equation}
\theta=\cos^{-1}\bigl(\,|\hat{\mathbf{A}}\cdot\mathbf{e}_n|\,\bigr)\in[0^\circ,90^\circ]\,.
\end{equation}

\begin{figure}[ht]
    \centering
    \includegraphics[width=\columnwidth]{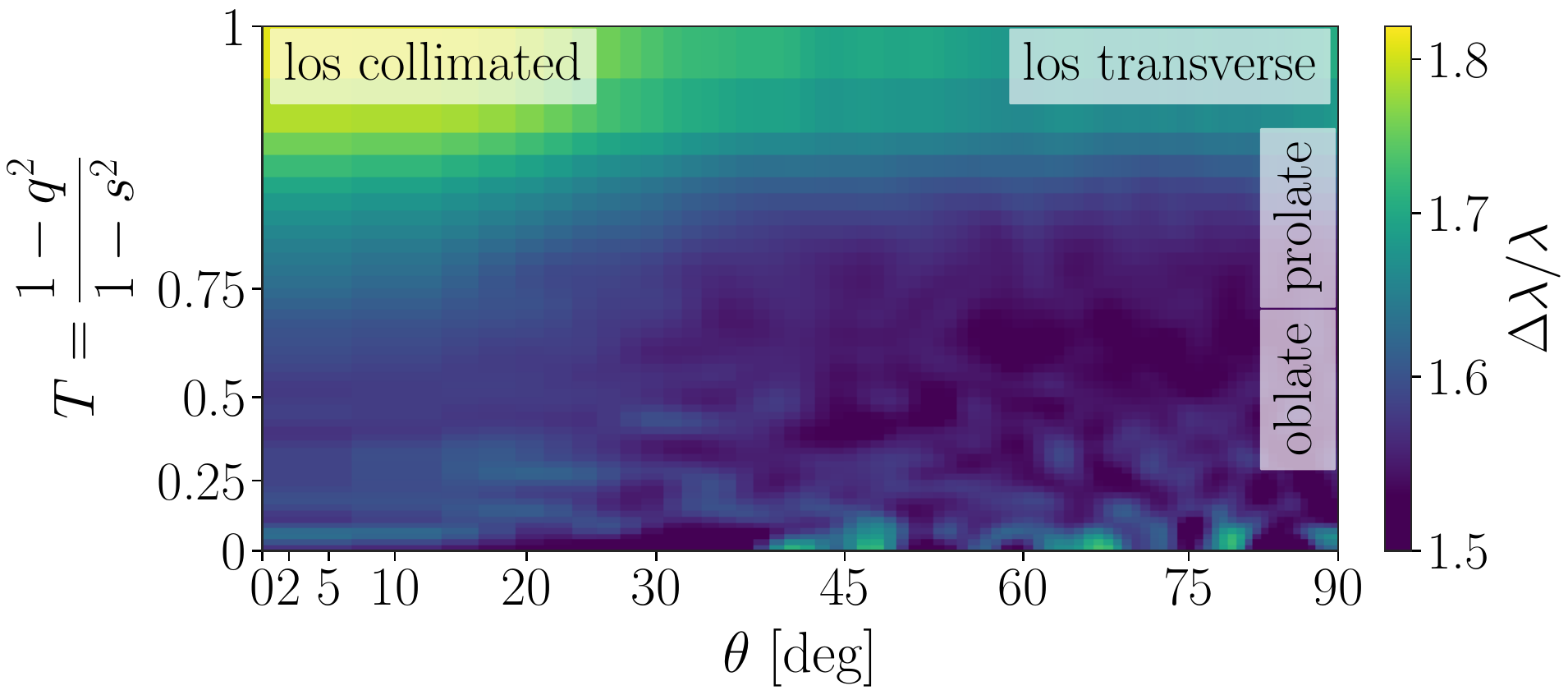}
    \caption{\label{fig:triaxial} 2D histogram, as a function of inclination angle, $\theta$, and triaxiality parameter, $T$, colour-coded by the richness boost, $\Delta\lambda$. The sample corresponds to the full fiducial analysis and includes only up-scattered systems with $\Delta\lambda/\lambda > 0$.}
\end{figure}

For completeness, a reduced tensor-of-inertia variant is available that follows the same iterative envelope but replaces the PCA step with reduced tensor-of-inertia principal axes. This option is retained for cross-checks and diagnostics; in practice, the approach based on the tensor-of-inertia is disfavoured because it is noticeably slower than the PCA path. In \cref{fig:orientation-bias} we study the selection bias on density profiles as a function of inclination angle $\theta$. Binning in orientation we isolate the orientation bias whereby systems whose richness is artificially enhanced by projection effects are over-represented at small $\theta$ that is, preferentially aligned with the LoS. This is even more evident in \cref{fig:triaxial}, where we focus on up-scattered objects exhibiting a positive richness boost,
\(
(\lambda_{\mathrm{obs}}-\lambda_{\mathrm{true}})/\lambda_{\mathrm{true}}
\equiv \Delta\lambda/\lambda,
\)
and analyse how this quantity correlates with both the LoS inclination of the halo and its triaxiality parameter
\(
T=(1-q^2)/(1-s^2),
\)
which provides a measure of the halo shape by quantifying deviations from spherical symmetry in terms of the intermediate-to-major ($q$) and minor-to-major ($s$) axis ratios. From this analysis it emerges that prolate ($T>2/3$) objects show the largest boost when aligned to the LoS due to the larger amount of contaminants along the LoS. Oblate objects ($T<1/3$) are associated to lower boost values, with exception of few sets of object at $\theta>\ang{45}$ close to $T=0$. These classes of object are probably enhanced by contaminants with small radial separation entering the richness computation due to the object being relative spread on the plane perpendicular to the LoS.

\section{Definition and calibration for the baryonification scheme}
\label{app:calib}

This appendix presents a single, end-to-end account of the baryonification model and its calibration, written to be implementation-ready but also to clarify how each modelling choice is tied to the \texttt{Magneticum} hydrodynamical simulations (see \cref{sc:Magneticum}). Throughout we use the same symbols and centring conventions as in the main text. All the distances hereafter are comoving.

\paragraph{Stellar components and baryon fractions (multi-cosmology; interpolated inputs):}
We start from the multi-cosmology \texttt{Magneticum} suite by fitting the total stellar and central-galaxy stellar mass fractions as functions of host-halo mass and redshift. To produce a baryonic correction for our sample in the \texttt{PICCOLO} simulations we need calibrated relations for $z<2$. However given the availability of \texttt{Magneticum} simulations we extended the sampling of the redshift evolution from $z\simeq 0$ to $z\simeq 6$ using all 14 snapshots in the multi-cosmology suite. These fits are used to build interpolators that provide the stellar fractions used in the subsequent calibration steps.

The central-galaxy and satellite-associated density profiles used in the baryonification model are
\begin{equation}
\rho_{\rm cga}(r)=\frac{\fcga\,\Mvir}{4\pi^{3/2}\,\rh\,r^2}\;\,
\exp\biggl[-\frac12\,\biggl(\frac{r}{\rh}\biggr)^3\biggr],\quad
\rho_{\rm sga}(r)=\frac{\fsga}{4\pi r^2}\,\frac{\mathrm{d}M_{\rm dm}(r)}{\mathrm{d}r}\,,
\end{equation}
with $f_{\rm sga}=f_{\star}-f_{\rm cga}$ and $f_{\rm gas}=\fb-f_{\star}$.
The stellar and central-galaxy mass fractions are fitted as smooth functions of mass and redshift,
\begin{align}
f_{\star}(M,z) &= A_{\star}\left(\frac{M}{10^{13}\,h^{-1}\,M_\odot}\right)^{\eta_{\star}(z,\fb)}
+\Delta_{\star}(z,\fb)\:, \label{eq:fractions_star}\\
f_{\rm cga}(M,z)  &= A_{\rm cga }\left(\frac{M}{10^{13}\,h^{-1}\,M_\odot}\right)^{\eta_{\rm cga}(z,\fb)}
+\Delta_{\rm cga}(z,\fb)\,. \label{eq:fractions_cga}
\end{align}
We place safeguard checks in the interpolator for the cosmological and redshift dependence of the fractions to ensure they are always in the interval $[0,f_{\rm b}]$ and their sum adds up to $1$.

To propagate the fitted ${\eta_{\star},\eta_{\rm cga},\Delta_{\star},\Delta_{\rm cga}}$ to arbitrary $(z,\fb)$, we build smooth 2D interpolants over the sampled $(z,\fb)$ points. We compute the Delaunay triangulation and evaluate values inside the convex hull via piecewise-linear (barycentric) interpolation on each triangle; any remaining undefined locations (e.g.\ thin gaps) are filled using a nearest-neighbour fallback. This approach preserves the local trends of the discrete fits and avoids spurious oscillations.

In practice, we first fit \cref{eq:fractions_star,eq:fractions_cga} on each simulation snapshot, thereby capturing the redshift evolution across the full snapshot grid. We then fix the mass-fraction normalisations to the median of the snapshot best-fit values and perform a second fit in which only the mass-trend parameters are left free, that is the power-law slopes and additive offsets. This yields $A_{\star}=0.04119$ and $A_{\rm cga}=0.01084$. The posterior means from this second fitting stage are used to build the interpolators, which are subsequently employed in the profile calibrations. For the gas calibration, the corresponding dependence on the stellar mass fraction enters through $f_{\rm gas}=\fb-f_{\star}$ (in the fitting implementation this relation is used in a direct, explicit form).

\paragraph{Hot gas profiles fitting:} In four simulation snapshots, corresponding to the redshifts $z\in\{1.17,\,0.47,\,0.25,\,0.00\}$, we selected halos with $M_{\rm vir}\ge10^{13}\,h^{-1}M_\odot$ and built spherically averaged density profiles, $\rho^{\rm data}$, for the hot gas ($T>10^6\,{\rm K}$), stars, dark matter, and for their total matter distribution. Halos are grouped in logarithmic mass bins (4 bins per snapshot) and stacked on a common radial grid. Radial uncertainties are estimated from the stacking procedure and propagated into the fits.

For each stack (i.e.\ at fixed snapshot $z$ and mass bin $b$) we fit the hot-gas density profile of \cref{eq:hot-gas-1,eq:hot-gas-2,eq:hot-gas-3} using a Gaussian least-squares objective. Defining
\begin{equation}
\chi^2_b(\boldsymbol\vartheta)
=
\sum_j
\left[
\frac{
\rho^{\rm data}_{b,j}
-
\rho^{\rm model}_{b,j}(\boldsymbol\vartheta)
}{
\sigma_{b,j}
}
\right]^2\;,
\end{equation}
where $j$ indexes the available radial points and $\sigma_{b,j}$ is the corresponding uncertainty, we use the equivalent log-likelihood
\begin{equation}
\ln\widetilde{\mathcal{L}}_b(\boldsymbol\vartheta)
=
-\frac{1}{2}\chi^2_b(\boldsymbol\vartheta)\;.
\end{equation}
Here $\widetilde{\mathcal{L}}_b$ denotes an unnormalised Gaussian likelihood: all parameter-independent normalisation terms, including the usual factors involving $2\pi$ and the data covariance, are omitted. Since the uncertainties are fixed and are not part of the parameter model, these terms do not affect the calibration.

At fixed snapshot, the combined calibration objective is implemented as a mass-weighted sum over bins,
\begin{equation}
\ln\widetilde{\mathcal{L}}_{\rm snap}(\boldsymbol\vartheta)
=
\sum_{b=1}^{B}
w(M_b)\,
\ln\widetilde{\mathcal{L}}_b(\boldsymbol\vartheta),
\qquad
w(M_b)=
\frac{M_b}{10^{14}\,h^{-1}\,M_\odot}\,,
\end{equation}
with $M_b$ as the median mass within each bin. The mass weights $w(M_b)$ are introduced as a calibration choice in the construction of the snapshot-level objective, not as a Bayesian prior. The resulting quantity should therefore be interpreted as a weighted fitting objective, or equivalently as an unnormalised pseudo-likelihood, rather than as a fully normalised probability density. The parametric form of the hot-gas profile is physically motivated and has been validated primarily over a limited mass regime; therefore, we deliberately anchor the global calibration to the bins where the underlying assumptions are expected to hold best. In practice, in hydrodynamical simulations higher-mass halos yield a cleaner and more stable characterisation of the hot component, so up-weighting their contribution prevents the fit from being dominated by low-mass bins where the model is not intended to be equally accurate.

We adopt uninformative priors and sample the posterior using the affine-invariant ensemble MCMC sampler implemented in \texttt{emcee} \citep{2013PASP..125..306F}.
Convergence is monitored using the integrated auto-correlation time, $\tau$, estimated from the chains: we require the total chain length to satisfy $N_{\rm step}\,\gtrsim\,50\,\max(\tau)$ and the $\tau$ estimates to be stable with fractional changes $\lesssim 1\,\%$ between successive checks. We discard an initial burn-in of twice $\,\max(\tau)$ and verify trace-plot stability by eye. As a basic sampler health check, we require the mean acceptance fraction to lie in the range $\sim 0.2$--$0.5$.
Posterior means are reported in \Cref{tab:gas-snapshot-fits,tab:ac-fits}.

The hot gas is modelled as the sum of a compact core (``one-halo'') component and a diffuse tail,
\begin{equation}
\rho_{\rm gas}(r)=\rho_{\rm core}(r)+\rho_{\rm tail}(r)\,.
\label{eq:hot-gas-1}
\end{equation}
The core controls the inner distribution and fixes the gas mass budget, while the tail sets the onset of the diffuse regime and the large-radius decay. In the fitting implementation, radial scalings are expressed relative to the halo size provided by each stack (denoted as $R_{\rm vir}$ in the code; we keep $R_{\rm vir}$ below for consistency).

The core uses two characteristic radii,
\begin{equation}
r_{\rm ej}=\theta_{\rm ej}\,r_{\rm vir}\;,\qquad r_{\rm co}=\theta_{\rm core}\,r_{\rm vir}\,,
\end{equation}
and the core profile is written as
\begin{equation}
\rho_{\rm core}(r)=\frac{\rho_0}{\left(1+r/r_{\rm co}\right)^{\beta}\,\left[1+(r/r_{\rm ej})^2\right]^{(7-\beta)/2}}\;.
\label{eq:hot-gas-2}
\end{equation}
The parameter $\beta$ controls the slope for core profile in inner and intermediate region; its value is allowed to vary smoothly with halo mass and to remain bounded between 0 and 3. The central density $\rho_0$ is fixed by the gas mass budget as described below.

The tail component of the gas density profile is implemented as a smooth rise to an outer normalisation followed by a softened power-law decay,
\begin{equation}
\rho_{\rm tail}(r)=
\left[f_{\rm in}+\frac{\rho_{\rm norm}-f_{\rm in}}{1+{\rm e}^{-s\,(r-x_{\rm t})}}\right]
\left[1+\left(\frac{r}{x_{\rm cut}}\right)^k\right]^{-\gamma/k}\,,
\label{eq:hot-gas-3}
\end{equation}
with fixed \((f_{\rm in},s,k)=(10^{7},10,5)\). The first factor in \cref{eq:hot-gas-3} describes a smooth transition from an inner floor to an outer normalisation. In the implementation, \(f_{\rm in}\) is a fixed, dimensionless baseline value that prevents the tail component from vanishing at small radii, \(x_{\rm t}\) sets the radial location of the transition, and \(s\) controls its sharpness through the width of the logistic rise. The second factor governs the large-radius behaviour: \(x_{\rm cut}\) defines the characteristic cut-off scale, \(k\) controls how gradual the turnover is around \(x_{\rm cut}\), and \(\gamma\) fixes the asymptotic outer slope, \(\rho_{\rm tail}\propto r^{-\gamma}\) for \(r\gg x_{\rm cut}\).

For a fixed snapshot, the gas-profile parameters fitted to the stacked profiles are collected in the vector
\begin{equation}
\boldsymbol\Theta=
\bigl[\mathcal{M}_{c,0},\mathcal{M}_{c,1},\mu,\theta_0,\theta_1,\theta_2,\theta_{{\rm ej},{\rm br}},x_{t,0},
\gamma_0,\gamma_1,\gamma_2,\gamma_{{\rm br}}\bigr]\,,
\end{equation}
where \(\mathcal{M}_{c,0}\) and \(\mathcal{M}_{c,1}\) control the mass dependence of the core-pivot scale (defined below), \(\mu\) sets the sharpness of the transition in the core slope, \((\theta_0,\theta_1,\theta_2,\theta_{{\rm ej},{\rm br}})\) parametrise the broken-line mass dependence of the ejection scale \(\theta_{\rm ej}\), \(x_{t,0}\) fixes the transition location of the tail component, and \((\gamma_0,\gamma_1,\gamma_2,\gamma_{{\rm br}})\) define the corresponding broken-line dependence of the outer slope \(\gamma\).

The mass dependence is expressed in terms of the dimensionless variable
\begin{equation}
x_M \equiv \logten\!\left(\frac{M}{10^{14}\,h^{-1}\,M_\odot}\right)\,,
\end{equation}
and we introduce a characteristic pivot mass \(M_c(M)\) through its logarithm (in units of \(h^{-1}\,M_\odot\)),
\begin{equation}
\logten\!\left(\frac{M_c(M)}{h^{-1}\,M_\odot}\right)
\equiv \mathcal{M}_c(M)
= \mathcal{M}_{c,0}+\mathcal{M}_{c,1}\,x_M\,.
\label{eq:Mc_def}
\end{equation}
With this definition, \(\mathcal{M}_c\) is dimensionless and the corresponding mass scale is
\begin{equation}
M_c(M)=10^{\mathcal{M}_c(M)}\,h^{-1}\,M_\odot\,.
\end{equation}
The inner-slope parameter \(\beta\) in \cref{eq:hot-gas-2} is then written as a smooth function of the ratio between the halo mass and the pivot mass,
\begin{equation}
x \equiv \frac{M_c(M)}{M}\,,
\qquad
\beta(M)=\frac{3\,x^\mu}{1+x^\mu}\,,
\label{eq:beta_of_M}
\end{equation}
so that \(x\) is dimensionless and \(0\le\beta(M)\le 3\). Varying \(\mathcal{M}_c\) shifts the characteristic mass scale at which the core steepens, while \(\mu\) controls how rapidly the transition in \(\beta(M)\) occurs with halo mass.

The ejection scale $\theta_{\rm ej}$ and the outer slope $\gamma$ follow a continuous broken-line dependence in $x_M$,
\begin{equation}
\mathcal{B}(x_M;p_0,p_1,p_2,x_{\rm br})=
\begin{cases}
p_0+p_1\,(x_M-x_{\rm br}), & x_M<x_{\rm br}\,,\\
p_0+p_2\,(x_M-x_{\rm br}), & x_M\ge x_{\rm br}\,,
\end{cases}
\end{equation}
so that $\mathcal{B}(x_{\rm br})=p_0$ and the dependence is continuous at the break. In the fit,
\begin{align}
&\theta_{\rm ej}(M) = \mathcal{B}\!\left(x_M;\theta_0,\theta_1,\theta_2,\theta_{{\rm ej},{\rm br}}\right)\,,\\
&\gamma(M) = \mathcal{B}\!\left(x_M;\gamma_0,\gamma_1,\gamma_2,\gamma_{{\rm br}}\right)\,,
\end{align}
with the additional implementation safeguard $\gamma\ge 0.2$ when evaluating the model. The tail transition location is kept mass-independent and set to $x_{\rm t}=x_{t,0}$.

The mass-dependence of the core-size parameter is set by a piecewise prescription,
\begin{equation}
\small
\theta_{\rm core}(M,z)=
\begin{cases}
0.07\,(1+z)^{1/2},& M/(h^{-1}\,M_\odot)<1.2\times10^{13}\,,\\
0.10\,(1+z)^{1/2},& 1.2\times10^{13}\leq M /(h^{-1}\,M_\odot)<7\times10^{13}\,,\\
0.025\,(1+z)^{1/2},& M/(h^{-1}\,M_\odot)\ge7\times10^{13}\,,
\end{cases}
\end{equation}
which fixes $r_{\rm co}=\theta_{\rm core}\,r_{\rm vir}$. In the shown fitting implementation, this prescription is evaluated at fixed $z$ within a snapshot fit.

The core amplitude is fixed by enforcing that the total core mass equals the gas mass fraction times the halo mass. In the fitting implementation, the gas fraction entering this normalisation is supplied by an explicit fit $f_{\rm gas}(M)$, and we impose
\begin{equation}
f_{\rm gas}(M)\,M = 4\pi\int_0^{\infty}\rho_{\rm core}(r)\,r^2\,{\rm d}r\,,
\end{equation}
where the integral is evaluated numerically.

For the tail component we set the cut-off scale proportional to the ejection radius,
\begin{equation}
x_{\rm cut}=1.25\,\theta_{\rm ej}\,r_{\rm vir}\,,
\end{equation}
which places the transition between the intermediate and asymptotic regimes slightly beyond the gas ejection scale.

The outer normalisation is defined relative to a reference density scale,
\begin{equation}
\rho_{\rm norm}
=\Omega_{\rm m}\,\rho_{\rm ref}
\left[
f_{\rm gas}(M)
+
1.25\,\frac{\rho_0\,r_{\rm vir}^3}{M}
\right]\,,
\end{equation}
where $\rho_{\rm ref}$ denotes the reference density used in the implementation (e.g. the critical or mean matter density, depending on the adopted convention). The second term accounts for the residual contribution of the core component to the outer gas distribution and is written in terms of the dimensionless ratio $\rho_0 r_{\rm vir}^3/M$, ensuring that the bracketed quantity remains dimensionless.

The fitted model entering the likelihood is therefore
\begin{equation}
\rho^{\rm model}(r)=\rho_{\rm core}(r)+\rho_{\rm tail}(r)\;,
\end{equation}
evaluated on the same radial grid used for the stacked profiles.

\begin{table*}[!ht]
\caption{Hot-gas best-fit parameters.}
\label{tab:gas-snapshot-fits}
\centering
\apptabletune
\resizebox{\textwidth}{!}{
\begin{tabular}{ccccccccccccc}
\toprule
$z$ & $\mathcal{M}_{c,0}$ & $\mathcal{M}_{c,1}$ & $\mu$
& $\theta_0$ & $\theta_1$ & $\theta_2$ & $\theta_{\rm br}$
& $x_{\rm t}$ & $\gamma_0$ & $\gamma_1$ & $\gamma_2$ & $\gamma_{\rm br}$ \\
\midrule
0.00 & $11.1^{+1.7}_{-2.2}$ & $1.71^{+0.62}_{-0.43}$ & $0.11^{+0.15}_{-0.05}$
& $1.157^{+0.016}_{-0.016}$ & $-0.080^{+0.019}_{-0.020}$
& $-1.3^{+0.8}_{-1.9}$ & $0.36^{+0.04}_{-0.14}$
& $0.363^{+0.042}_{-0.051}$ & $1.642^{+0.080}_{-0.079}$
& $-0.18^{+0.14}_{-0.13}$ & $-4.50^{+0.54}_{-0.35}$
& $0.072^{+0.043}_{-0.033}$ \\
0.25 & $10.9^{+1.9}_{-2.0}$ & $2.19^{+0.86}_{-0.67}$ & $0.08^{+0.13}_{-0.03}$
& $1.196^{+0.015}_{-0.017}$ & $-0.015^{+0.022}_{-0.024}$
& $-1.2^{+0.8}_{-2.0}$ & $0.25^{+0.05}_{-0.15}$
& $0.296^{+0.054}_{-0.072}$ & $1.79^{+0.11}_{-0.10}$
& $0.00^{+0.16}_{-0.17}$ & $-3.75^{+0.72}_{-0.78}$
& $0.042^{+0.052}_{-0.064}$ \\
0.47 & $10.9^{+1.7}_{-1.9}$ & $0.94^{+0.24}_{-0.28}$ & $0.09^{+0.11}_{-0.04}$
& $1.152^{+0.016}_{-0.016}$ & $-0.053^{+0.022}_{-0.022}$
& $-1.1^{+0.6}_{-1.9}$ & $0.23^{+0.07}_{-0.19}$
& $0.317^{+0.050}_{-0.073}$ & $1.72^{+0.10}_{-0.11}$
& $0.04^{+0.14}_{-0.17}$ & $-3.1^{+0.9}_{-1.2}$
& $0.08^{+0.08}_{-0.10}$ \\
1.17 & $11.7^{+1.0}_{-1.7}$ & $-0.09^{+0.62}_{-0.89}$ & $0.078^{+0.073}_{-0.035}$
& $0.942^{+0.040}_{-0.043}$ & $-0.29^{+0.20}_{-0.06}$
& $1.1^{+1.4}_{-0.7}$ & $-0.1^{+1.5}_{-0.2}$
& $0.14^{+0.11}_{-0.09}$ & $1.47^{+0.17}_{-0.15}$
& $-1.53^{+0.42}_{-0.42}$ & $2.8^{+1.4}_{-1.9}$
& $-0.28^{+0.09}_{-0.11}$ \\
\bottomrule
\end{tabular}}
\tablefoot{Parameters are defined in \cref{eq:hot-gas-1,eq:hot-gas-2,eq:hot-gas-3}. $\mathcal{M}_{c,0}$ and $\mathcal{M}_{c,1}$ define the mass pivot, $\mu$ controls the sharpness of the transition, $(\theta_0,\theta_1,\theta_2,\theta_{\rm br})$ govern the mass trend of the ejection scale, $x_{\rm t}$ marks the onset of the diffuse tail, and $(\gamma_0,\gamma_1,\gamma_2,\gamma_{\rm br})$ determine the outskirts slope. Units follow the definitions in \Cref{app:calib}.}
\end{table*}

\paragraph{3. Stellar profiles fits:}
For the stellar profiles we keep the functional forms fixed and use the interpolated fractions from step (1), with $f_{\rm sga}=f_{\star}-f_{\rm cga}$ and $f_{\rm gas}=\fb-f_{\star}$. To account for the \texttt{Magneticum} brightest central galaxies (BCG) being more massive than expected \citep[e.g.][]{2019MNRAS.483.3545G,2020AAS...23613904A,2025A&A...694A.207M}, we use a central-galaxy scale radius $rh$ five times larger than the fiducial choice:
\begin{equation}
\rho_{\rm cga}(r)=\frac{\fcga\,\Mvir}{4\pi^{3/2}\,\rh\,r^2}\,
\exp\bigl[-\tfrac12\,(r/r_{\rm h})^3\bigr],\,\,\,\,\rh=0.075\,\rvir\;,
\end{equation}
and the satellite-associated component follows the local dark matter slope,
\begin{equation}
\rho_{\rm sga}(r)=\frac{\fsga}{4\pi r^2}\,\frac{\mathrm{d}M_{\rm dm}(r)}{\mathrm{d}r}\;.
\end{equation}
The total baryonic density and enclosed mass are then expressed as
\begin{align}
&\rho_{\rm b}(r) =\rho_{\rm gas}(r)+\rho_{\rm cga}(r)+\rho_{\rm sga}(r)\;,\\
&M_{\rm b}(<r)=4\pi\int_0^r r'^2\,\rho_{\rm b}(r')\,{\rm d}r'\;.
\end{align}

\paragraph{Adiabatic contraction:} Contraction is calibrated by comparing the dark matter density profile in the hydrodynamical run to the DMO profile obtained by applying the \cite{2025JCAP...12..043S} AC mapping to the matched gravity-only run, while using the simulated baryonic components as inputs. We define: (i) $\rho_{\rm DMO}$ and $M_{\rm DMO}$ as the density and enclosed mass of the gravity-only dark matter; (ii) $\rho_{\rm dm}^{\rm hydro}$ and $M_{\rm dm}^{\rm hydro}$ as the corresponding quantities for the DM component in the  hydrodynamical simulations; (iii) a ``central'' baryonic component $\rho_{\rm cga}$ given by stars plus cold gas from the simulation, and a hot-gas component $\rho_{\rm hga}$ given by the simulated hot gas. We denote by $M_{\rm cga}$ and $M_{\rm hga}$ the corresponding enclosed masses. The cold dark matter (CDM) fraction is fixed to $f_{\rm CDM}=1-\fb$.

We model the mapping between initial and final radii through $\xi \equiv r_{\rm i}/r_{\rm f}$. Numerically, we first construct a provisional mapping on the native radial grid by defining
\begin{equation}
\xi(r)=1+K(r),\qquad K(r)=K_0(r)+K_1(r)+K_2(r)\,,
\end{equation}
with
\begin{align}
K_0(r) &= \frac{Q_0(z)}{1+\bigl(r/r_{\rm step}\bigr)^{3/2}}\,,\\
K_1(r) &= Q_1(z)\,
\frac{M_{\rm cga}(<r)}{M_{\rm dm}^{\rm hydro}(<r)}\,,\\
K_2(r) &= Q_2(z)\,
\frac{M_{\rm hga}(<r)}{M_{\rm dm}^{\rm hydro}(<r)}\,,
\end{align}
and redshift evolution
\begin{equation}
Q_i(z)=q_i(1+z)^{p_i},\qquad i=0,1,2\,.
\end{equation}
The radial pivot in $K_0$ is linked to the virial scale and to the peak height $\nu$, which provides a dimensionless measure of halo mass relative to the characteristic fluctuation amplitude of the linear density field at redshift $z$. We define
\begin{equation}
\nu(M,z)\equiv \frac{\delta_{\rm c}(z)}{\sigma(M,z)}\,,
\end{equation}
where $\sigma(M,z)$ is the root-mean-square fluctuation of the matter density field smoothed with a top-hat filter enclosing mass $M$ (in Lagrangian space) and evaluated at redshift $z$. The quantity $\delta_{\rm c}(z)$ is the linear overdensity threshold for spherical collapse; in $\Lambda$CDM it has a weak redshift dependence and is often approximated as $\delta_{\rm c}\simeq 1.686$. In this notation, the step radius is written as
\begin{equation}
r_{\rm step}=r_{\rm vir}\,\frac{\varepsilon_0+\varepsilon_1\,\nu}{\varepsilon_0}\,,
\end{equation}
so that, at fixed $r_{\rm vir}$, the pivot shifts to larger radii for higher peak height.

The provisional relation above defines $r_{\rm f}(r)=r/\xi(r)$. We then re-express the mapping as a function of final radius by inverting this relation (using a monotonic interpolation), yielding $\xi(r_{\rm f})$ on the same radial grid. This $\xi(r_{\rm f})$ is used to compute the contracted dark matter density profile predicted from the gravity-only run,
\begin{equation}
\rho_{\rm dm}^{\rm AC}(r_{\rm f})
=\frac{1}{4\pi r_{\rm f}^2}\,
\frac{\mathrm{d}}{\mathrm{d}r_{\rm f}}\Bigl[f_{\rm CDM}\,M_{\rm DMO}\bigl(\xi(r_{\rm f}),r_{\rm f}\bigr)\Bigr]\,.
\end{equation}

We fit $\rho_{\rm dm}^{\rm AC}(r)$ to the hydrodynamical dark matter density profile $\rho_{\rm dm}^{\rm hydro}(r)$ in $\logten$-space on the same radial grid as the gas fits. The parameters
$\boldsymbol\phi_{\rm AC}=[q_0,p_0,q_1,p_1,q_2,p_2,\varepsilon_0,\varepsilon_1]$
are inferred jointly across snapshots; the resulting posterior means are reported in \Cref{tab:ac-fits}. Together with $\boldsymbol\Theta$ they determine the contracted dark matter component and hence the total density,
\begin{equation}
\rho_{\rm tot}(r)=\rho_{\rm dm}^{\rm AC}(r)+\rho_{\rm gas}(r)+\rho_{\rm cga}(r)\,.
\end{equation}

\begin{table}
\caption{Best-fitting global adiabatic-contraction parameters.}
\label{tab:ac-fits}
\centering
\begingroup
\apptabletune
\small
\setlength{\tabcolsep}{6pt}
\renewcommand{\arraystretch}{1.25}
\begin{tabular*}{\columnwidth}{@{\extracolsep{\fill}}cccc@{}}
\toprule
$q_0$ & $p_0$ & $q_1$ & $p_1$ \\
\midrule
$0.0225^{+0.0126}_{-0.0095}$ &
$0.68^{+0.91}_{-0.97}$ &
$0.158^{+0.073}_{-0.056}$ &
$1.05^{+0.66}_{-0.69}$ \\
\midrule
$q_2$ & $p_2$ & $\varepsilon_0$ & $\varepsilon_1$ \\
\midrule
$-0.215^{+0.088}_{-0.098}$ &
$0.23^{+0.85}_{-1.03}$ &
$4.09^{+0.84}_{-0.82}$ &
$0.48^{+0.56}_{-0.54}$ \\
\bottomrule
\end{tabular*}
\endgroup
\tablefoot{These are the global parameters $\boldsymbol{\phi}_{\rm AC}$ of the \cite{2025JCAP...12..043S} AC model, calibrated on matched \texttt{Magneticum} halos. The calibration matches the hydrodynamical dark matter density profile to the contracted DMO profile obtained by applying the AC model to the corresponding gravity-only run. It is performed jointly across snapshots and mass bins using a Gaussian likelihood. Values are posterior medians with $1\sigma$ ($16$--$84\,\%$) credible intervals.}
\end{table}

\paragraph{5. Model validation:}
Model predictions are compared to surface-density observables using the standard 3D-to-2D explained in text (in the context of surface quantities such as $\Sigma$ and $\Delta \Sigma$ the $r$ variable must be interpreted as projected radius perpendicular to the LoS). As shown in the main text in \cref{fig:deltasigma-vs-magneticum}, after the conversion to $\Delta \Sigma(r)$ we retain an agreement within $1$--$2\,\%$ between simulation and calibrated model. Furthermore residuals are examined as a function of mass, redshift, and radius. The adopted settings were selected to mitigate residuals mass-dependent trends in the stacks while retaining numerical stability and physical monotonicity across the full radial range.

\section{\label{app:flagship}\texttt{Flagship2} calibrations}

\begin{figure}[b]
  \centering
  \includegraphics[width=\linewidth]{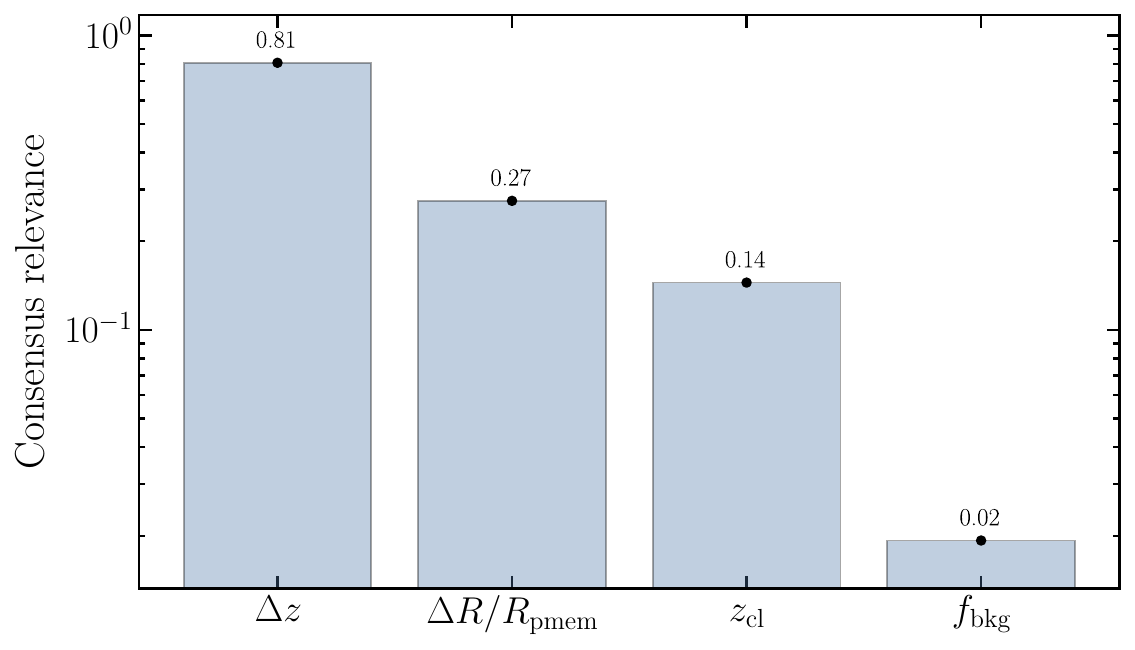}
  \vspace{-0.5em}
  \caption{Consensus relevance of the four inputs \((\Delta z,\,\Delta r/r_{\mathrm{cl}},\,z_{\mathrm{cl}},\,f_{\mathrm{bkg}})\) obtained by combining saliency, integrated gradients, and conditional permutation importance.}
  \label{fig:relevance}
\end{figure}

Here we detail the calibrations of the HOD model and of the neural-network membership emulator used in the main analysis.

\begin{figure*}[!ht]
  \centering
  \includegraphics[width=\textwidth]{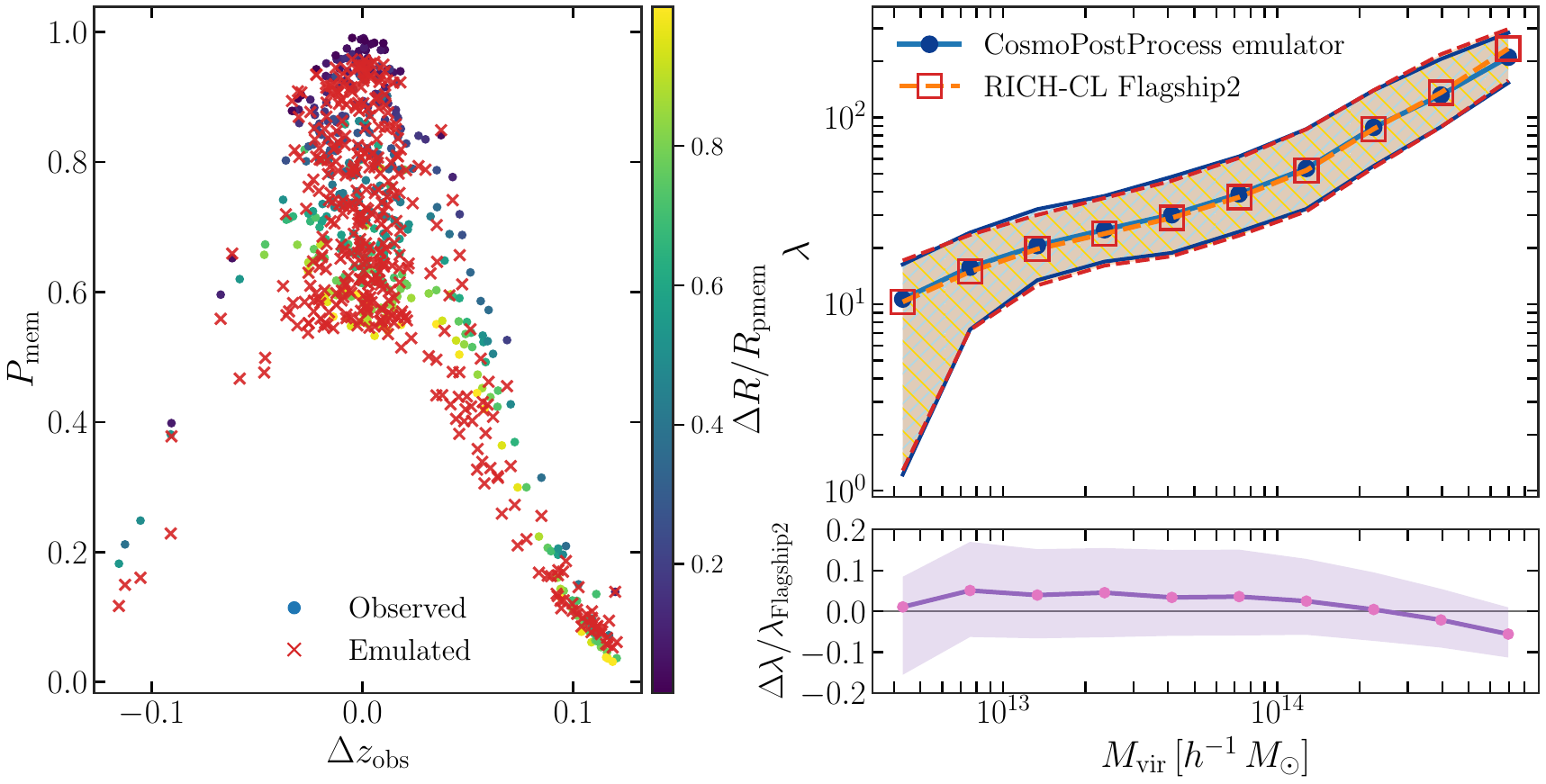}
  \vspace{-0.5em}
  \caption{\textit{Left panel:} Comparison between observed and emulated $P_{\mathrm{mem}}$ as a function of $\Delta z$. \textit{Right, top:} Richness versus $M_{\mathrm{vir}}$ for the reference pipeline (squares) and for the emulator (circles) with the filled regions marking the $1\sigma$ scatter in the relation. \textit{Right, bottom:} Fractional residuals relative to the reference calibration.}
  \label{fig:membership-panels}
\end{figure*}

\paragraph{HOD calibration:} We use an eight-tile subsample covering a total of $400\,\mathrm{deg}^2$ of the {\tt Flagship2} $5200\,\mathrm{deg}^2$ catalogue as our reference set. To produce the final catalogue used to fit the model described in \cref{sc:galaxy-painting} we match halos to detected clusters according to their major contribution to the richness. We then apply a conservative mass threshold of $10^{13}\,h^{-1}\,M_\odot$ and fix the pivot redshift to $z_{\rm piv}=1.25$. The mean relation between richness, mass, and redshift is as defined in \cref{sc:galaxy-painting}. To capture extra dispersion beyond counting noise we adopt a gamma-Poisson (negative binomial) model, with latent rate $\Lambda$ drawn from a gamma distribution of shape $\kappa$ and scale $\theta$, and observed richness $N$ drawn from a Poisson distribution conditional on $\Lambda$,
\begin{equation}
\begin{aligned}
&\Lambda \sim \mathrm{Gamma}(\kappa,\theta)\,,\\
&N \mid \Lambda \sim \mathrm{Poisson}(\Lambda)\,.
\end{aligned}
\end{equation}
We choose $\theta$ and $\kappa$ so that $\mathbb{E}[\Lambda]=\bar{\lambda}(M,z)$ and $\mathrm{Var}[\Lambda]=\big[\bar{\lambda}(M,z)\,s\big]^2$, implying
\begin{equation}
\begin{aligned}
&\theta = \frac{\bar{\lambda}(M,z)}{\kappa}\,, \qquad
\kappa = \frac{1}{s^{2}}\,.
\end{aligned}
\end{equation}
After marginalisation over $\Lambda$ the richness follows a negative binomial distribution with probability mass function
\begin{equation}
\begin{aligned}
P_{\lambda}\left(N=n \mid \bar{\lambda},\kappa\right)
&= \frac{\Gamma(n+\kappa)}{\Gamma(\kappa)\,n!}\,
\left(\frac{\kappa}{\kappa+\bar{\lambda}}\right)^{\kappa}
\left(\frac{\bar{\lambda}}{\kappa+\bar{\lambda}}\right)^{n}\,,
\end{aligned}
\end{equation}
with mean and variance
\begin{equation}
\begin{aligned}
&\mathbb{E}[N] = \bar{\lambda}\,,\qquad
\mathrm{Var}[N] = \bar{\lambda}+\big(\bar{\lambda}\,s\big)^{2}\,.
\end{aligned}
\end{equation}
The total log-likelihood over the calibration set is
\begin{equation}
\begin{aligned}
\ln \mathcal{L}
&= \sum_{i} \ln P_{\lambda}\!\left(N_i \mid \bar{\lambda}_i,\kappa_i\right) \\
&= \sum_{i} \Bigg[
\ln\Gamma(N_i+\kappa_i)
- \ln\Gamma(\kappa_i)
- \ln(N_i!) \\
&\hspace{2.5em}
+ \kappa_i \ln\!\left(\frac{\kappa_i}{\kappa_i+\bar{\lambda}_i}\right)
+ N_i \ln\!\left(\frac{\bar{\lambda}_i}{\kappa_i+\bar{\lambda}_i}\right)
\Bigg]\;,
\end{aligned}
\end{equation}
with $\bar{\lambda}_i=\bar{\lambda}(M_i,z_i)$ and $\kappa_i=1/s^{2}$. The posterior for the mass--richness relation used to calibrate the HOD is obtained with the {\tt Nautilus} nested sampler \citep{2023MNRAS.525.3181L}. The adopted priors and resulting constraints are listed in \cref{tab:hod_priors}. The HOD variations displayed in \cref{fig:C0_HODtails_bias} are chosen starting from the DES+SPT joint posterior for the mass--richness relation. Specifically, the HOD parameters are selected by mapping the original model in \cite{2021PhRvD.103d3522C} to ours, and selecting the $2\,\sigma$ extremal points along the direction orthogonal to the parameters $\logten{(M_{1,\mathrm{sat}}\,h/M_{\odot})}$ and $\alpha$ degeneracy.

\begin{figure*}[ht!]
    \centering
    \includegraphics[width=1.02\textwidth]{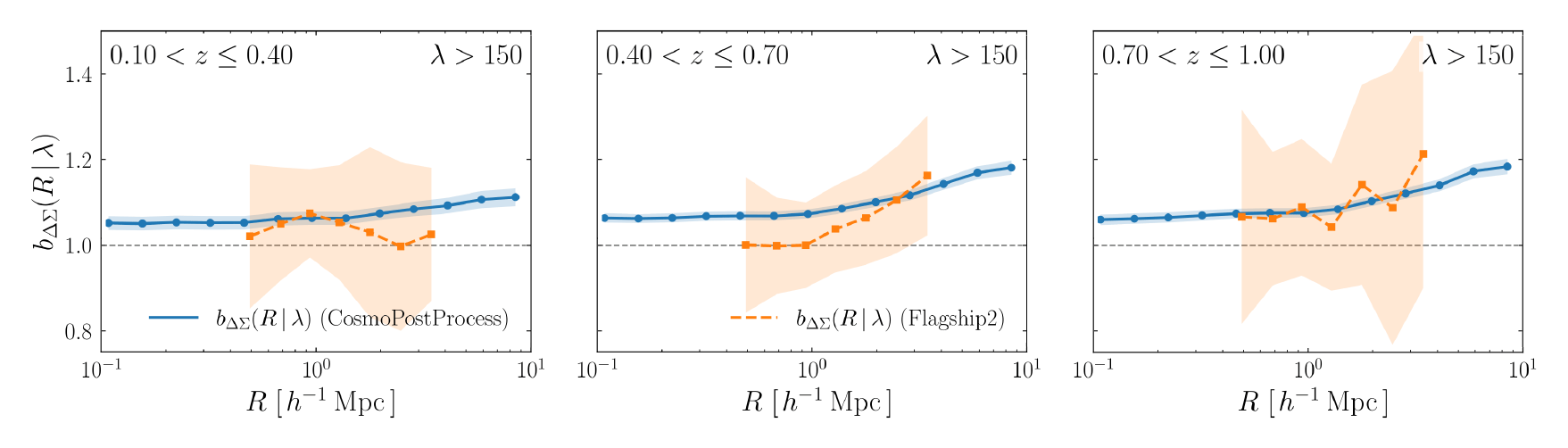}
    \caption{\label{fig:DelatSigmaCPPvsFS2}Comparison of the selection bias on $\Delta\Sigma$ profiles for the richest objects. In blue the bias calculated with \texttt{CosmoPostProcess} using the fiducial \texttt{PICCOLO} boxes as input for all redshift bins, in orange the results from \texttt{Flagship2} stacking the profiles obtained with \texttt{COMB-CL} according to \texttt{RICH-CL} richness, and the bands around making the $1\sigma$ uncertainty estimated via jackknife.}
\end{figure*}

\begin{table*}
\caption{HOD priors and posterior constraints.}
\label{tab:hod_priors}
\centering
\apptabletune
\begin{tabularx}{\linewidth}{@{} l X c c @{}}
\toprule
Parameter & Meaning & Prior range & Posterior (68\,\% CL) \\
\midrule
$\logten{(M_{1,\mathrm{sat}}\,h/M_{\odot})}$ & Characteristic mass scale for satellites & $[10,\,16]$
& $12.191_{-0.008}^{+0.008}$ \\
$\alpha$ & Slope of the satellite term & $[0.5,\,1.5]$
& $0.879_{-0.004}^{+0.004}$ \\
$\epsilon$ & Index of the redshift evolution & $[-2,\,5]$
& $0.955_{-0.013}^{+0.013}$ \\
$\sigma_{\rm intr}$ & Intrinsic fractional scatter & $[0,\,0.3]$
& $0.208_{-0.003}^{+0.003}$ \\
$\logten{(M_{\min,\mathrm{sat}}\,h/M_{\odot})}$ & Minimum mass scale & $[11,\,13]$
& $11.073_{-0.056}^{+0.062}$ \\
\bottomrule
\end{tabularx}
\tablefoot{Mass-related parameters are sampled in base-ten logarithmic space, with $M_{1,\mathrm{sat}}$ and $M_{\min,\mathrm{sat}}$ expressed in units of $h^{-1}\,M_{\odot}$. The physical constraint $10 \le M_{1,\mathrm{sat}}/M_{\min,\mathrm{sat}} \le 100$ is imposed during sampling as a likelihood veto to avoid unphysical regimes.}
\end{table*}

\paragraph{2. Membership emulator:}
We train a compact fully connected emulator for $P_{\mathrm{mem}}$ on the same eight tiles. The inputs are $\Delta z$, $\Delta r/r_{\mathrm{cl}}$, $z_{\mathrm{cl}}$, and $f_{\mathrm{bkg}}$ (the same quantities used in \cref{sc:membership}); the architecture has three dense layers with Leaky ReLU and dropout, followed by a sigmoid output. Targets and training tuples are built by matching halos to detections on the eight tiles with the standard membership-based association.

Hyperparameters are selected with \texttt{Optuna} \citep{Optuna}. To interpret the tuning results, we apply functional ANOVA, which decomposes the variance of a surrogate model of the loss over the hyperparameter space into additive contributions from each hyperparameter. A functional ANOVA \citep{Hutter2014,2017arXiv171004725V} indicates that the hidden size contributes $<1\,\%$ of the loss variance; we therefore reduce it from $756$ to $300$ to accelerate training with negligible impact. We also verify that $P_{\mathrm{mem}}$ is insensitive to the hidden size provided it is at least 100 (see \cref{fig:relevance} for input relevance).

\begin{table}[H]
\centering
\begin{threeparttable}
\caption{Training hyperparameters from \texttt{Optuna}}
\label{tab:hyperparams}
\begin{tabularx}{\columnwidth}{@{} l >{\centering\arraybackslash}X @{}}
\toprule
Hyperparameter & Value \\
\midrule
Batch size           & $128$ \\
Learning rate        & $2 \times 10^{-4}$ \\
Drop rate            & $14\,\%$ \\
Number of layers     & $3$ \\
Hidden size          & $300$\tnote{*} \\
Max number of epochs & $191$ \\
\bottomrule
\end{tabularx}
\begin{tablenotes}
\footnotesize
\item[*] Largest value that keeps training under one hour.
\end{tablenotes}
\end{threeparttable}
\end{table}

In \cref{fig:membership-panels}, we show that the emulator reproduces the dependence of $P_{\mathrm{mem}}$ on $\Delta z$ and projected separation for an individual object and, at the catalogue level, preserves the mass-richness relation with residuals consistent with zero within the uncertainty band.

To assess which inputs are most informative, we carry out a consensus relevance test that combines saliency-based importance, integrated gradients, and conditional permutation importance. The scores are normalised and aggregated to produce a single ranking, evaluated both globally and in bins of $z_{\mathrm{cl}}$ and $\Delta r/r_{\mathrm{cl}}$. As anticipated in \cref{sc:membership}, \cref{fig:relevance} identifies $\Delta z$, $\Delta R/R_{\mathrm{cl}}$, and $z_{\mathrm{cl}}$ as the leading predictors, with $f_{\mathrm{bkg}}$ contributing little. This supports the simplified proxy used for $f_{\mathrm{bkg}}$ and explains why the emulator retains the calibrated mass--richness relation. Nevertheless, runs where $f_{\mathrm{bkg}}$ was omitted showed larger residuals compared to \cref{fig:membership-panels}, so we kept it in the input vector. As a final consistency test, \cref{fig:DelatSigmaCPPvsFS2} compares the selection bias computed with \texttt{CosmoPostProcess} on stacked $\Delta\Sigma$ profiles to that extracted from the \texttt{RICH-CL} and \texttt{COMB-CL} runs on \texttt{Flagship2}. The $\Delta\Sigma(r)$ profiles are obtained from $\Sigma(r)$ via \cref{eq:sigma_conv}, and the bias is then computed using \cref{eq:def_bias_results}.

\end{appendix}

\label{LastPage}
\end{document}